\documentclass[a4paper, oneside]{article}
\usepackage[T2A]{fontenc} 
\usepackage[cp1251]{inputenc} 
\usepackage[tbtags]{amsmath}
\usepackage{amsfonts,amssymb,mathrsfs,amscd,comment}
\usepackage{amsthm}
\usepackage{graphicx}
\usepackage{cite}
\usepackage{hyperref}
\hypersetup{
	colorlinks=true,
	linkcolor=blue,
	filecolor=blue,
	urlcolor=blue,
}

\usepackage{bm}
\usepackage{bbm}
\usepackage[usenames]{color}

\makeatletter \@addtoreset{equation}{section}
\renewcommand\theequation
{\ifnum \c@section>\z@ \thesection.\fi \@arabic\c@equation}
\renewcommand{\@biblabel}[1]{#1.}
\makeatother
\newtheorem{definition}{Definition}[section]
\newtheorem{remark}{Remark}[section]
\newtheorem{lemma}{Lemma}[section]
\newtheorem{theorem}{Theorem}[section] 
 
\newcommand{\Tr}{{\rm Tr}}
\begin{document}
	
\title{\bf\large\MakeUppercase{Krotov Method for Optimal Control in Closed Quantum Systems}}
 
\author{O.\,V.~Morzhin$^1$ and A.\,N.~Pechen$^2$}
\date{April 30, 2019}

\vspace{1cm}

\maketitle

$^1$ Steklov Mathematical Institute of Russian Academy of Sciences (Moscow), 
Senior Researcher. {\it E-mail}: \href{mailto:morzhin.oleg@yandex.ru}{morzhin.oleg@yandex.ru}

$^2$ Steklov Mathematical Institute of Russian Academy of Sciences (Moscow), Head of the 
Department of Mathematical Methods for Quantum Technologies;
National University of Science and Technology ``MISIS'' (Moscow), Department of Mathematics, Leading Researcher.
{\it E-mail}: \href{mailto:apechen@gmail.com}{apechen@gmail.com}  (corresponding author)
\vspace{.5cm}

\makeatletter
\renewcommand{\@makefnmark}{}
\makeatother
\footnotetext{Sections 1, 3, 4, and 5 were preformed by both authors
	in Steklov Mathematical Institute of Russian Academy of Sciences
	within the project No.~17-11-01388 of the Russian Science Foundation.
	Subsections~2.1,~2.2,~2.3 were performed by both authors within government project
	of Steklov Mathematical Institute of Russian Academy of Sciences.
	Subsection~2.4 and Section~6 were performed by the first author
	within government project of Steklov Mathematical Institute of Russian Academy of Sciences and by the second author in MISiS within the Project No.~1.669.2016/FPM of the Ministry of Science and Higher Education
	of the Russian Federation. Subsections~2.5~and~2.6 were performed by the second author
	in MISiS within the Project No.~1.669.2016/FPM of the Ministry of Science and Higher Education
	of the Russian Federation.}

\rightline{\it Dedicated to the bright memory of Prof. Vadim F. Krotov}

\vspace{0.7cm}

\leftline {AMS 2010 Mathematics Subject Classification. Primary 81Q93; Secondary 49Mxx,} 
\leftline{35Q40, 93C15}

\vspace{0.3cm}
 
\leftline {UDC 517.958}
 
\begin{abstract}
Mathematical problems of optimal control in quantum systems attract high interest 
in connection with fundamental questions and existing and prospective applications. 
An important problem is the development of methods for constructing controls 
for quantum systems. One of the commonly used methods is the Krotov method 
initially proposed beyond quantum control
in the articles by V.F.~Krotov and I.N.~Feldman (1978, 1983). The method was used to develop a novel approach for finding optimal controls for quantum systems in [D.J. Tannor, V. Kazakov, V. Orlov, In: Time-Dependent Quantum
Molecular Dynamics, Boston, Springer, 347--360 (1992)] and
[J.~Soml\'{o}i, V.A.~Kazakov, D.J.~Tannor, Chem. Phys., 172:1, 85--98 (1993)], and in many works of various scientists, as described in details in this review. 
The review discusses mathematical aspects of this method 
for optimal control of closed quantum systems. It outlines various modifications
with respect to defining the improvement function (which in most cases is linear or linear-quadratic), 
constraints on control spectrum and on the states of a quantum system, regularizers, etc.
The review describes applications of the Krotov method to control of molecular 
dynamics, manipulation of Bose-Einstein condensate, quantum gate generation.
We also discuss comparison with GRAPE (GRadient Ascent Pulse Engineering), 
CRAB (Chopped Random-Basis), the Zhu---Rabitz and the Maday---Turinici methods. 

Bibliography: 154 titles. 
	
\vspace{0.3cm} {\bf Key words}: quantum control, coherent control, Krotov method, closed quantum systems, quantum technology.
	
\end{abstract}

\tableofcontents

\section{Introduction}

Optimal control theory considers optimal control problems (OCP) for 
dynamical systems described by ordinary differential equations (ODE),
partial differential equation (PDE), etc. Optimal control theory began to develop
in the middle of the 20th century starting from such fundamental results
as Pontryagin maximum principle developed by L.S.~Pontryagin, V.G.~Boltyansky, 
R.W.~Gamkrelidze, and E.F.~Mishchenko~\cite{Pontryagin_et_al_book_1962} 
and R.~Bellman optimality principle~\cite{Bellman_book_1957}. Now optimal control theory is one of the 
leading areas of mathematics with numerous applications in flight dynamics, 
robotics, economics, quantum technologies, etc.   

A branch of applications of optimal control theory is quantum control --- an advanced 
interdisciplinary direction devoted to studying control of quantum systems, 
i.e. of individual electrons, atoms, molecules, photons. Control is implemented 
by shaped laser pulses, modulating electromagnetic radiation, engineered 
environment, or other actions. Modern technology allows to make laser pulses 
of an ultra-small duration on the order of femtoseconds ($10^{-15}$~sec) 
and attoseconds ($10^{-18}$~sec). The high interest to mathematical problems 
of quantum control is motivated by progress in experiments on manipulation 
by quantum systems. In 1997, the Nobel Prize in Physics was awarded 
to S.~Chu, C.~Cohen-Tannoudji, and W.D.~Phillips 
``for development of methods to cool and trap atoms with 
laser light''$^1$\footnote{$^1$\url{https://www.nobelprize.org/prizes/physics/1997/summary/}}.  
In 2001, the Nobel Prize in Physics was awarded to E.A.~Cornell, W.~Ketterle, 
and C.~Wieman, who in 1995 made Bose-Einstein condensate in the
laboratories$^2$\footnote{$^2$\url{https://www.nobelprize.org/prizes/physics/2001/summary/}}.
In 2012, the Nobel Prize in Physics was awarded to S.~Haroche and D.J.~Wineland 
``for ground-breaking experimental methods that enable measuring and manipulation 
of individual quantum systems''$^3$\footnote{$^3$\url{https://www.nobelprize.org/prizes/physics/2012/summary/}}. 
The Nobel Prize in Physics 2018 was awarded ``for groundbreaking inventions in the field 
of laser physics'' to A.~Ashkin ``for the optical tweezers and their application 
to biological systems'', G.~Mourou and D.~Strickland ``for their method 
of generating high-intensity, ultra-short optical 
pulses''$^4$\footnote{$^4$\url{https://www.nobelprize.org/prizes/physics/2018/summary/}}.

Mathematical problems of quantum control are actively studied since the 1980s. 
Quantum control is important for existing and prospective technologies, including  
control of atomic and molecular dynamics (for example, laser cooling of molecules);
manipulation of Bose-Einstein condensate; implementation of quantum computing  
(e.g., for optimal generation of qubits and quantum gates), designing atomic chips,
laser-assisted isotope separation, laser chemistry, nuclear magnetic resonance, 
dynamic nuclear polarization, and magnetic resonance imaging. The theoretical 
and experimental results on quantum control are summarized in the books and reviews by: 
A.G.~Butkovskiy and Yu.I.~Samoilenko (1984)~\cite{Butkovsky_Samoilenko_book_1984_1990}; 
I.V.~Krasnov, N.Ya.~Shaparev, and I.M.~Shkedov (1989)~\cite{Krasnov_Shaparev_Shkedov_book_1989}; 
S.A.~Rice and M.~Zhao (2000)~\cite{Rice_Zhao_book_2000}; 
A.D.~Bandrauk, M.C.~Delfour, and C.~Le~Bris as editors 
(2003)~\cite{Bandrauk_Delfour_Bris_Edt__book_AMS_2003}; 
D.~D’Alessandro (2003)~\cite{DAlessandro_Directions_2003}; 
P.~Brumer and M.~Shapiro (2003)~\cite{Brumer_Shapiro_book_2003}; 
D.J.~Tannor (2007)~\cite{TannorD_book_IntroQM_2007}; 
V.S.~Letokhov (2007)~\cite{Letokhov_book_2007}; 
D.~D'Alessandro (2007)~\cite{DAlessandro_book_2007}; 
A.L.~Fradkov (2007)~\cite{Fradkov_book_2007}; 
C.~Brif, R.~Chakrabarti, and H.A.~Rabitz (2010)~\cite{Brif_Chakrabarti_Rabitz_article_2010}; 
D.~Dong and I.R.~Petersen (2010)~\cite{Dong_Petersen_2010}; 
H.M.~Wiseman and G.J.~Milburn (2010)~\cite{Wiseman_Milburn_book_2010}; 
C.~Altafini and F.~Ticozzi (2012)~\cite{Altafini_Ticozzi_article_IEEE_2012}; 
B.~Bonnard and D.~Sugny (2012)~\cite{Bonnard_Sugny_2012};  
J.E.~Gough (2012)~\cite{Gough_article_2012}; 
S.~Cong (2014)~\cite{CongS_book_2014}; 
W.~Dong, R.~Wu, X.~Yuan, C.~Li, T.-J.~Tarn (2015)~\cite{Dong_Wu_Yuan_Li_Tarn_2015}; 
S.J.~Glaser, U.~Boscain, T.~Calarco, C.P.~Koch, W.~K\"{o}ckenberger, R.~Kosloff, 
I.~Kuprov, B.~Luy, S.~Schirmer, T.~Schulte-Herbr\"{u}ggen, D.~Sugny, 
F.K.~Wilhelm (2015)~\cite{Glaser_Boscain_Calarco_et_al_2015};
C.P.~Koch (2016)~\cite{CPKoch_2016_OpenQS}; 
A.~Borz\`\i, G.~Ciaramella, and M.~Sprengel (2017)~\cite{Borzi_book_2017}.

As in optimal control theory, quantum optimal control considers {\it program} control, 
when control function depends on time 
and {\it feedback control}, when control function depends on time and on the system state. 
The article~\cite{Belavkin_article_AiT_1983} considers quantum control 
for discrete-observable quantum systems 
which between observations evolve according to the Schr\"{o}dinger equation. 
The article~\cite{Wiseman_Milburn_1993} contains 
the theory of real-time feedback control for physical models of quantum optics.
Methods used for program quantum control include genetic and evolutionary 
algorithms~\cite{Judson_Rabitz_1992, Pechen_Rabitz_2006},
Pontryagin maximum principle and geometric control, e.g., for minimal-time quantum 
control~\cite{Boscain_Charlot_Gauthier_et_al_2002, Boscain_Chambrion_Charlot_2005, 
Assemat_Lapert_Sugny_Glaser_2013, Boscain_Gronberg_Long_Rabitz_2014, 
Salamon_Hoffmann_Tsirlin_2012, Carlini_Hosoya_Koike_Okudaira_2006, 
Boscain_Mason_2006, Romano_2014, Albertini_DAlessandro_2016}, 
gradient flows~\cite{Schulte-Herbruggen_Glaser_Dirr_Helmke_2010}, 
GRAPE (GRadient Ascent Pulse Engineering)~\cite{Khaneja_Reiss_Kehlet_SchulteHerbruggen_Glaser_2005, 
Jager_Reich_Goerz_et_al_2014}, 
CRAB (Chopped Random-Basis)~\cite{Caneva_Calarco_Montangero_2011}, 
Zhu---Rabitz~\cite{Zhu_Rabitz_article_1998} and 
Maday---Turinici~\cite{Maday_Turinici_New_formulations__2003} methods, 
dynamic programming~\cite{Gough_Belavkin_Smolyanov_2005, Pechen_Trushechkin_2015},
time-parallelized algorithms~\cite{Riahi_Salomon_Glaser_Sugny_2016}, 
speed gradient method~\cite{Ananevskii_Fradkov_2005, Pechen_SpeedGrad_2016}, 
Ho---Rabitz TBQCP (Two-point Boundary-value Quantum Control Paradigm) method
\cite{Ho_Rabitz_PhysRev_2010}, gradient projection method~\cite{Morzhin_Pechen_2019}, etc. 
Machine learning is also used in research on quantum systems and technologies: 
reinforcement learning is applied for constructing 
quantum controls~\cite{Dong_Chen_Tarn_Pechen_Rabitz_2008,  
Niu_Boixo_Smelyanskiy_Neven_arXiv_2018};
auto-encoder is applied for reducing the dimensionality of data describing quantum 
dynamics~\cite{Zauleck_VivieRiedle_2018}; the restricted Boltzmann machine is used 
for quantum tomography~\cite{Palittapongarnpim_Sanders_NaturePhysics_2018, 
Torlai_Mazzola_Carrasquilla_et_al_NaturePhysics_2018}. Quantum machine learning is considered~\cite{quant_machine_learning_Nature_Biamonte_2017}.

One of the commonly used methods for constructing program controls for quantum systems
is the Krotov$^5$\footnote{$^5$Vadim 
F.~Krotov (1932 -- 2015) --- famous scientist, Honored worker of science of Russia, an author of the 
fundamental results in optimal control theory. In 1962 V.F.~Krotov defended the candidate dissertation ``A new method 
of variational calculus and some its applications'' in Steklov Mathematical Institute of the USSR Academy 
of Sciences, and in 1963 he defended the doctoral dissertation ``Some new methods of 
variational calculus and their application to the flight dynamics'' at Moscow Aviation 
Institute. He was the head of departments in Moscow Aviation Technological Institute and Moscow 
Institute of Economics and Statistics. In 1982 --- 2015 he worked in the Institute 
of Control Sciences of Russian Academy of Sciences (ICS RAS) as the head of the 45th 
laboratory, which now has his name. The website of ICS RAS has the page devoted to V.F.~Krotov: 
\url{http://www.ipu.ru/node/32378} (In Russian.)} 
method. This method was initially proposed beyond quantum control
by V.F.~Krotov and I.N.~Feldman~\cite{Krotov_Feldman_1978, Krotov_Feldman_IzvAN_article_1983} 
(1978, 1983) based on the Krotov optimality 
principle~\cite{Krotov_Gurman_book_1973, KrotovVF_book_NewYork_1996} and further developed
by A.I.~Konnov and V.F.~Krotov~\cite{Konnov_Krotov_article_1999} (1999). 
An example with control of an open (i.e. interacting with the environment) quantum system was analyzed by V.A.~Kazakov and V.F.~Krotov in 1987 in the article~\cite{Kazakov_Krotov_AiT_article_1987} 
(also in~\cite{Krotov_1989}). Crucial step in development to quantum systems was done in 1992--1993, when D.J.~Tannor and coauthors used the 1st order Krotov method to develop a general approach for finding optimal controls 
for quantum systems~\cite{Tannor_Kazakov_Orlov_1992, Somloi_Kazakov_Tannor_article_1993}.
In 2002, the 2nd order Krotov 
method~\cite{Krotov_Feldman_IzvAN_article_1983, Konnov_Krotov_article_1999}
was adapted by S.E.~Sklarz and D.J.~Tannor for modeling of optimal control for Bose-Einstein condensate, whose
dynamics is defined in terms of a controlled Gross--Pitaevskii equation~\cite{Sklarz_Tannor_article_2002}. In 2008, J.P. Palao, R. Kosloff, and C.P. Koch developed the method to optimal control for the problem of obtaining an objective in a subspace of the Hilbert space while avoiding population transfer to other subspaces~\cite{Palao_Kosloff_Koch_article_2008}. The Krotov method was applied with various modifications and taking into account 
specific details of quantum optimal control problems, for atomic and molecular 
dynamics~\cite{Szakacs_Amstrup_Gross_Kosloff_Rabitz_Lorincz_1994,
Sola_Santamaria_Tannor_article_1998, Bartana_Kosloff_Tannor_2001, Maday_Turinici_New_formulations__2003,
Koch_Palao_Kosloff_et_al_article_2004, TannorD_book_IntroQM_2007, Palao_Kosloff_Koch_article_2008, 
Caneva_Murphy_Calarco_et_al_2009, Ndong_Koch_article_2010,
Kumar_Malinovskaya_2011, Eitan_Mundt_Tannor_article_2011, 
Palao_Reich_Koch_PhysRev_2013, Trushkova_SaratovUniv_article_2013, 
Ndong_Koch_Sugny_article_2014, CongS_book_2014};
qubits, quantum gates, quantum 
networks~\cite{Palao_Kosloff_2002, Palao_Kosloff_article_2003, Treutlein_Microwave_article_2006, Chiara_Calarco_et_al_article_2008, Gollub_Kowalewski_VivieRiedle_article_2008, Koike_Okudaira_2010,
Singer_et_al_article_Colloquim_2010, Goerz_Calarco_Koch_2011,  Muller_Reich_Murphy_et_al_article_2011, Reich_Ndong_Koch_article_2012, Goerz_Gualdi_Reich_Koch_et_al_article_2015,
Goerz_Whaley_Koch_article_2015, Goerz_dissertation_Kassel_2015, Goerz_Motzoi_Whaley_Koch_2017,
Basilewitsch_Schmidt_Sugny_et_al_2017, Basilewitsch_Marder_Koch_2018, Goerz_Jacobs_article_2018}; 
manipulation of Bose--Einstein condensate~\cite{Sklarz_Tannor_article_2002, 
Jager_Reich_Goerz_et_al_2014, JagerG_thesis_2015, Sorensen_Aranburu_Heinzel_Sherson_2018};
nuclear magnetic resonance, dynamic nuclear polarization, and magnetic resonance 
imaging~\cite{Maximov_et_al_article_2008, Maximov_Salomon_Turinici_Nielsen_article_2010, 
Vinding_Maximov_et_al__article_2012}. The program 
tools$^6$\footnote{$^{6}$ \url{https://www.qdyn-library.net/} and \url{https://github.com/qucontrol/}}  
were developed in Fortran and Python languages 
by the research group of C.P.~Koch and coauthors. They include
the implementation of both first and second order Krotov methods, 
also including constraints on quantum states and on the control spectrum.
 
There are regularly defended dissertations on quantum optimal control which use the iterative Krotov 
method. For example, dissertation~\cite{Reich_dissertation_Kassel_2015} by M.D.~Reich (2015)
is devoted to foundations of quantum optimal control for open quantum systems,
dissertation~\cite{Goerz_dissertation_Kassel_2015} by M.H.~Goerz (2015) to
optimization of robust quantum gates for open quantum systems, 
and dissertation~\cite{JagerG_thesis_2015} by G.~J\"{a}ger (2015) 
to optimal control of Bose--Einstein condensate.

This review outlines mathematical results and applications of the Krotov method for closed, 
i.e., not interacting with the environment, quantum systems evolved under coherent control 
entering in the perturbed part of the Hamiltonian. Open quantum systems will be considered elsewhere.  
The review does not intend to be a complete overview of all modifications of the method 
for quantum systems, only basic results are provided. We use the name ``Krotov method'' 
following the tradition established in quantum optimal control. 
At the same time, the publications~\cite{Krotov_Feldman_1978, 
Krotov_Feldman_IzvAN_article_1983} (V.F.~Krotov, I.N.~Feldman) and~\cite{Konnov_Krotov_article_1999} 
(A.I.~Konnov, V.F.~Krotov) were in co-authorship that can be reflected in the name of the method. 

The structure of the review is the following. Section~2 provides formulation of various OCPs for 
closed quantum systems. It also includes a brief discussion of controllability and control landscapes
for closed quantum systems. Section~3 is devoted to the Krotov method for OCPs with real-valued states. 
Section~4 discusses the first-order Krotov method, Zhu-Rabitz, and Madey-Turinici methods, 
for systems governed by Schr\"odinger and Liouville--von Neumann equations. 
Section~5 considers the generation of target unitary transformations and control of ensembles 
of quantum states. Section~6 discusses applications of Krotov and GRAPE methods 
for manipulation by Bose-Einstein condensate whose dynamics is governed by the controlled 
Gross--Pitaevski equation. The Conclusions section summarizes the review. 
 
\section{Classes of optimal control problems for closed quantum systems}

In this section we consider formulation of various optimal control problems for closed quantum systems. Formulation of an optimal control problem includes defining dynamical equation, 
space of controls, cost functional to be minimized, and constraints on controls and states.

Each quantum system is associated with some Hilbert space ${\cal H}$:
for example, ${\cal H}=\mathbb C^n$
for a system with $n$ states; Hilbert space for a quantum particle moving in some domain $\Omega \in \mathbb R^d$ 
is ${\cal H}=L^2(\Omega;\mathbb C)$. Pure states of the system 
are unit norm vectors $\psi \in{\cal H}$, $\|\psi\|^2=1$. 
Some models consider the Hilbert space $L^2(\Omega; \mathbb{C}^M)$, $\Omega \subseteq \mathbb{R}^q$. 
Here the state is a vector function $\psi = (\psi_1, \dots, \psi_M)$. 
Most general states of a quantum system are described by density matrices.
A density matrix is a self-adjoint trace class operator $\rho$ 
in the Hilbert space $\cal H$ which satisfies the conditions
$\rho \geq 0$, ${\rm Tr} \rho = 1$. 

The dynamics of the system state in the absence of controls is determined by free system 
Hamiltonian ${\bf H}_0(t)$, which is a self-adjoint operator in $\cal H$. 
There is a dynamical invariant: $\|\psi(t)\|^2_{\mathcal{H}} = 1$. 
In the case of an electron in $\Omega \subset \mathbb{R}^3$
this invariant means that the probability of detecting an electron 
in $\Omega$ at any time $t$ is~1. In the case of a qubit,
$n=2$ and $\psi(t) = \alpha(t) |0\rangle + \beta(t) |1\rangle 
\in \mathbb{C}^2$ 
(here $|0\rangle = \begin{pmatrix}
1 \\
0
\end{pmatrix}$
and $|1 \rangle = 
\begin{pmatrix}
0 \\
1
\end{pmatrix}$ 
in the Dirac notation), $|\alpha(t)|^2 + |\beta(t)|^2 = 1$, what provides 
$\| \psi(t) \|^2_{\mathbb{C}^2} = 1$. The value of $|\alpha(t)|^2$ is the probability 
of finding the system in the pure state $|0\rangle$, and the value $|\beta(t) |^2$ is 
the probability of obtaining $|1\rangle$. We consider $m$
controls, so that interaction with control $u_l$ ($l= \overline{1, m}$) is described 
by a self-adjoint interaction Hamiltonian ${\bf H}_l$. An observable $O$ 
of the system is a self-adjoint operator in $\cal H$. If the system is in
pure state $\psi$, then average observed value of $O$ is 
$\langle O \rangle = \langle \psi, O \psi \rangle$. If the system is in
the state with density matrix $\rho$, then $\langle O \rangle = \Tr (\rho O$).

\subsection{Schr\"{o}dinger equation with controlled Hamiltonian and cost criteria} 

Based on, in particular~\cite{Butkovsky_Samoilenko_book_1984_1990, DAlessandro_book_2007, 
Borzi_book_2017}, we formulate the following definition.

\begin{definition}
A system with Hilbert space $\cal H$ governed by the Schr\"{o}dinger equation with a linearly 
controlled Hamiltonian ${\bf H}$ is a quantum system whose state $ \psi(t) \in \mathcal{H}$ 
satisfies 
\begin{eqnarray} 
\dfrac{d\psi(t)}{dt} &=& -\dfrac{i}{\hbar} {\bf H}[u(t)] \psi(t), \qquad
\psi(0) = \psi_0,  \label{ch2_f1} 
\end{eqnarray}
where
\begin{eqnarray} 
{\bf H}[u(t)] &=& {\bf H}_0 + \sum\limits_{l=1}^m {\bf H}_l u_l(t), \label{ch2_f2} \\
u \in \mathcal{U} &=& PC([0,T]; Q), \qquad Q \subseteq \mathbb{R}^m. \label{ch2_f3}
\end{eqnarray} 
The initial state $\psi_0 \in \mathcal{H}$ and the final time moment $T$ are fixed; 
$\hbar$ is the Planck constant; $u(t) = (u_l(t))_{l = \overline{1, m}}$, is a vector control 
function; $\mathcal{U}$ is the class of admissible controls; $Q$ is a convex set; 
the Hamiltonian ${\bf H}$ is a self-adjoint operator in the Hilbert space 
${\cal H}$; the operator ${\bf H}_0$ is the unperturbed part of ${\bf H}$; the
operator ${\bf H}_l$ characterizes the interaction 
of the quantum system with external control $u_l(t)$.
\end{definition}

As the class of admissible controls $\mathcal{U}$, we consider 
the space of piecewise-continuous functions $PC([0,T]; Q)$. 
$T$ can be orders of femtoseconds, picoseconds, etc.

The following theorem is a corollary of the Caratheodory's theorem on 
the existence and uniqueness of the solution of a differential equation 
with discontinuous right-hand side (r.h.s.).

\begin{theorem}
Let $\mathcal{H} = \mathbb{C}^n$. If $u \in L^1([0,T]; Q)$, 
then the solution of the equation (\ref{ch2_f1}) 
exists in the class of absolutely continuous functions on the 
interval $[0, T]$ and is unique. 
\end{theorem}

Since $PC([0,T]; Q) \subset L^1([0,T]; Q)$, then for $\mathcal{H} = \mathbb{C}^n$ for each piecewise 
continuous control $u$ the solution of (\ref{ch2_f1}) exists and is unique.  

\begin{definition} 
The process $v = (\psi(t), u(t)~|~ t \in [0,T])$ is called admissible, 
if it satisfies the conditions (\ref{ch2_f1}) --- (\ref{ch2_f3}). 
\end{definition}

Denote by $\mathcal{D}$ the set of all admissible processes.

\begin{remark}
In the general case, control $u$ can be a complex-valued 
function~\cite{Kosloff_Rice_Gaspard_Tersigni_Tannor_article_1989, 
Tannor_Kazakov_Orlov_1992, TannorD_book_IntroQM_2007}. 
Further we will consider real-valued controls as specified in (\ref{ch2_f3}). In the general case,  
Hamiltonian ${\bf H}$ can depend nonlinearly on the control 
$u$~\cite{Lapert_Tehini_2008}. Further we will consider only the linear 
case commonly used.  
\end{remark}  

\begin{definition}
For the system (\ref{ch2_f1}) --- (\ref{ch2_f3}), the following OCP
is called the problem of maximizing the mean 
$\left\langle O \right\rangle = \langle \psi(T), O \psi(T) \rangle$ for a Hermitian operator $O$: 
\begin{eqnarray} 
J(v) &=& \mathcal{F}_O(\psi(T)) + \lambda_u \int\limits_0^T \dfrac{\| u(t) \|^2} {S(t)} dt + \nonumber \\
&& + \lambda_{\psi} \int\limits_0^T \langle \psi(t), D(t) \psi(t) \rangle dt 
\to \min\limits_{v \in \mathcal{D}},  \qquad J : \mathcal{D} \to \mathbb{R}, 
\label{ch2_f4}
\end{eqnarray}
where the cost criterion has the terminant $\mathcal{F}_O = -\left\langle O \right\rangle$,
parameters $\lambda_u \geq 0$, $\lambda_{\psi} \leq 0$; operator $D(t)$ ($t \in [0,T]$) 
is self-adjoint positive semi-defined; and $S$ is some shape function. 
\end{definition}

For the first term in the r.h.s. of (\ref{ch2_f4}), the condition $O \geq 0$ can be considered, what
is essential for sequential improvements of controls using the methods
given in the articles~\cite{Tannor_Kazakov_Orlov_1992} (D.J.~Tannor, V.A.~Kazakov,~V.N.~Orlov, 1992), 
\cite{Zhu_Rabitz_article_1998} (W.~Zhu, H.A.~Rabitz, 1998), 
\cite{Maday_Turinici_New_formulations__2003} (Y.~Maday, G.~Turinici, 2003) 
for OCPs of the type (\ref{ch2_f1}) --- (\ref{ch2_f4})
with $S(t) \equiv 1$ and $\lambda_{\psi} = 0$. 

The second term in the r.h.s. of (\ref{ch2_f4}) can be understood 
as a condition of energy minimization, which is important for avoiding  
non-physical values of coherent control or as a possibility to simplify 
application of optimization methods. This term can significantly change 
the original OCP by affecting values of the terminant 
$\mathcal{F}(\psi(T))$ (that is why adjusting $\lambda_u > 0$ is required).
Along with $S(t) \equiv 1$~\cite{Tannor_Kazakov_Orlov_1992, 
Somloi_Kazakov_Tannor_article_1993, Zhu_Rabitz_article_1998, 
KrotovVF_Dokl_AN_2008, KrotovVF_AiT_article_2009}, non-constant function $S$ can be used 
to make smooth start of a laser field 
from $t = 0$ and its smooth shutdown to the moment $t = T$. 
Examples include $S(t) = \sin^2(\pi t/T)$~\cite{Sundermann_VivieRiedle_article_1999}
(K.~Sundermann, R.~de~Vivie-Riedle, 1999),
$S(t) = \exp \left[ -32(t/T-1/2)^2 \right]$~\cite[p. 5]{Palao_Kosloff_Koch_article_2008} 
(J.P.~Palao, R.~Kosloff, C.P.~Koch, 2008), etc. 

The third term with the operator $D(t)$ allows to specify forbidden or allowed subspaces in $\mathcal{H}$~\cite{Palao_Kosloff_Koch_article_2008}. 
If the operator $D(t)$ is positive semi-definite, then it describes an allowed subspace. 
The role of this term is discussed in details in~\cite{Palao_Kosloff_Koch_article_2008}. 
The requirement $\lambda_{\psi} \leq 0$ is chosen to satisfy a special condition for non-decreasing 
of the cost functional using the 1st order Krotov method considered in Section~4.

Moreover, a constraint of the type $\int\limits_0^T u^2(t) dt \{=, \leq\} E$ ($E$ is some fixed value) or,  
in the complex-valued case, of the type 
$\int\limits_0^T u(t) u^{\ast}(t)dt \{=, \leq\} E$~\cite{Kazakov_Krotov_AiT_article_1987, 
Kosloff_Rice_Gaspard_Tersigni_Tannor_article_1989, 
KrotovVF_AiT_article_2009} can be used. In this review we do not consider such constraints.  
 
Along with $\mathcal{F}_O$ one considers terminants 
with a given target state $\psi_{\rm target} \in \mathcal{H}$
(e.g., \cite[p. 20]{Butkovsky_Samoilenko_book_1984_1990}, \cite[p. 4]{Schirmer_Fouquieres_2011}):
\begin{eqnarray}
\mathcal{F}_{\psi_{\rm target}} &=& 1 - \left| \langle \psi(T), \psi_{\rm target} \rangle \right|^2, \label{ch2_f5}\\
\mathcal{F}_{\psi_{\rm target}} &=& 1 - {\rm Re} \left\langle \psi(T), \psi_{\rm target} \right\rangle = 
\dfrac{1}{2} \left\| \psi(T) - \psi_{\rm target} \right\|^2. 
\label{ch2_f6}
\end{eqnarray} 
Minimization of (\ref{ch2_f5}) means maximization of the probability that the final 
state $\psi(T) \in \mathcal{H}$ is the target state $\psi_{\rm target}$.  

For the criterion (\ref{ch2_f4}), the terminant
(\ref{ch2_f5}) represents a particular case, where $O = P_{\rm target}$ is the projector
onto the target state $\psi_{\rm target}$. For $M = 1$, $x \in \mathbb{R}^3$ one considers as $\psi_{\rm target}$, for example,  
a sum of Gaussian functions \cite[с. 6]{Eitan_Mundt_Tannor_article_2011}:
\begin{eqnarray*}
\psi_{\rm target}(x) &=& 
A \left[ \exp\left( -\sum\limits_{j=1}^3 \alpha_j (x_j - x^{\alpha}_j)^2 \right) +
\exp\left( -\sum\limits_{j=1}^3 \beta_j (x_j - x^{\beta}_j)^2 \right) \right],
\end{eqnarray*} 
$\alpha_j$, $\beta_j > 0$. The following term   
\cite[p. 4]{Schirmer_Fouquieres_2011}, which describes optimization 
of the entire trajectory, is sometimes considered in the cost functional $J$:
\begin{eqnarray*}
\mathcal{F}_{\psi_{\rm target}(t)} &=&
\dfrac{1}{2} \int\limits_0^T \left\| \psi(t) - \psi_{\rm target}(t) \right\|^2 dt =
T - \int\limits_0^T {\rm Re} \left\langle \psi_{\rm target}(t), \psi(t) \right\rangle dt.
\end{eqnarray*}

\begin{definition}
For OCP (\ref{ch2_f1}) --- (\ref{ch2_f4}), an admissible
process $v^{\ast}$ is a solution if this process provides the global
minimum $J(v^{\ast}) = \min\limits_{v \in \mathcal{D}} J(v)$ of the cost functional $J$.   
\end{definition}

An OCP for the Schr\"{o}dinger equation in an infinite dimensional Hilbert space
in some cases can be reduced to an approximate OCP for a finite-dimensional system, and for the last OCP 
one can consider the corresponding OCP with real-valued states.
The articles~\cite{Boussaid_Caponigro_Chambrion_IEEE_2012, Trushkova_SaratovUniv_article_2013} 
consider the equation
\begin{eqnarray*}
i\dfrac{\partial \psi(\theta, t)}{\partial t} &=& 
\left( -\dfrac{\partial^2}{\partial \theta^2} + 
u(t) \cos\theta \right) \psi(\theta,t), 
\qquad \psi(0) = \psi_0,\quad \theta \in \Omega 
\end{eqnarray*}
describing rotations of a planar molecule, and the following 
approximate OCP:
\begin{eqnarray} 
\dfrac{dz(t)}{dt} &=& \left(A + u(t) B \right) z(t), \qquad z(0) = z_0, \qquad z(t) \in \mathbb{C}^n, \label{ch2_f7_add} \\
J(z,u) &=& |z_2(T)|^2 \to \min. 
\label{ch2_f7} 
\end{eqnarray} 
Here $\theta$ is the angle between the polarization direction and the molecular axis; 
$\Omega$ is a one-dimensional torus; $A$ and $B$ are ($n \times n$) matrices obtained with approximation
by the Galerkin method; $n = 22$.

\subsection{Liouville--von Neumann equation with control and cost criteria}

The evolution of the density matrix of a closed quantum system  
under the control is described by the Liouville--von Neumann equation 
with controlled Hamiltonian:
\begin{eqnarray}
\dfrac{d \rho}{dt} &=& -\dfrac{i}{\hbar} \Big[ {\bf H}[u(t)], \rho \Big],\qquad
\rho(0) = \rho_0, 
\label{ch2_f9} 
\end{eqnarray}
where $[,]$ is commutator ($[A,B] = AB - BA$); the initial density matrix $\rho_0$ is given. 

If a quantum system is open, that is, interacting with the environment 
(reservoir), then the evolution of the density matrix $\rho (t)$ under the 
influence of coherent control will not be unitary and 
can be described by the equation 
\begin{eqnarray}
\dfrac{d \rho}{dt} &=& -\dfrac{i}{\hbar} \Big[ {\bf H}[u(t)], \rho \Big] 
+ \mathcal{L}(\rho),\qquad \rho(0) = \rho_0, 
\label{ch2_f10} 
\end{eqnarray}
where dissipator $\mathcal{L}(\rho)$ can have, for example, the Lindblad form
$\mathcal{L}(\rho) = \sum\limits_k \gamma_k \Big( A_k \rho A_k^{\dagger} - 
\dfrac{1}{2} \Big\{ A_k^{\dagger} A_k, \rho \Big\} \Big)$; 
$A_k$ are the Lindblad operators which model different channels of dissipation; 
$\{,\}$ --- anticommutator ($\{A,B\} = AB + BA$); parameters $\gamma_k\ge 0$. 
If $\mathcal{L}(\rho) \equiv 0$, then the equation (\ref{ch2_f10}) 
becomes the Liouville--von Neumann equation (\ref{ch2_f9}).
 
\begin{definition}
For the system (\ref{ch2_f9}), the problem 
of maximization of the mean $\left\langle \rho_{\rm target} \right\rangle = 
{\rm Tr} \left( \rho(T) \rho_{\rm target} \right)$~\cite{Butkovsky_Samoilenko_book_1984_1990, 
Schirmer_Fouquieres_2011} between the final density matrix $\rho(T)$ and
the target density matrix $\rho_{\rm target}$ 
is defined with the following cost criterion:    
\begin{eqnarray} 
J(v) &=& \mathcal{F}_{\rho_{\rm target}}(\rho(T)) + \lambda_u \int\limits_0^T 
\dfrac{\| u(t) \|^2}{S(t)} dt + 
\lambda_{\rho} \int\limits_0^T \Tr\left( \rho(t) D(t) \right) dt  
\to \min\limits_{v \in \mathcal{D}}, 
\label{ch2_f12} 
\end{eqnarray}
where the terminant $\mathcal{F}_{\rho_{\rm target}} = 
-\left\langle \rho_{\rm target} \right\rangle$; 
process $v = (\rho, u)$, parameters
$\lambda_u \geq 0$, $\lambda_{\rho} \leq 0$; and operator $D(t)$
is self-adjoint positive semi-definite.
\end{definition}
 
\subsection{Cost criteria for unitary transformations and for ensemble of solutions of the Schr\"{o}dinger equation}

Potential applications of quantum technologies include quantum computation which could  significantly increase 
the speed of solving complex problems such as 
factorization of a large number or search in an unsorted 
database~\cite{Kitaev_article_1997, Kitaev_Shen_Vyalyi_book_AMS_2002, 
Valiev_PhysicsUspekhi_2005, Nielsen_Chuang_2011_10thEd_book}. Basic objects 
in quantum computation are {\it qubit} (quantum bit), which is a two-state 
quantum system, and {\it quantum gate} --- elementary operation which transforms 
input states of one or several qubits into some output states.

Mathematically quantum gate is a unitary 
matrix $W$~\cite{Holevo_book_De_Gruyter_2012, Nielsen_Chuang_2011_10thEd_book, Ohya_book_2011}. 
For example, the Hadamard gate is the following one-qubit ($n = 2$)
unitary matrix:
\begin{eqnarray*} 
W_{\rm H} &=& \dfrac{1}{\sqrt{2}}
\begin{pmatrix}
1 & 1   \\
1 & -1  \\ 
\end{pmatrix}.  
\end{eqnarray*}  
Two-qibit gates ($n = 2^2$) include, for example, such gates 
as Controlled NOT (CNOT), Quantum Fourier Transform (QFT),
Controlled Phase gate (CPHASE), and BGATE: 
\begin{eqnarray*} 
	W_{\rm CNOT} &=&  
	\begin{pmatrix}
		1 & 0 & 0 & 0 \\
		0 & 1 & 0 & 0 \\
		0 & 0 & 0 & 1 \\
		0 & 0 & 1 & 0
	\end{pmatrix}, \quad
	W_{\rm QFT} = \dfrac{1}{2}
	\begin{pmatrix}
		1 & 1 & 1 & 1 \\
		1 & i & -1 & -i \\
		1 & -1 & 1 & -1 \\
		1 & -i & -1 & i
	\end{pmatrix}, \\
	W_{\rm CPHASE} &=&  
	\begin{pmatrix}
		1 & 0 & 0 & 0 \\
		0 & 1 & 0 & 0 \\
		0 & 0 & 1 & 0 \\
		0 & 0 & 0 & -1
	\end{pmatrix}, \quad
	W_{\rm BGATE} = 
	\begin{pmatrix}
		\cos\dfrac{\pi}{8} & 0 & 0 & i \sin\dfrac{\pi}{8} \\
		0 & \cos\dfrac{3\pi}{8} & i \sin\dfrac{3\pi}{8} & 0 \\
		0 & i\sin\dfrac{3\pi}{8} & \cos\dfrac{3\pi}{8} & 0 \\
		i\sin\dfrac{\pi}{8} & 0 & 0 & \cos\dfrac{\pi}{8}
	\end{pmatrix}.
\end{eqnarray*}

Examples of three-qubit gates are Toffoli and Fredkin gates.

Implementation of quantum gate $W$ in a quantum processor 
means realization of a suitable controlled 
physical process 
(e.g.,~\cite{Palao_Kosloff_2002, Palao_Kosloff_article_2003})
which produces the evolution 
of the system which equals to the unitary matrix $W$.  

\begin{definition}
The operator $U(t)$ satisfying the Cauchy problem
\begin{eqnarray}
\dfrac{d U(t)}{dt} &=& 
-\dfrac{i}{\hbar} \left({\bf H}_0 + \sum\limits_{l=1}^m {\bf H}_l u_l(t) \right) U(t), 
\qquad U(0) = {\mathbb I}, 
\label{ch2_f21}
\end{eqnarray} 
is called the unitary evolution operator of a quantum system with Hilbert space $\mathcal{H}$ and with control $u(t) = (u_1(t), \dots, u_m(t))$, where ${\bf H}_l$ ($l = \overline{0, m}$) are Hermitian operators
and ${\mathbb I}$ is the identity operator.
\end{definition}

Similarly to the Theorem 1 (see also Lemma~2.1 in~\cite{Jurdjevic_Sussmann_1972} (1972)), 
the solution of the equation (\ref{ch2_f21}) exists, is absolutely continuous 
matrix function and unique for controls from the space $L^1([0,T]; Q)$. 

We have $\psi(t) = U(t)\psi_0$. The evolution of the density matrix is given by
\begin{eqnarray}
\rho(t) &=& U(t) \rho_0 U^{\dagger}(t). \label{ch2_f22}
\end{eqnarray}
Density matrix $\rho(t)$ satisfies Liouville--von Neumann equation (\ref{ch2_f9}).
From (\ref{ch2_f22}) it follows that 
\begin{eqnarray*}
\dfrac{1}{2} \left\| \rho(T) - \rho_{\rm target} \right\|^2 &=& C - \Tr\left(\rho_{\rm target} \rho(T)  \right), 
\end{eqnarray*} 
where $C = \dfrac{1}{2}\Tr\left(\rho_0^2 \right) + \dfrac{1}{2}\Tr\left( \rho^2_{\rm target} \right)$.
For trajectory optimization the following functional is used
\cite[p. 4]{Schirmer_Fouquieres_2011}:
\begin{eqnarray*}
\dfrac{1}{2} \int\limits_0^T \left\| \rho(t) - \rho_{\rm target}(t) \right\|^2 dt &=& CT - 
\int\limits_0^T \Tr\left(\rho_{\rm target}(t) \rho(t) \right)dt. 
\end{eqnarray*} 

\begin{definition}
The OCP for the unitary operator $U(t)$ is the problem of minimizing the cost functional
\begin{eqnarray}
J_X(v) &=& \mathcal{F}_{X}(U(T)) 
+ \lambda_u \int\limits_0^T \dfrac{\| u(t) \|^2}{S(t)} dt, \qquad \lambda_u \geq 0,
\qquad v = (U, \rho)
\label{ch2_f24}
\end{eqnarray}
for the system (\ref{ch2_f21}), where the terminant $\mathcal{F}_{X}$, 
$X \in \{W,~O,~A\}$, is defined by one of the following ways:
\begin{eqnarray}
\mathcal{F}_W(U(T))  &=& 
-\dfrac{1}{n^2}\left| \Tr ( W^{\dagger} U(T)  ) \right|^2, \label{ch2_f25} \\
\mathcal{F}_O(U(T))  &=& 
-\Tr( O \rho(T) ) = 
-\Tr( O U(T) \rho_0 U^{\dagger}(T) ), \label{ch2_f26} \\
\mathcal{F}_A(U(T))  &=& 
-{\rm Re} \left[ \Tr( A^{\dagger} \rho(T)  ) \right] = 
-{\rm Re} \left[ \Tr( A^{\dagger}  U(T) \rho_0 U^{\dagger}(T) ) \right], 
\label{ch2_f27} \\
\mathcal{F}_A(U(T)) &=& - | \Tr ( A^{\dagger} \rho(T) ) |^2 = 
- | \Tr ( A^{\dagger} U(T) \rho_0 U^{\dagger}(T)  ) |^2; \label{ch2_f28}  
\end{eqnarray}
$W$ is unitary, $O$ is self-adjoint, $A$ is non-self-adjoint given operator;
$S$ is some shape function.
\end{definition}
 
Problems of this kind are considered, for example, in connection with nuclear magnetic resonance, 
magnetic resonance imaging, dynamic nuclear 
polarization~\cite{Maximov_et_al_article_2008, Maximov_Salomon_Turinici_Nielsen_article_2010},
quantum gate generation (e.g.,~\cite{Pechen_Ilin_PhysRevA_2012, Pechen_Ilin_2014}).
 
Following the articles~\cite{Palao_Kosloff_article_2003, Palao_Kosloff_Koch_article_2008, 
Singer_et_al_article_Colloquim_2010, Muller_Reich_Murphy_et_al_article_2011, 
Reich_Ndong_Koch_article_2012, Palao_Reich_Koch_PhysRev_2013, 
Goerz_Whaley_Koch_article_2015, Goerz_Gualdi_Reich_Koch_et_al_article_2015, 
Reich_dissertation_Kassel_2015}, consider
the definition.

\begin{definition}
The ensemble of solutions of the Schr\"{o}dinger equation is a 
set of functions $\{ \psi_j(t)~ | ~ t \in [0, T], ~ j = \overline{1, n} \}$ such that 
$j$th element of this set is the solution of the Schr\"{o}dinger equation (\ref{ch2_f1}) 
with a controlled Hamiltonian (\ref{ch2_f2}), where $\psi = \psi_j$, $\psi_0 = \psi_{j, 0}$,
the initial states $\{ \psi_{j, 0}~ | ~ j = \overline{1, n} \}$ are given,
and the control $u$ is the same for all $j = \overline{1, n}$.
\end{definition}

Instead of a unitary transformation $U(t)$ 
we can study the corresponding ensemble of solutions of the Schr\"{o}dinger equation 
due to $\psi_j(T) = U(T) \psi_{0,j}$, $\psi_{{\rm target},j} = W \psi_{0,j}$, $j = \overline{1,n}$,
where $W$ is a target unitary transformation. For the ensemble of solutions,
one can consider the corresponding OCP for controlled simultaneous transitions of the system 
from the set of initial states $\psi_{0, j}$ to the set 
of target states $\psi_{{\rm target},j}$, $j = \overline{1, n}$.

\begin{definition}
The following problem is called OCP for an ensemble of solutions of the Schr\"{o}dinger
equation: 
\begin{eqnarray}
\dfrac{d \psi_j(t)}{dt} &=& -\dfrac{i}{\hbar} {\bf H}[u(t)] \psi_j(t), 
\qquad \psi_j(0) = \psi_{0,j}, \quad j = \overline{1,n}, 
\label{ch2_f34} \\
J(v) &=& \mathcal{F}\left(\{ \psi_j(T), j= \overline{1,n} \} \right) + 
\lambda_{\psi}  \int\limits_0^T \sum\limits_{j=1}^n \left\langle \psi_j(t), 
{\bf D}(t) \psi_j(t) \right\rangle dt \to \min, 
\label{ch2_f33} 
\end{eqnarray}
where $v = \left( \{ \psi_j \}_{j = \overline{1,n} }, u \right)$ is the
control process; the terminant $\mathcal{F}$ 
is defined on the set of final states of the ensemble of solutions; parameter 
$\lambda_{\psi} \leq 0$. 
\end{definition}
 
The specific details of an OCP (\ref{ch2_f34}), (\ref{ch2_f33})  
can be different~\cite{Palao_Kosloff_article_2003, Palao_Kosloff_Koch_article_2008, Muller_Reich_Murphy_et_al_article_2011, 
Reich_Ndong_Koch_article_2012, Palao_Reich_Koch_PhysRev_2013, Goerz_Whaley_Koch_article_2015, 
Goerz_Gualdi_Reich_Koch_et_al_article_2015, Reich_dissertation_Kassel_2015}.
The articles~\cite{Palao_Kosloff_2002, Palao_Kosloff_article_2003} (J.P.~Palao, R.~Kosloff, 2002, 2003)
consider control of lithium and sodium molecules with two electronic states (ground and excited). 
The first $n = 2^q$ levels of the ground state are considered as 
registers for $q$ qubits, and the goal is to find such control $u$
which provides the realization of the target gate $W$. The article~\cite{Palao_Kosloff_2002}
considers OCPs for implementing the quantum gates $W_{\rm H}$ and $W_{\rm QFT}$ by 
realizing simultaneous transitions between electronic surfaces 
for solutions $\psi_j$ of the Schr\"{o}dinger equation. The goal is expressed as minimization 
of the terminant 
\begin{eqnarray*}
\mathcal{F}(U(T)) &=& -\dfrac{1}{n^2} \Big| \Tr\{ W^{\dagger}  U(T) P_n \}  \Big|^2 = 
-\dfrac{1}{n^2} \sum\limits_{j=1}^n \sum\limits_{j'=1}^n \langle j | W^{\dagger} U(T) | j \rangle 
\langle j' | U(T)^{\dagger} W | j' \rangle,
\end{eqnarray*}  
where $P_n$ is the projector on the subspace where a unitary transformation 
$W$ is considered. 

The theory of quantum gates uses the {\it Cartan decomposition 
on the $SU(4)$} group, 
{\it Weyl chamber}, and {\it local invariants} related to 
the {\it equivalence classes}~\cite{Muller_Reich_Murphy_et_al_article_2011, 
Goerz_dissertation_Kassel_2015, Goerz_Motzoi_Whaley_Koch_2017} 
(T.~Calarco, C.P.~Koch, M.M.~M\"{u}ller, D.M.~Reich, J.~Vala, and others).
The Cartan decomposition for a two-qubit operator $U \in SU(4)$ 
is~\cite{Muller_Reich_Murphy_et_al_article_2011}
\begin{eqnarray}
U &=& k_1 A k_2, \text{ where } A = \exp\left(\dfrac{i}{2} \Big( c_1 \sigma_x \otimes \sigma_x + 
c_2 \sigma_y \otimes \sigma_y + c_3 \sigma_z \otimes \sigma_z \Big) \right).
\label{ch2_f36}
\end{eqnarray}
Here $\sigma_x$, $\sigma_y$, $\sigma_z$ are the Pauli matrices,
$k_1,~k_2 \in SU(2) \otimes SU(2)$ are some local operations,
real numbers $c_1$, $c_2$, $c_3$ are coordinates in the 
Weyl chamber. $W_{\rm CNOT}$ can be
represented by (\ref{ch2_f36}) 
as $W_{\rm CNOT} = (\mathbb{I} \otimes W_{\rm H})~ W_{\rm CPHASE}~ (\mathbb{I} 
\otimes W_{\rm H})$, where $\mathbb{I}$ is the identity 
matrix~\cite{Goerz_dissertation_Kassel_2015}. $W_{\rm CNOT}$ 
and $W_{\rm CPHASE}$ differ from each other
only by local operations. Hence they are locally equivalent 
($W_{\rm CNOT} \sim W_{\rm CPHASE}$) and belong to the same equivalence class 
$[W_{\rm CNOT}]$, which corresponds to the point $(c_1, c_2, c_3) = (\pi/2, 0, 0)$ 
in the Weyl chamber. The equivalence classes are related 
to local invariants that can be written using the following 
representation of $U$ via the Bell basis:  
\begin{eqnarray*}
g_1 = \dfrac{1}{16}{\rm Re} \{ (\Tr \widehat{m})^2 \}, \qquad
g_2 = \dfrac{1}{16}{\rm Im} \{ (\Tr \widehat{m})^2 \}, \qquad
g_3 = \dfrac{1}{4} \left[ (\Tr \widehat{m} )^2 - \Tr \widehat{m}^2 \right].
\end{eqnarray*}  
Here
\begin{eqnarray*}
\widehat{m} = U_B^{\dagger} U_B, \qquad U_B = B U B^{\dagger}, \qquad
B = \dfrac{1}{2} 
\begin{pmatrix}
	1 & 0 & 0 & i \\
	0 & i & 1 & 0 \\
	0 & i & -1 & 0 \\
	1 & 0 & 0 & -i
\end{pmatrix}.
\end{eqnarray*}  
The papers~\cite{Muller_Reich_Murphy_et_al_article_2011, 
Goerz_dissertation_Kassel_2015, Goerz_Motzoi_Whaley_Koch_2017}
consider OCPs where the cost functional is formulated in terms of distances 
in the space of coordinates $(c_1, c_2, c_3)$ or in the space of local invariants 
$(g_1, g_2, g_3)$ for a given target gate $W_{\rm target}$. One can 
consider the terminant~\cite{Muller_Reich_Murphy_et_al_article_2011}  
\begin{eqnarray}
\mathcal{F}_{\rm LI} &=& \sum\limits_{i=1}^3 \left(\Delta g_i \right)^2 + 1 - 
\dfrac{1}{n} \Tr \left\{ U_n(T) U_n^{\dagger}(T) \right\},
\quad \Delta g_i = \left| g_i(W_{\rm target}) - g_i(U_n(T)) \right|, 
\label{ch2_f37}
\end{eqnarray} 
where $W_{\rm target}$ is the target gate, and $U_n(T) = P_n U(T) P_n$ is the result
of projecting to a subspace. The explicit form of (\ref{ch2_f37}) 
in terms of $\{ \psi_j(T) \}_{j=1}^n$ is a polynomial of 8th degree, and the corresponding 
optimization results~\cite{Reich_Ndong_Koch_article_2012} are described in Section~5.

\subsection{Gross--Pitaevskii equation and optimal control of Bose--Einstein condensate}

The linear Schr\"{o}dinger equation is often suitable for describing quantum systems, 
but in some situations nonlinear equations appear, such as the Gross--Pitaevskii equation 
suggested by E.P.~Gross and L.P.~Pitaevskii in 1961. This equation describes dynamics of 
Bose--Einstein condensate, which is important both from a theoretical point of view 
and for creation of new technologies, e.g., atomic chips~\cite{Frank_Bonneau_et_al_article_2016}. 
Modelling of controlled dynamics of a Bose-Einstein 
condensate is done using the Gross--Pitaevskii equation 
with control~\cite{Sklarz_Tannor_article_2002, Hohenester_Rekdal_Borzi_article_2007, Bucker_Berrada_vanFrank_et_al_2013,	Hintermuller_et_al_SIAM_2013, Frank_Negretti_Berrada_et_al__article_2014, Jager_Reich_Goerz_et_al_2014, JagerG_thesis_2015, Hocker_Yan_Rabitz_article_2016, Frank_Bonneau_et_al_article_2016, Borzi_book_2017, Sorensen_Aranburu_Heinzel_Sherson_2018}.

\begin{definition}
	The equation
	\begin{eqnarray}
	\dfrac{\partial \psi(x,t)}{\partial t} &=& 
	-\dfrac{i}{\hbar} \Big( K + V(x, u(t)) + \kappa \left| \psi(x,t) \right|^2 \Big) \psi(x,t),  
	\label{ch2_f13}
	\end{eqnarray} 
	with initial condition $\psi(x, 0) = \psi_0(x)$ is called the Gross--Pitaevskii 
	equation with control $u \in \mathcal{U}$ in the potential
	(e.g., \cite[p. 336]{Borzi_book_2017}). Here 
	$x \in \Omega \subseteq \mathbb{R}^d$, $\psi(\cdot,t) \in L^2(\Omega; \mathbb{C})$;
	$K = -\dfrac{\hbar^2}{2m} \nabla^2$ is the kinetic energy operator; $m$ is atomic mass;
	$V(x, u(t))$ is the controlled potential; $\kappa$ is a coefficient (for example, 
	$\kappa = U_0(N_a-1)$~\cite{Jager_Reich_Goerz_et_al_2014} where $N_a$ is the number of atoms, 
	and $U_0$ is the strength of interaction between atoms under one-dimensional $x$); 
	the class $\mathcal{U}$ is defined in (\ref{ch2_f3}).
\end{definition}  

Meaning $\psi(\cdot,t)$, we will use $\psi(t)$ for shortness.

Potentials of various form are used~\cite{Sklarz_Tannor_article_2002, Hohenester_Rekdal_Borzi_article_2007, 
	Bucker_Berrada_vanFrank_et_al_2013, Hintermuller_et_al_SIAM_2013, 
	Frank_Negretti_Berrada_et_al__article_2014, Jager_Reich_Goerz_et_al_2014, 
	JagerG_thesis_2015, Hocker_Yan_Rabitz_article_2016, Frank_Bonneau_et_al_article_2016, Borzi_book_2017, 
	Sorensen_Aranburu_Heinzel_Sherson_2018}. In the article~\cite{Sklarz_Tannor_article_2002} 
(S.E.~Sklarz, D.J.~Tannor, 2002) the Gross--Pitaevskii equation (\ref{ch2_f13}) with the following
potential for one-dimensional $x$ is considered: 
\begin{eqnarray}
V(t, x, u(t)) &=& u(t) x^2 + s(t) V_0 \cos^2(k x). 
\label{ch2_f14}
\end{eqnarray}
Here control $u$ characterizes the strength of the trap potential; 
$V_0$ characterizes the lattice intensity; $s(t)$ is the
switching-on function of the field; $k$ is the laser field wave number. 
These wave packets represent quantum bits for quantum information. Initially the Bose-Einstein
condensate is located in the ground state of the potential. The OCP with the criterion
\begin{eqnarray}
J(v) &=& \left\langle \psi(T), (\cos \theta(T))^2 ~\psi(T) \right\rangle - 
\big\langle \psi(T), \cos \theta(T) ~\psi(T) \big\rangle^2 \to \min 
\label{ch2_f15}
\end{eqnarray} 
is considered, where $\theta(T) = \theta(\cdot,T)$ is the phase of the wave packet at the time $T$;
$\cos\theta = \dfrac{1}{2} \dfrac{\psi + \psi^{\ast}}{|\psi|} = \dfrac{{\rm Re}\psi}{|\psi|}$;
$v = (\psi,u)$.

In the article~\cite{Hohenester_Rekdal_Borzi_article_2007}  the following potentials for
one-dimensional $x$ are considered:
\begin{eqnarray}
V(x, u(t)) &=& \dfrac{1}{2} \left( x - u(t) x_0 \right)^2, 
\label{ch2_f16} \\
V(x, u(t)) &=& 
\begin{cases}
(1/2) \left( |x| - u(t) d/2 \right)^2, & \qquad |x| > u(t) d /4, \\
(1/2) \left( (u(t))^2/8 - x^2 \right), & \qquad |x| \leq u(t) d /4.
\end{cases} 
\label{ch2_f17}
\end{eqnarray}

In the articles~\cite{Bucker_Berrada_vanFrank_et_al_2013, Frank_Bonneau_et_al_article_2016, 
Sorensen_Aranburu_Heinzel_Sherson_2018} the following polynomial potential for one-dimensional $x$ is used:
\begin{eqnarray}
V(t, x, u(t)) &=& p_2 \left(x - u(t) \right)^2 + p_4 \left( x - u(t)\right)^4 + 
p_6 \left( x - u(t) \right)^6. 
\label{ch2_f18}
\end{eqnarray}
Here control $u$ defines moving along the axis $Ox$ for shaking the
condensate, numbers $p_2$, $p_4$, $p_6$ are fitting parameters. 
The condensate  
is initially prepared in the ground state $V(x, 0, u(0))$ with interacting bosons, and 
$\psi_{\rm target}$ is set as the first excited state for $V$. 

The goal of control is often expressed as maximization of the probability  
$\big| \langle \psi_{\rm target}, \psi(T) \rangle_{L^2}\big|^2$
of transition to the target state $\psi_{\rm target}$. This control goal is described by
the terminants (\ref{ch2_f5}) and (\ref{ch2_f6}) for 
(\ref{ch2_f13}). Sometimes, together with the GRAPE 
method~\cite{Khaneja_Reiss_Kehlet_SchulteHerbruggen_Glaser_2005,
Jager_Reich_Goerz_et_al_2014}, which will be described in Section~6, 
the following $H_1$-regularizer is included in the cost functional for
smoothing the control: 
\begin{eqnarray}
\lambda_{du} \int\limits_0^T 
(u(t))^2 dt, \qquad \lambda_{du} > 0.
\label{ch2_f20}
\end{eqnarray}
The general cost criterion for (\ref{ch2_f13}) is:
\begin{eqnarray}
J(v) &=& \mathcal{F}(\psi(T)) + \lambda_u \int\limits_0^T 
\dfrac{\| u(t) \|^2}{S(t)} dt + \lambda_{du} \int\limits_0^T (\dot u(t))^2 dt \to \min, 
\label{ch2_f19}
\end{eqnarray} 
where $\mathcal{F}$ is defined, for example, using (\ref{ch2_f5});
$\lambda_u$, $\lambda_{du} \geq 0$; $S$ is a shape function.

The rigorous formulation of the optimal control problem for the Gross--Pitaevskii equation for the potential $V(x,u)=U(x)+u(t)\tilde V(x)$, where $U$ and $\tilde V$ are given and $u(t)$ is the control, is provided in~\cite{Hintermuller_et_al_SIAM_2013}. In this formulation the potential $V\in W^{1,\infty}(\mathbb R^d)$ (the space of Lipschitz functions), the potential $U$ satisfies
\[
U\in C^\infty(\mathbb R^d),\qquad \partial^k U\in L^\infty(\mathbb R^d)
\qquad \text{for any multi-index } k\text{ such that }|k|>2.
\]
The initial state  $\psi_0(x)$ belongs to the subspace
\[
\Upsilon=\{\psi\in H^1(\mathbb R^d):\,\, x\psi\in L^2(\mathbb R^d)\}.
\]
The objective functional has the form 
\[
J(\psi,u)=\langle\psi(T,\cdot), A\psi(T,\cdot)\rangle_{L^2(\mathbb R^d)}+\int\limits_0^T
(\dot u(t))^2 \left[\gamma_1\left(\int\limits_{\mathbb R^d} \tilde V(x)|\psi(x,t)|^2dx\right)^2 + 
\gamma_2\right] dt,
\]
where $T>0$, $\gamma_1\ge 0$, $\gamma_2>0$, and $A:\Upsilon \to L^2(\mathbb R^d)$ is a (possibly unbounded) essentially self-adjoint on $L^2(\mathbb R^d)$ operator. The integral terms in the objective are introduced to avoid highly oscillating control. For this control problem, well--posedness and existence of an optimal control were proved.

\subsection{Controllability of closed quantum systems}

Creating quantum optimal control is closely related to the notions and criteria for controllability 
of quantum systems which are discussed, for example, in the 
articles~\cite{Butkovsky_Samoilenko_article_1980, Huang_Tarn_Clark_1983,
Altafini_2002, Turinici_Rabitz_2003, Kurniawan_Dirr_Helmke_2012, 
DAlessandro_Albertini_Romano_SIAM_2015, Boscain_Gauthier_Rossi_Sigalotti_2015, 
Agrachev_et_al_IEEE_CDC_2017} and monographs~\cite{DAlessandro_book_2007, Borzi_book_2017}. 
Before starting search for an optimal control, it is desirable to know whether such a
control exists at all. The answer to this question is given by the controllability 
criteria, which are well-known for closed quantum systems. We will give 
the basic concepts and results without details.

For closed systems one of fundamental notions is the notion of projective state 
controllability or equivalent state controllability.

\begin{definition}\label{Def2_12}
A quantum system (\ref{ch2_f1}) --- (\ref{ch2_f3}) with states  
$\psi(t) \in \mathbb{C}^n$ is called to be projective state controllable,
if for any initial state $\psi_0$ and final equivalence class
$[\psi_{\rm target}] := \left\{ e^{i \phi} \psi_{\rm target}~:~ \phi \in [0, 2\pi) \right\}$ 
with some $\psi_{\rm target}$ there are $T>0$ and control $u \in \mathcal{U}$ 
such that the system can be moved from $\psi_0$ to $[\psi_{\rm target}]$ 
during time $T$.    
\end{definition}

Also important is the definition of controllability on the special unitary group.

\begin{definition}
A system (\ref{ch2_f21}) describing the evolution of a unitary operator
$U(t) \in \mathbb{C}^{n \times n}$ is called to be controllable on the group $SU(n)$,
if for any unitary operator $W \in SU(n)$ there exist a time instant $T > 0$ and control
$u \in \mathcal{U}$ such that $W = e^{i \phi} U(T)$, where $\phi \in [0,2 \pi)$ is some phase.
\end{definition}

The analysis of controllability of quantum systems, including for systems 
(\ref{ch2_f1}) --- (\ref{ch2_f3}), is crucial for quantum control, 
since the presence or absence of controllability determines the solvability 
of an optimal control problem. For example, the terminant 
$\mathcal{F}_{\psi_{\rm target}}$ (\ref{ch2_f5}) determines the probability 
for a system state transfer on the sphere by the time $T$. Realization of this
transfer between arbitrary initial and target states is impossible
for uncontrollable systems.
	
The control criteria are based on the analysis of the Lie algebra $\cal L$ 
generated by all possible commutators of the operators ${\bf H}_0, {\bf H}_1,\dots, {\bf H}_m$. 
Without loss of generality we can put ${\rm Tr}{\bf H}_i=0$ for $i= \overline{0,m}$. 
Indeed, the dynamics of a system with operators having nonzero traces will differ 
from the dynamics with operators ${\bf H}_i -{\rm Tr}{\bf H}_i/n$ on 
a physically nonessential phase factor. A consequence of the analysis 
in~\cite{Jurdjevic_Sussmann_1972} is the following 
theorem~\cite{Fu_Schirmer_Solomon_2001, Albertini_Alessandro_2003}.

\begin{theorem}
Let $\mathcal{H} = \mathbb{C}^n$. The system (\ref{ch2_f21}) 
is projective state 
controllable if and only if the Lie algebra 
${\rm Lie}\left\{ -i {\bf H}_0,\dots, -i {\bf H}_m \right\}$,
generated by all commutators of operators $-iH_0, \dots, -iH_m$, is 
isomorphic to the Lie algebra $\mathfrak{sp}(n/2)$ 
or $\mathfrak{su}(n)$ for even $n$, or $\mathfrak{su}(n)$ for odd $n$.
\end{theorem}

Verification of the controllability for a particular quantum system is one 
of the basic questions before search for optimal controls. 
Projective state controllability of the system (\ref{ch2_f1}) --- (\ref{ch2_f3}) 
means that there exist time instant $T$ and control $u \in \mathcal{U}$ such that 
$\mathcal{F}_{\psi_{\rm target}} = 0$ (exact controllability). For some sets of 
$Q$ and $T$, the system can be not controllable. On other hand, we can look for such $Q$ and $T$, 
that $\mathcal{F}_{\psi_{\rm target}}(\psi(T)) = 0$ and some 
additional condition is satisfied, for example, such that $T$ takes the minimum possible 
value.

\subsection{Landscapes of control problems for closed quantum systems}

The goal of application of a particular numerical method to quantum control is to find globally optimal (for the minimization problem) controls, those that 
deliver the global minimum of the cost functional. If all minima of the cost 
functional would be global, the most natural numerical methods for finding optimal controls 
would be gradient type. If the system is controllable, but the landscape is replete with traps (i.e., local but not global minima), then stochastic algorithms (e.g., genetic algorithms)  would be the proper choice to step around or out of such traps. Therefore theoretical analysis of minima of the objective functionals is important for choosing a proper numerical strategy for finding optimal controls. Here we briefly outline some results on this topic.

The problem of analysis 
of extrema of cost functionals for quantum systems was posed in~\cite{Rabitz_Hsieh_Rosenthal_article_2004} 
(H.A.~Rabitz, M.~Hsieh, C.M.~Rosenthal, 2004), where the conjecture 
was made that all minima of target functionals are global (i.e., control landscapes are free of traps). Since then, gradient-based methods were found to frequently succeed in quantum control numerical simulations and experiments. Large number of simulations have shown that even if the landscapes for optimal control problems may contain singular critical points and traps, the conditions necessary to yield these points are sufficiently strict so that many control landscapes may lack traps~\cite{Moore_Rabitz_2012,Riviello_Brif_etal_2014}.

A considerable effort was made to understand the surprise ease of gradient-based optimal control simulations. However, despite this effort the trap-free assumption has been rigorously proven only for $n=2$ -- level quantum 
systems (Theorem~ \ref{TheoremQCL})~\cite{Pechen_Ilin_PhysRevA_2012, Pechen_Ilin_2014}
(A.N.~Pechen, N.B.~Il'in, 2012, 2014) and for the problem of controlling 
transmission of a quantum particle through a one-dimensional 
potential barrier (Theorem~\ref{TheoremQCLTransmission})~\cite{Pechen_Tannor_2014}
(A.N.~Pechen, D.J.~Tannor, 2014). For dimensions more than two there were 
found systems with trapping 
properties~\cite{Pechen_Tannor_article_2011} (A.N.~Pechen, D.J.~Tannor, 
2011),~\cite{Fouquieres_Schirmer_2013} (P.~de~Fouquieres, S.G.~Schirmer, 2013).  
The numerical evidence of the trap-free behavior for various cases was emphasized in~\cite{Rabitz_Ho_etal_2012} 
(H.~Rabitz, T.-S.~Ho, R.~Long, R.~Wu, C.~Brif, 2012).

Let in (\ref{ch2_f24}) $\lambda_u = 0$. Then the cost functional
$J_X(u) = \mathcal{F}_X(U(T))$, where terminant $\mathcal{F}_X: SU(n) \to \mathbb{R}$ is a function on the
special unitary group and
$U(T)$ is the solution of the
Schr\"{o}dinger equation at time $T$ for control $u$. For example, for (\ref{ch2_f25})
one has $\mathcal{F}_W(U) = -\dfrac{1}{n^2} \left| \Tr \left(W^{\dagger} U\right) \right|^2$.

\begin{definition}
The graph of the functional $J_X(u)$ is called the control landscape
of the control problem.  
The control $u$ is called trap if it gives a local but not 
a global minimum for $J_X(u)$.
\end{definition}
 
For single-qubit 
gates, the works~\cite{Pechen_Ilin_PhysRevA_2012, Pechen_Ilin_2014, Pechen_Ilin_2015, 
Ilin_Pechen_article_2018} consider control landscapes for 
system of the form (\ref{ch2_f21}) with 
$U(t) \in \mathbb{C}^{2 \times 2}$, scalar control ($m=1$) and cost functionals 
$J_O$ and $J_W$, which describe 
the mean value of an observable and implementation of single-qubit gates. 
The main result is the following theorem~\cite{Pechen_Ilin_2014}.

\begin{theorem}\label{TheoremQCL}
Let for the system (\ref{ch2_f21}) with $\mathcal{H} = \mathbb{C}^2$ 
and the scalar control ($m = 1$) the operators ${\bf H}_0$, ${\bf H}_1$ and the time instant
$T$ are such that $[{\bf H}_0, {\bf H}_1] \neq 0$ and
\begin{eqnarray*}
T &\geq& \frac{\pi }{ \| {\bf H}_0 - \Tr{\bf H}_0/2 + u_0 ({\bf H}_1-{\rm Tr}{\bf H}_1/2 \|},
\end{eqnarray*}
where
\begin{eqnarray*}
u_0 &:=& \frac{- \Tr({\bf H}_0) \Tr({\bf H}_1) +
2 \Tr({\bf H}_0 {\bf H}_1)}{ (\Tr({\bf H}_1))^2 - 2 \Tr({\bf H}_1^2)}. 
\end{eqnarray*}
Then all the minima of the cost functionals 
$J_O(u) = \langle O \rangle_T = \Tr\{ U(T) \rho_0 U^{\dagger}(T) O \}$ and
$J_W(u) = -\dfrac{1}{4} \left| \Tr(W^{\dagger} U(T)) \right|^2$ 
are global minima for any Hermitian observable $O$,
unitary operator $W$ and the density matrix $\rho_0$.  
\end{theorem}  

The problem of controlling the tunneling of a quantum particle through a potential barrier
is described by the stationary Schr\"{o}dinger equation
\begin{eqnarray*} 
\left(-\frac{d^2}{dx^2}+V(x)\right)\Psi(x) &=& E\Psi(x),\qquad E\in\mathbb R.  
\end{eqnarray*}
The potential $V(x)$ is assumed to have compact support 
($V(x)=0$ if $|x|>a$ for some $a>0$). The solution that has the 
following asymptotic is considered:
\begin{eqnarray*} 
\Psi(x) &=&
\left\{
\begin{array}{l}
e^{i \sqrt{E} x}+A_Ee^{-i \sqrt{E}x},\qquad x<-a,\\
B_E e^{i \sqrt{E} x},\qquad x>a. 
\end{array}\right.  
\end{eqnarray*}
This solution describes a particle falling from the left on the potential barrier, 
which with probability $|A_E|^2$ is reflected from the barrier, and with probability 
$T_E(V)=|B_E|^2$ passes  to the right through the barrier. 
The potential is considered as control. Transmission coefficient is the cost functional that
should be maximized. The main result is the following theorem~\cite{Pechen_Tannor_2014}.

\begin{theorem}\label{TheoremQCLTransmission}
All extrema of the transmission coefficient $T_E(V)=|B_E|^2$ are global maxima corresponding 
to the value $T_E(V)=1$ (i.e. to the full tunneling). 
\end{theorem}

Thus, the absence of traps for quantum systems is proved for 
two-dimensional and infinite-dimensional Hilbert spaces. In the general case 
of an arbitrary dimension, the problem remains open. As a result, it is important to develop 
effective methods for obtaining optimal controls for quantum systems of any dimension.

\section{Krotov method for systems with states in $\mathbb{R}^n$}  

\subsection{Optimal control problems and Krotov Lagrangian}

Consider the following class 
of OCPs with real-valued states.

\begin{definition}
The following problem is called OCP with free final state for a dynamical system 
defined by ODE with control:
\begin{eqnarray}
\dfrac{dy(t)}{dt} &=& f(t, y(t), u(t)), \qquad y(0) = y_0, 
\label{ch3_f2} \\
u \in \mathcal{U} &=& PC([0,T]; Q), \qquad Q \subseteq \mathbb{R}^m, 
\label{ch3_f3} \\
J(v) &=& \mathcal{F}(y(T)) + 
\int\limits_{0}^{T} f^0(t,y(t),u(t))dt \to \inf\limits_{v \in \mathcal{D}}, \qquad
v = \left(y, u \right), 
\label{ch3_f1} 
\end{eqnarray}
where $y$ is
continuous, piecewise differentiable function; $\mathcal{D}$ is the set of admissible processes $v = (y, u)$. 
The state $y_0$ 
and the moment $T$ are fixed. For $\mathcal{F},~f^0,~f$ there are the following
traditional conditions: 
\par 1) vector function $f(t,y,u) = (f_1(t,y,u), \dots, f_n(t,y,u))$
and scalar function $f^0(t,y,u)$ are defined 
and are continuous on the set of variables $(t,y,u) \in [0,T] \times \mathbb{R}^n \times Q$
together with their partial derivatives in $y$, $u$; $f$ satisfies the Lipschitz condition 
with respect to $y$, i.e. 
\[
\| f(t, y + \Delta y, u) \| \leq L \| \Delta y\| \qquad \forall u \in Q, \qquad t \in [0,T];
\]
\par 2) function $\mathcal{F}(y)$ is continuously-differentiable in $\mathbb{R}^n$.
\end{definition}

As is known (for example,~\cite{Srochko_book_2000}), 
the Cauchy problem (\ref{ch3_f2}) for any control $u \in \mathcal{U}$ has a unique  
solution $y$ in the class of piecewise differentiable functions.
Continuous differentiability of $f^0$ and $f$ with respect to $u$ is also necessary for the 
further consideration. 

\begin{definition}
The solution of the problem (\ref{ch3_f2}) --- (\ref{ch3_f1}) 
is understood as the {\it minimizing sequence}, i.e. a sequence of processes 
$\{v^{(k)}\}_{k \geq 0} \subset \mathcal{D}$ that satisfies 
$\lim\limits_{k \to \infty} J(v^{(k)}) = \inf\limits_{v \in \mathcal{D}} J(v)$. 
If there exists an element $\overline{v} \in \mathcal{D}$ such that   
$J(\overline{v}) = \min\limits_{v \in \mathcal{D}} J(v)$, then the process $\overline{v}$ 
and control $\overline{u}$ are called (globally) {\it optimal}.
\end{definition}

\begin{definition}
For the problem (\ref{ch3_f2}) --- (\ref{ch3_f1}) consider a process 
$v^{(k)} \in \mathcal{D}$ ($k \geq 0$ is the iteration number). 
The problem of finding a process $v^{(k+1)} \in \mathcal{D}$ for 
which $J(v^{(k+1)}) < J(v^{(k)})$ is called the problem 
of improvement of $v^{(k)}$. A sequence 
$\{ v^{(k)} \}_{k=0}^K \subset \mathcal{D}$ such that $J(v^{(k+1)}) < J(v^{(k)})$, $k = \overline{0, K}$
is called improving sequence. 
\end{definition}

In the early 1960s V.F.~Krotov formulated the {\it generalized Lagrangian} 
({\it Krotov Lagrangian}) and {\it sufficient conditions for 
optimality}~\cite{Krotov_Gurman_book_1973, KrotovVF_book_NewYork_1996}. 

\begin{definition}
For the problem (\ref{ch3_f2}) --- (\ref{ch3_f1}), the Krotov Lagrangian is the following functional
\begin{eqnarray}
L^{\varphi}(v) &=& G^{\varphi}(y(T)) - \int\limits_0^T R^{\varphi}(t,y(t),u(t))dt, \qquad
v \in \mathcal{E}, 
\label{ch3_f4} 
\end{eqnarray}
where $\mathcal{E}$ is an extension of the set $\mathcal{D}$,
\begin{eqnarray} 
G^{\varphi}(y(T)) &=& \mathcal{F}(y(T)) + 
\varphi(T,y(T)) - \varphi(0,y(0)), 
\label{ch3_f5} \\
R^{\varphi}(t,y,u) &=& \left\langle \dfrac{\partial \varphi(t,y)}{\partial y}, 
f(t,y,u)\right\rangle - f^0(t,y,u) + \dfrac{\partial \varphi(t,y)}{\partial t}.  
\label{ch3_f5_add}
\end{eqnarray}
Here $\varphi$ belongs to the set $\Phi$ of functions each of them has
continuous partial derivatives for all $t,y$ except, maybe, a finite number 
of values of $t$ over $[0, T]$.  
\end{definition} 
 
As is known~\cite{Krotov_Gurman_book_1973, KrotovVF_book_NewYork_1996}, 
$L^{\varphi}(v) \equiv J(v)$ on $\mathcal{D}$ for any $\varphi \in \Phi$, and that is why
$L(v)$ can be considered as a special representation of $J(v)$.
Partial derivatives of $\varphi(t,y)$ in $y$ are Lagrange multipliers.  
The function $R$ defined in (\ref{ch3_f5_add}) can be written as  
\begin{eqnarray*}
R^{\varphi}(t,y,u) &=& H\left(t, \dfrac{\partial \varphi(t, y)}{\partial y}, 
y, u\right) + \dfrac{\partial \varphi(t, y)}{\partial t},
\end{eqnarray*}
where $H(t,q,y,u) = \langle q, f(t,y,u) \rangle - f^0(t,y,u)$ is the Pontryagin function,
$y \in \mathbb{R}^n$, $u \in Q \subseteq \mathbb{R}^m$.

In the frames of (\ref{ch2_f1}) --- (\ref{ch2_f3}) we consider

\begin{definition}
The following problem is called OCP for bilinear system with regularization on control 
(e.g.,~\cite{Krotov_Bulatov_Baturina_AiT_2011}):
\begin{eqnarray}
\dfrac{dy(t)}{dt} &=& \left(A(t) + \sum\limits_{j=1}^m u_j(t) B_j(t) \right) y(t), \qquad y(0)=y_0, 
\label{ch3_f7} \\
J(v) &=& \mathcal{F}(y(T)) + \lambda_u\int\limits_0^T 
\sum\limits_{j=1}^m u^2_j(t)dt \to \inf\limits_{v \in \mathcal{D}},
\qquad \lambda_u \geq 0, 
\label{ch3_f6}  
\end{eqnarray} 
where $u$ belongs to the class $\mathcal{U}$ defined in (\ref{ch3_f3}). 
\end{definition}

The problem (\ref{ch2_f1}) --- (\ref{ch2_f4}) with states $\psi(t) \in \mathbb{C}^n$  	 
can be described by the corresponding OCP with states $y(t) \in \mathbb{R}^{2n}$,
if the vector $y(t)$ represents both real and imaginary parts 
of complex-valued $\psi(t)$: $y_j(t) = {\rm Re} \psi_j(t)$ 
and $y_{n+j}(t) = {\rm Im} \psi_j(t)$ for $j = \overline{1,n}$.
For example, the problem (\ref{ch3_f7}), (\ref{ch3_f6}) 
with $\mathcal{F}(y(T)) = -\left\langle y(T), M y(T) \right\rangle$, $M \geq 0$
can describe some quantum OCPs. The terminant 
$\mathcal{F}(y(T)) = -\left\langle y(T), M y(T) \right\rangle$ 
is bounded below due to the invariant 
\[
\| y(t) \|^2_{\mathbb{R}^{2n}} = 
\| \psi(t)\|^2_{\mathbb{C}^n} =1.
\]

\subsection{Krotov method in the general form} 

For the problem (\ref{ch3_f2}) --- (\ref{ch3_f1}), consider the following
iterative process, where $v^{(k)} = (y^{(k)}, u^{(k)})$ and 
$v^{(k+1)} = (y^{(k+1)}, u^{(k+1)})$ are input and output admissible processes correspondingly
at $k$th iteration of the method. 
\begin{enumerate} 
	\item Compute the function $\varphi^{(k)} \in \Phi$ which satisfies the conditions$^7$\footnote{$^7$Here
		$R^{\varphi^{(k)}}(t, y^{(k)}(t), u^{(k)}(t)) = $
		\begin{eqnarray*}
			= \left\langle \dfrac{\partial \varphi^{(k)}(t,y^{(k)}(t))}{\partial y}, 
			f(t,y^{(k)}(t),u^{(k)}(t))\right\rangle  
			- f^0(t,y^{(k)}(t),u^{(k)}(t)) + \dfrac{\partial \varphi^{(k)}(t,y^{(k)}(t))}{\partial t},
		\end{eqnarray*}
		where
		$\dfrac{\partial \varphi^{(k)}(t,y^{(k)}(t))}{\partial y} = \dfrac{\partial \varphi^{(k)}(t,y)}{\partial y}\Big|_{y = y^{(k)}(t)}$ and 
		$\dfrac{\partial \varphi^{(k)}(t,y^{(k)}(t))}{\partial t} = \dfrac{\partial \varphi^{(k)}(t,y)}{\partial t}\Big|_{y = y^{(k)}(t)}$. In the review, for shortness we use notations like to $\dfrac{\partial \varphi^{(k)}(t,y^{(k)}(t))}{\partial y}$ instead of $\dfrac{\partial \varphi^{(k)}(t,y)}{\partial y}\Big|_{y = y^{(k)}(t)}$.}
	\begin{eqnarray} 
	G^{\varphi^{(k)}}(y^{(k)}(T)) &=& \max\limits_{y \in \mathbb{R}^n} G^{\varphi^{(k)}}(y), 
	\label{ch3_f11} \\
	R^{\varphi^{(k)}}(t, y^{(k)}(t), u^{(k)}(t)) &=& \min\limits_{y \in \mathbb{R}^n} R^{\varphi^{(k)}}(t, y, u^{(k)}(t)), 
	\qquad t \in [0, T). 
	\label{ch3_f10} 
	\end{eqnarray} 
	\item Find the solution $y^{(k+1)}$ of the Cauchy problem 
	\begin{eqnarray}
	\dfrac{dy^{(k+1)}(t)}{dt} &=& f\left(t, y^{(k+1)}(t), \widetilde{u}^{(k)}(t, y^{(k+1)}(t)) \right),
	\qquad y^{(k+1)}(0) = y_0   
	\label{ch3_f12}
	\end{eqnarray}
	and find the control $u^{(k+1)}$ defined by 
	\begin{eqnarray}
	u^{(k+1)}(t) &=& \widetilde{u}^{(k)}(t, y^{(k+1)}(t)) \nonumber \\
	&:=& {\rm arg}\max\limits_{u \in Q} R^{\varphi^{(k)}}\left(t, y^{(k+1)}(t), u \right), 
	\qquad t \in [0,T]. 
	\label{ch3_f9}
	\end{eqnarray} 	  
\end{enumerate}

\begin{theorem} \label{theorem:3.1}
For the problem (\ref{ch3_f2}) --- (\ref{ch3_f1}),
the method (\ref{ch3_f11}) --- (\ref{ch3_f9}) 
provides sequence of processes $\{v^{(k)} \} \subset \mathcal{D}$
such that $J(v^{(k+1)}) \leq J(v^{(k)})$. Moreover, if 
\begin{eqnarray}
\int\limits_0^T \max\limits_{u \in Q}R^{\varphi^{(k)}}(t,y^{(k)}(t), u)~dt &\neq&
\int\limits_0^T R^{\varphi^{(k)}}(t,y^{(k)}(t),u^{(k)}(t))~dt,   
\label{ch3_f13}
\end{eqnarray}
then the sequence $\{v^{(k)}\}$ is improving.  
\end{theorem} 

{\bf Proof.} Following~\cite{Krotov_Feldman_IzvAN_article_1983}, to prove the theorem, 
consider the increment $J(v^{(k)}) - J(v)$ for a given process 
$v^{(k)} \in \mathcal{D}$, arbitrary process $v \in \mathcal{D}$ and $\varphi \in \Phi$:
\begin{eqnarray}
J(v^{(k)}) - J(v) &=& L^{\varphi}(v^{(k)}) - L^{\varphi}(v) =
G^{\varphi}(y^{(k)}(T)) - G^{\varphi}(y(T)) + \nonumber\\
&&+ \int\limits_0^T \Big(R^{\varphi} ( t, y(t), u(t) ) - 
R^{\varphi} ( t, y^{(k)}(t), u^{(k)}(t)) \Big)dt. 
\label{ch3_f14}
\end{eqnarray} 
To satisfy the inequality $J(v^{(k)}) \geq J(v)$ for a process $v \in \mathcal{D}$ it is 
enough to find such a function $\varphi = \varphi^{(k)} \in \Phi$ that its values 
$\varphi^{(k)}(t,y^{(k)}(t))$, $\varphi^{(k)}(t,y(t))$ on $[0, T]$ allow to have 
\begin{eqnarray} 
G^{\varphi^{(k)}}(y^{(k)}(T)) - G^{\varphi^{(k)}}(y(T)) &\geq& 0, 
\label{ch3_f15} \\ 
R^{\varphi^{(k)}}(t, y(t), u(t)) - 
R^{\varphi^{(k)}}(t, y^{(k)}(t), u^{(k)}(t)) &\geq& 0 
\label{ch3_f16}
\end{eqnarray} 
and for $J(v^{(k)}) > J(v)$ it is sufficient to 
have strict inequality in (\ref{ch3_f16}) on some subset of positive measure in $[0, T]$.
Dividing the left-hand side of (\ref{ch3_f16}) to the sum of two particular
increments, we obtain the following conditions: 
\begin{eqnarray}   
R^{\varphi^{(k)}}(t,y(t),u(t)) - R^{\varphi^{(k)}}(t,y(t),u^{(k)}(t)) &\geq& 0, 
\label{ch3_f18} \\
R^{\varphi^{(k)}}(t,y(t),u^{(k)}(t)) - R^{\varphi^{(k)}}(t,y^{(k)}(t),u^{(k)}(t)) &\geq& 0. 
\label{ch3_f19}
\end{eqnarray}
For computing $\varphi^{(k)}$,  consider the conditions
(\ref{ch3_f11}), (\ref{ch3_f10})  obtained from (\ref{ch3_f15}) and (\ref{ch3_f19}). 
According to the inequality (\ref{ch3_f18}), 
for each $t$ we consider  
the condition
\begin{eqnarray*}
R^{\varphi^{(k)}}(t, y(t), u(t)) &=& \max\limits_{u \in Q} R^{\varphi^{(k)}}(t, y(t), u)   
\end{eqnarray*}
which leads to the formula (\ref{ch3_f9}), where the 
process $v=(y,u)$, satisfied to the mentioned above conditions,
is named $v^{(k+1)} = (y^{(k+1)}, u^{(k+1)})$. Thus, for finding the function $y^{(k+1)}$ 
we integrate the system (\ref{ch3_f12})  
derived from (\ref{ch3_f2}) where $u(t) = \widetilde{u}^{(k)}(t,y^{(k+1)}(t))$.  
Because of possible discontinuities of the function $\widetilde{u}^{(k)}(t,y)$ in $y$,
integration of the equation (\ref{ch3_f12}), in general, is done using the theory of ODEs with discontinuous r.h.s.~\cite{FilippovAF_book_1988}. 
We suppose that there exists a triple $\left(\varphi^{(k)},~ y^{(k+1)}, 
u^{(k+1)} \right)$ satisfying (\ref{ch3_f10}) --- (\ref{ch3_f12}). 
Then the process $v^{(k+1)}$ provides either $J(v^{(k+1)}) < J(v^{(k)})$
(improvement) or $J(v^{(k+1)}) = J(v^{(k)})$. If (\ref{ch3_f13}) 
is not satisfied, then $J(v^{(k+1)}) = J(v^{(k)})$.  This finishes the proof.

Solving the sequence of improvement problems by solving
(\ref{ch3_f11}) --- (\ref{ch3_f9}), we obtain the control sequence:
$u^{(0)} \to u^{(1)} \to \dots \to u^{(k)} \to \dots \to u^{(K)}$.

Consider the problem (\ref{ch3_f7}), (\ref{ch3_f6}). The Pontryagin function is
\begin{eqnarray*}
H(t, q, y, u) &=& \left\langle q,  A(t) + \sum\limits_{j=1}^m u_j B_j(t)  \right\rangle -
\lambda_u \sum\limits_{j=1}^m u_j^2,
\qquad q = \dfrac{\partial\varphi}{\partial y}(t,y).  
\end{eqnarray*}

Set $m=1$, $Q = [a, b]$, and $\lambda_u = 0$. In this case, the Pontryagin function is linear in $u$
and the function $\widetilde{u}^{(k)}$ defined in (\ref{ch3_f9}) is
\begin{eqnarray}
\widetilde{u}^{(k)}(t,y) &=&
\begin{cases}
a, &\qquad \text{$g^{(k)}(t,y) < 0$},\\
u_{\rm sing}^{(k)}(t,y) \in [a, b], &\qquad \text{$g^{(k)}(t,y) = 0$},\\
b, &\qquad \text{$g^{(k)}(t,y) > 0$},
\end{cases} 
\label{ch3_f20}
\end{eqnarray}
where $g^{(k)}(t,y) = \left\langle \dfrac{\partial \varphi^{(k)} (t,y)}{\partial y}, 
B(t) y \right\rangle$ is called the switching function; index 
\textquotedblleft {\rm sing}\textquotedblright\, means ``singular''. 
By substituting (\ref{ch3_f20}) in (\ref{ch3_f7}) we obtain 
the system of ODEs, in general, discontinuous in $y$. 
Therefore, the conditions for the existence and uniqueness 
of the solution of the Cauchy problem are violated~\cite{FilippovAF_book_1988}. 

Further, consider $m=1$, $Q = [a, b]$, and $\lambda_u > 0$.
The function $\widetilde{u}^{(k)}$ is
\begin{eqnarray}
\widetilde{u}^{(k)}(t,y) &=& 
\begin{cases}
a, & \qquad u_{\rm st}^{(k)}(t,y) < a, \\
u_{\rm st}^{(k)}(t,y), & \qquad u_{\rm st}^{(k)}(t,y) \in [a,b], \\
b, & \qquad u_{\rm st}^{(k)}(t,y) > b,
\end{cases} 
\label{ch3_f21}
\end{eqnarray}
where $u_{\rm st}^{(k)}(t,y)$ is a stationary point obtained from the condition
$\dfrac{\partial H}{\partial u} = 0$ for the Pontryagin function; index
\textquotedblleft {\rm st}\textquotedblright\ means ``stationary point''. 
The function (\ref{ch3_f21}) is continuous in $y$. 

\subsection{Krotov method with linear-quadratic function $\varphi$}

Based on the articles~\cite{Krotov_Feldman_IzvAN_article_1983, Konnov_Krotov_article_1999},
we describe the 2nd order ($\varphi(t,y)$ is considered 
in the class of linear-quadratic functions) Krotov method 
for the problem (\ref{ch3_f2}) --- (\ref{ch3_f1}).    

\begin{definition}
\label{Def3.6}
The function $\varphi(t,y)$ is called linear-quadratic, if it has the form:
\begin{eqnarray}
\varphi(t,y) &=& \left\langle p(t), y \right\rangle + 
\dfrac{1}{2} \left\langle y - y^{(k)}(t), 
\Sigma(t) (y - y^{(k)}(t)) \right\rangle 
\label{ch3_f24}
\end{eqnarray}
with some continuous, piecewise-differentiable vector function  
$p = (p_1, \dots, p_n)$ and matrix function
$\Sigma = (\Sigma_{i,j})_{i,j = \overline{1,n}}$. 
\end{definition} 
 
Based on~\cite{Konnov_Krotov_article_1999} we have the following statements.

\begin{lemma}
Let the following conditions be satisfied for the problem 
(\ref{ch3_f2}) --- (\ref{ch3_f1}): 

\par 1) functions $\mathcal{F}(y)$, $f^0(t,y,u)$ satisfy the conditions 
$\exists K, M < \infty$ $:$ $\forall y \in R^n$ and $\| y \| \geq M$ $\Rightarrow$ 
$\mathcal{F}(y) \leq K \| y \|^2$ and $f^0(t,y,u) \leq K \| y \|^2$ $\forall (t,u) \in [0, T] \times Q$;

\par 2) function $f(t,y,u)$ satisfies the condition 
$\exists K, M < \infty$ $:$ $\quad \forall (t,y,u) \in [0, T] \times \mathbb{R}^n \times Q$ 
and $\| y \| \geq M$ $\Rightarrow$ $f(t,y,u) \leq K \| y \|$;

\par 3) for the Jacobi matrix written for $f = (f_1, \dots, f_n)$, the 
condition $\exists A < \infty  \quad \forall (t,y,u) \in [0, T] \times R^n \times Q$ $:$
$\left\| \dfrac{\partial f_i(t,y,u)}{\partial y_j} \right\| \leq A$ is valid, where 
$\| \cdot \|$ is the matrix norm.

\par Then for the process $v^{(k)} = \left( y^{(k)}, u^{(k)} \right) \in \mathcal{D}$ 
there is the solution $\varphi^{(k)}(t,y)$ of the problem (\ref{ch3_f11}), (\ref{ch3_f10})  
in the form (\ref{ch3_f24}), where the function $p = p^{(k)}$  
is defined as the solution of the Cauchy problem (the system is integrated ``from right to left'')
\begin{eqnarray}
\dfrac{d p^{(k)}(t)}{dt} &=& - \dfrac{\partial H}{\partial y}(t, p^{(k)}(t), y^{(k)}(t), u^{(k)}(t)), \qquad
p^{(k)}(T) = -\dfrac{\partial \mathcal{F}}{\partial y}(y^{(k)}(T)),  
\label{ch3_f25}
\end{eqnarray}
and the matrix function $\Sigma = \Sigma^{(k)}$ is defined as 
\begin{eqnarray}
\Sigma^{(k)}(t) &=& 
\left(  \alpha\left(e^{\gamma(T-t)} - 1\right) + \beta \right) {\mathbb I}_n, 
\label{ch3_f26}
\end{eqnarray}
where the values $\alpha,~\beta < 0$, $\gamma>0$ are given; and ${\mathbb I}_n$ is the
identity matrix.
\end{lemma}   

For the problem (\ref{ch3_f1}) --- (\ref{ch3_f3}), consider the following
iterative process, where $v^{(k)} = (y^{(k)}, u^{(k)})$ and 
$v^{(k+1)} = (y^{(k+1)}, u^{(k+1)})$ are input and output admissible processes, correspondingly,
at $k$th iteration of the method.	     
\begin{enumerate}
	\item Consider a linear-quadratic function $\varphi^{(k)}$ 
	with $\Sigma^{(k)}$ defined
	by (\ref{ch3_f26}) with some  
	values of $\alpha, \beta < 0$, $\gamma > 0$, and find $p^{(k)}$ as
	the solution of the Cauchy problem (\ref{ch3_f25}).
	\item Find the solution $y^{(k+1)}$ of the Cauchy problem
	(\ref{ch3_f12}). Find the control $u^{(k+1)}$ defined by
	\begin{eqnarray}
	u^{(k+1)}(t) &=& \widetilde{u}^{(k)}(t, y^{(k+1)}(t)) := {\rm arg}\max\limits_{u \in Q} 
	H\Big(t, p^{(k)}(t) + \nonumber \\
	&& + \Sigma^{(k)}(t)(y^{(k+1)}(t) - y^{(k)}(t)), y^{(k+1)}(t), u\Big). 
	\label{u_tilde}
	\end{eqnarray}  
\end{enumerate} 
 
\begin{theorem} 
Let for the problem (\ref{ch3_f2}) --- (\ref{ch3_f1}) the conditions 1) --- 3)
from Lemma~3.1 are satisfied. The method (\ref{ch3_f12}),  
(\ref{ch3_f24}) --- (\ref{u_tilde}) provides sequence of processes
$\{v^{(k)} \} \subset \mathcal{D}$ such that $J(v^{(k+1)}) \leq J(v^{(k)})$. Moreover, if 
\begin{eqnarray}
\int\limits_0^T \max\limits_{u \in Q}H(t, p^{(k)}(t), y^{(k)}(t), u)~dt &\neq&
\int\limits_0^T H(t, p^{(k)}(t), y^{(k)}(t), u^{(k)}(t))~dt,   
\label{ch3_f27}	
\end{eqnarray}
then the sequence $\{v^{(k)} \}$ is improving.  
\end{theorem} 

If for some selected values 
$\alpha,~\beta < 0$, $\gamma>0$ there is no improvement,
then one has to adjust these parameters by decreasing $\alpha$, $\beta$ and 
increasing $\gamma$ (also it is possible to change only one of these
parameters), and then, without changing the iteration index, 
calculate the output process $v^{(k + 1)}$ with the updated values
of these parameters~\cite{Konnov_Krotov_article_1999}. If several re-selections do not give 
an improvement, then we stop iterations.   

In (\ref{ch3_f11}), (\ref{ch3_f10}) we use global $\min$ or $\max$.
Nevertheless, in order to simplify the general Krotov method,
it was suggested to use the 1st and 2nd order necessary conditions 
for {\it local} extrema in  (\ref{ch3_f11}), (\ref{ch3_f10})~\cite{Krotov_Feldman_IzvAN_article_1983}.
Consider $R_y^{\varphi^{(k)}}(t, y^{(k)}, u^{(k)}(t)) = 0$,
$G_y^{\varphi^{(k)}}(y^{(k)}(T)) = 0$, which lead to the equation (\ref{ch3_f25}). 
Note that (\ref{ch3_f25}) is the conjugate system
from the theory of Pontryagin maximum principle~\cite{Pontryagin_et_al_book_1962}.
Thus, it shows how to obtain the function $p^{(k)}$ specified in (\ref{ch3_f24}).

For obtaining the function $\Sigma^{(k)}$, consider
\begin{eqnarray} 
d^2 R^{\varphi^{(k)}}(t, y^{(k)}(t), u^{(k)}(t))  &=& 
\langle \Delta y, R_{yy}^{\varphi^{(k)}}(t, y^{(k)}(t), u^{(k)}(t)) \Delta y \rangle \geq 0, 
\label{fff1} \\
d^2 G^{\varphi^{(k)}}(y^{(k)}(T)) &=& \langle \Delta y(T), G_{yy}^{\varphi^{(k)}}(y^{(k)}(T)) \rangle \leq 0 \label{fff2}
\end{eqnarray} 
with the corresponding matrices
$R_{yy}^{\varphi^{(k)}}(t, y^{(k)}(t), u^{(k)}(t)) = {\rm diag}\{ \delta_1(t), \delta_2(t), \dots, \delta_n(t)\}$
and $G_{yy}^{\varphi^{(k)}}(y^{(k)}(T)) = {\rm diag}\{ \alpha_1, \alpha_2, \dots, \alpha_n \}$,
where $\delta_j(t) \geq 0$ and $\alpha_j \leq 0$, $j = \overline{1, n}$. 
Based on (\ref{fff1}), (\ref{fff2}), there is the following Cauchy problem for the function 
$\Sigma^{(k)}$: 
\begin{eqnarray} 
\dfrac{d\Sigma^{(k)}_{i,j}(t)}{dt} &=& \dfrac{\partial^2 f^0(t, y^{(k)}(t), u^{(k)}(t))}{\partial y_i \partial y_j}  
- \sum\limits_{l=1}^n \Big[ \Sigma^{(k)}_{l,i} \dfrac{\partial f_l(t, y^{(k)}(t), u^{(k)}(t))}{\partial y_j} + \nonumber \\
&& + \Sigma^{(k)}_{l,j} \dfrac{\partial f_l(t, y^{(k)}(t), u^{(k)}(t))}{\partial y_i}
+ p_l(t) \dfrac{\partial^2 f_l(t, y^{(k)}(t), u^{(k)}(t))}{\partial y_i \partial y_j}\Big] + \nonumber \\
&& + \begin{cases}
0, &\text{$i \neq j$,}\\
\delta_i(t) > 0, &\text{$i=j$},
\end{cases} 
\label{ch3_f28} \\
\Sigma^{(k)}_{i,j}(T) &=& - \dfrac{\partial^2 F(y^{(k)}(T))}{\partial y_i \partial y_j} - 
\begin{cases}
0, &\text{$i \neq j$,}\\
\alpha_i > 0, &\text{$i=j$},
\end{cases} 
\label{ch3_f29}
\end{eqnarray} 
where $i,j = \overline{1,n}$.

For the problem (\ref{ch3_f2}) --- (\ref{ch3_f1}), use the following iterative process, where $v^{(k)} = (y^{(k)}, u^{(k)})$ and 
$v^{(k+1)} = (y^{(k+1)}, u^{(k+1)})$ are input and output admissible processes, correspondingly, at $k$th iteration. 
 \begin{enumerate}
	\item Define the linear-quadratic function $\varphi^{(k)}$ by obtaining  
	the functions $p^{(k)}$ and $\Sigma^{(k)}$ as the solutions of the Cauchy problems (\ref{ch3_f25}) and
	(\ref{ch3_f28}), (\ref{ch3_f29}) with fixed $\delta_i(t)$, 
	$\alpha_i$ ($i  = \overline{1,n}$), correspondingly.
	\item Find the solution $y^{(k+1)}$ of the Cauchy problem
	(\ref{ch3_f12}). Find the control $u^{(k+1)}$ defined by  (\ref{u_tilde}). 
\end{enumerate} 

Based on~\cite{Krotov_Feldman_IzvAN_article_1983}, consider the following theorem. 
  
\begin{theorem}
Let for the problem (\ref{ch3_f2}) --- (\ref{ch3_f1}) the conditions 1) --- 3)
from Lemma~3.1 are satisfied. The method (\ref{ch3_f12}),  
(\ref{ch3_f24}), (\ref{ch3_f25}), (\ref{u_tilde}),
(\ref{ch3_f28}), (\ref{ch3_f29}) provides sequence of processes 
$\{v^{(k)}\} \subset \mathcal{D}$ such that 
$J(v^{(k+1)}) \leq J(v^{(k)})$. Moreover, if (\ref{ch3_f27}) is satisfied, the the sequence $\{ v^{(k)} \}$ is improving.
\end{theorem}  

\begin{remark} If the process $v^{(k)}$ 
satisfies the Pontryagin maximum principle,
then the Krotov method does not improve 
\cite[p. 64]{Gurman_book_1987} this process, including the case,
when $v^{(k)}$ is not optimal. 
\end{remark}

In (\ref{ch3_f28}), (\ref{ch3_f29}), the functions $\delta_i(t)$ and parameters 
$\alpha_i$ ($i  = \overline{1,n}$) are needed for regulation of improvement.  
The constructions (\ref{ch3_f25}), (\ref{ch3_f28}), (\ref{ch3_f29}) 
are obtained using the necessary conditions for local extrema, 
so the assignment $\delta_i (t)$, $\alpha_i$ is done for compensation 
as much as possible for the lack of global search extrema. 
As noted in~\cite{Krotov_Feldman_IzvAN_article_1983, Konnov_Krotov_article_1999},
if there is no improvement for the selected $\delta_i(t)$, $\alpha_i$ ($i = \overline{1,n}$), 
then (without changing the index $k$) we need to increase 
$\delta_i (t)>0$, reduce $\alpha_i<0$ and find the corresponding process $v^{(k+1)}$.
Taking into account symmetry $\Sigma_{i,j}(t) = \Sigma_{j,i}(t)$, 
it requires to consider $n (n+1)/2$ equations in 
(\ref{ch3_f28}), (\ref{ch3_f29}). 

\subsection{Krotov method for bilinear systems}

\begin{lemma}
For the problem (\ref{ch3_f7}), (\ref{ch3_f6}) with 
$\mathcal{F}(y(T)) = -\left\langle y(T), M y(T) \right\rangle$, $M \geq 0$, the conjugate 
system (\ref{ch3_f25}), considered at some process $v^{(k)} \in \mathcal{D}$, is
\begin{eqnarray}
\dfrac{d p^{(k)}(t)}{dt} &=& -\left(A^{\rm T}(t) + B^{\rm T}(t) u^{(k)}(t) \right) p^{(k)}(t),
\qquad p^{(k)}(T) = 2 M y^{(k)}(T). 
\label{ch3_f33}
\end{eqnarray} 
\end{lemma} 

\begin{lemma} 
Let in the problem (\ref{ch3_f7}), (\ref{ch3_f6}), $\lambda_u > 0$ and $Q = [a_1, b_1] \times \dots \times [a_m, b_m]$. Then for  
(\ref{ch3_f9}) with a linear-quadratic function $\varphi^{(k)}(t,y)$ 
(\ref{ch3_f24}) one has
\begin{eqnarray*} 
\widetilde{u}_l^{(k)}(t,y) &=& 
\begin{cases}
a_l, & \qquad u_{l,{\rm st}}^{(k)}(t,y) < a_l,\\
u_{l,{\rm st}}^{(k)}(t,y), &\qquad u_{l,{\rm st}}^{(k)}(t,y) \in [a_l, b_l],\\
b_l, & \qquad u_{l,{\rm st}}^{(k)}(t,y) > b_l,
\end{cases}   \\
u_{l,{\rm st}}^{(k)}(t,y) &=& 
\dfrac{\left\langle p^{(k)}(t) + \Sigma^{(k)}(t)\Delta y, B_l(t) y \right\rangle}{2\lambda_u}, 
\qquad l = \overline{1,m}, 
\end{eqnarray*}
where the function $p^{(k)}$ satisfies (\ref{ch3_f33}), and $\Sigma^{(k)}$ is defined by 
(\ref{ch3_f26}) or (\ref{ch3_f28}), (\ref{ch3_f29}).
Here $\widetilde{u}_l(t,y)$ are {\it continuous} in $y$.
\end{lemma}

Let the function $\varphi^{(k)}$ to be considered 
in the class of linear functions: $\varphi^{(k)}(t,y) = \left\langle p^{(k)}(t), y \right\rangle$. 
This choice corresponds to the 1st order Krotov method. 
For the maximization problem (\ref{ch3_f11}), where
\[
G^{\varphi^{(k)}}(y(T)) = -\langle y(T), M y(T) \rangle + 
\langle p^{(k)}(T), T\rangle - \langle p^{(k)}(0), x_0\rangle, 
\]
the condition $M \geq 0$, which provides a concave function 
$\mathcal{F}(y(T))$, is important.
 
Consider two cases with respect to $\lambda_u$: $\lambda_u = 0$ and $\lambda_u > 0$. Following~\cite{Krotov_Bulatov_Baturina_AiT_2011, Baturina_Morzhin_2011, Krotov_Morzhin_Trushkova_AiT_2013}, there is

\begin{lemma}
Let in the problem (\ref{ch3_f7}), (\ref{ch3_f6}), 
$m = 1$, $Q = [a, b]$. Then for (\ref{ch3_f9}) with linear function 
$\varphi^{(k)}(t,y) = \left\langle p^{(k)}(t), y \right\rangle$ 
we have:

\par a) if $\lambda_u = 0$, then the function $\widetilde{u}^{(k)}(t,y)$ 
is defined by (\ref{ch3_f20}), where the function $u_{\rm sing}^{(k)}(t,y)$ is
\begin{eqnarray}
u_{\rm sing}^{(k)}(t,y) &=& u^{(k)}(t) +
\dfrac{\langle p^{(k)}(t), \Big(A(t)B(t) - B(t)A(t) - \dfrac{d B(t)}{dt} \Big) y}
{\langle p^{(k)}(t), (B(t))^2 y \rangle};  
\label{ch3_f36}
\end{eqnarray} 

\par b) if $\lambda_u > 0$, then the function $\widetilde{u}^{(k)}(t,y)$
is defined by (\ref{ch3_f21}), where the function $u_{\rm st}^{(k)}(t,y)$ is 
\begin{eqnarray}
u_{\rm st}^{(k)}(t,y) &=& \dfrac{\langle p^{(k)}(t), B y \rangle}{2\lambda_u}. \label{ch3_f37}
\end{eqnarray} 
\end{lemma}

In contrast to the case $\lambda_u = 0$, for $\lambda_u > 0$ 
the function $\widetilde{u}^{(k)}(t,y)$ is continuous in $y$.
The formula (\ref{ch3_f36}) is obtained by differentiating the  
switching function $g^{(k)}(t,y) = \langle p^{(k)}(t), B y \rangle$. 
The function (\ref{ch3_f37}) is derived from 
the condition $\dfrac{\partial H}{\partial u} = 0$. 

\section{Krotov method of the 1st order for controlling quantum	systems governed by the Schr\"{o}dinger and Liouville--von~Neumann equations}

After the articles~\cite{Kazakov_Krotov_AiT_article_1987}
(V.F.~Krotov, V.A.~Kazakov, 1987) and \cite{Krotov_1989} (V.F.~Krotov, 1989),
the next important contribution for adaptation of the Krotov method  
for optimal quantum control are the articles 
\cite{Tannor_Kazakov_Orlov_1992} (D.J.~Tannor, V.A.~Kazakov, V.N.~Orlov, 1992), 
\cite{Somloi_Kazakov_Tannor_article_1993} (J.~Soml\'{o}i, V.A.~Kazakov, D.J.~Tannor, 1993)
which contain the 1st order (with linear function
$\varphi$) Krotov method applied for optimization of controls
in the Schr\"{o}dinger equation. 
Based on these and further publications,
this section outlines theoretical and numerical results on applications of
the 1st order Krotov method for quantum systems governed by the Schr\"{o}dinger and Liouville--von~Neumann equations.  

\subsection{Krotov method for the Schr\"{o}dinger equation}

Сonsider OCP (\ref{ch2_f1}) --- (\ref{ch2_f4}).
The Pontryagin function for this problem is  
\begin{eqnarray}
H(t, q, \psi, u) &=& 2{\rm Re} \left\langle q, 
-\dfrac{i}{\hbar} {\bf H}[u] \psi \right\rangle_{\mathcal{H}} -
\lambda_{\psi} \langle \psi, D(t) \psi \rangle_{\mathcal{H}} - 
\dfrac{\lambda_u \| u \|^2}{S(t)}, 
\label{ch4_f3}
\end{eqnarray} 
where $\psi, q \in \mathcal{H}$, $u \in Q \subseteq \mathbb{R}^m$. By analogy with (\ref{ch3_f14}),
the Krotov Lagrangian for this problem with a linear 
function $\varphi(t, \psi) = 2 {\rm Re} \langle \chi(t), \psi \rangle$ 
is the functional
\begin{eqnarray}  
L^{\varphi}(v) &=& G^{\varphi}(\psi(T)) - \int\limits_0^T R^{\varphi}(t, \psi(t), u(t))dt, \qquad
v = (\psi,u), 
\label{ch4_f1}
\end{eqnarray}
where
\begin{eqnarray} 
G^{\varphi} &=& -\langle \psi(T), O \psi(T) \rangle + 2 {\rm Re} \langle \chi(T), \psi(T) \rangle - 2 {\rm Re} 
\langle \chi(0), \psi(0) \rangle, 
\label{ch4_f2} \\
R^{\varphi} &=& 2 {\rm Re} \left[ \left\langle \chi(t), -\dfrac{i}{\hbar} {\bf H}[u] \psi \right\rangle + 
\left\langle \dfrac{d\chi(t)}{dt}, \psi \right\rangle \right] - 
\lambda_{\psi} \langle \psi, D(t) \psi \rangle - \lambda_u \|u\|^2.
\label{ch4_f2_add} 
\end{eqnarray}

Let operators $O \geq 0$ and $D(t) \geq 0$. 
By analogy with (\ref{ch3_f4}) --- (\ref{ch3_f5_add}) and relying on the work~\cite{Tannor_Kazakov_Orlov_1992} 
we write the 1st order Krotov method. Consider the following iterative process, where $v^{(k)} = (\psi^{(k)}, u^{(k)})$ and 
$v^{(k+1)} = (\psi^{(k+1)}, u^{(k+1)})$ are input and output admissible 
processes, correspondingly, at $k$th iteration of the method. 
\begin{enumerate}
	\item Compute the solution $\chi^{(k)}$ of the Cauchy problem 
\begin{eqnarray}
\dfrac{d \chi^{(k)}(t)}{dt} &=& 
-\dfrac{i}{\hbar} {\bf H}\left[u^{(k)}(t)\right] \chi^{(k)}(t) + 
\lambda_{\psi} D(t) \psi^{(k)}(t),  \nonumber \\
\chi^{(k)}(T) &=& O \psi^{(k)}(T) 
\label{ch4_f7} 
\end{eqnarray}
	(the system is integrated ``from right to left'').  
	\item Find the solution $\psi^{(k+1)}$ of the Cauchy problem
	\begin{eqnarray}
	\dfrac{d \psi^{(k+1)}(t)}{dt} &=& 
	-\dfrac{i}{\hbar} {\bf H}\left[ \widetilde{u}^{(k)}(t, 
	\psi^{(k+1)}(t))\right] \psi^{(k+1)}(t), 
	\qquad \psi^{(k+1)}(0) = \psi_0.  
	\label{ch4_f6} 
	\end{eqnarray}
	Find the control $u^{(k+1)}$ defined by 
	\begin{eqnarray}
	u^{(k+1)}(t) = \widetilde{u}^{(k)}(t, \psi^{(k+1)}(t)) &:=&   {\rm arg}\max\limits_{u \in Q}
	H(t, \chi^{(k)}(t), \psi^{(k+1)}(t), u). 
	\label{ch4_f5} 
	\end{eqnarray} 
\end{enumerate}  
 
By analogy with the proof of Theorem~\ref{theorem:3.1},
one can prove the following theorem.

\begin{theorem} \label{theorem4.1}
For the problem (\ref{ch2_f1}) --- (\ref{ch2_f4}) with operators $O \geq 0$ and $D(t) \geq 0$,
$\lambda_{\psi} \leq 0$ the method (\ref{ch4_f7}) --- (\ref{ch4_f5}) provides 
sequence of processes
$v^{(k)} \in \mathcal{D}$ such that
$J(v^{(k+1)}) \leq J(v^{(k)})$. Moreover, if 
\begin{eqnarray*}
\int\limits_0^T \max\limits_{u \in Q} H \left(t,\chi^{(k)}(t), \psi^{(k)}(t), u \right)dt
&\neq& \int\limits_0^T  H\left(t,\chi^{(k)}(t), \psi^{(k)}(t), u^{(k)}(t)\right)dt,  
\end{eqnarray*}
then the sequence $\{ v^{(k)} \}$ is improving.
\end{theorem}  
 
\begin{remark}
Realization of the Krotov method  
depends on whether there is a regularizer  
$\lambda_u \int\limits_0^T \dfrac{\| u(t) \|^2} {S(t)} dt$ ($\lambda_u > 0$) 
or not (i.e. $\lambda_u = 0$) in the cost criterion (\ref{ch2_f4}). 
For illustration let in the problem (\ref{ch2_f1}) --- (\ref{ch2_f4}) the set
$Q = [a_1, b_1] \times \dots \times [a_m, b_m]$. 

If $\lambda_u = 0$, then for (\ref{ch4_f5}) 
consider the function $\widetilde{u}^{(k)}(t,\psi)$ defined by
\begin{eqnarray*}
\widetilde{u}_l^{(k)}(t,\psi) &=& 
\begin{cases}
a_l, & g_l^{(k)}(t, \psi) < 0, \\
u_{l, {\rm sing}}^{(k)}(t, \psi) \in [a_l, b_l], & g_l^{(k)}(t, \psi) = 0, \\
b_l, & g_l^{(k)}(t, \psi) > 0, 
\end{cases}  
\end{eqnarray*}
where 
\begin{eqnarray*}
g_l^{(k)}(t, \psi) &=& 2{\rm Re} \left\langle \chi^{(k)}(t), 
-\dfrac{i}{\hbar}{\bf H}_l \psi \right\rangle = \dfrac{2}{\hbar} {\rm Im} \left\langle \chi^{(k)}(t), 
{\bf H}_l \psi \right\rangle, \qquad l = \overline{1,m}  
\end{eqnarray*}
are the components of the switching function; functions 
$u_{l, {\rm sing}}^{(k)}(t, \psi)$ can be obtained from the condition 
$\dfrac{d g_l^{(k)}(t, \psi(t))}{dt} = 0$.

If $\lambda_u > 0$, then for (\ref{ch4_f5}) the 
function $\widetilde{u}^{(k)}(t,\psi)$ is defined by 
\begin{eqnarray}
\widetilde{u}_l^{(k)}(t,\psi) &=& 
\begin{cases}
a_l, & u_{l, {\rm st}}^{(k)}(t, \psi) < a_l, \\
u_{l, {\rm st}}^{(k)}(t, \psi), & u_{l, {\rm st}}^{(k)}(t, \psi) \in [a_l, b_l], \\
b_l, & u_{l, {\rm st}}^{(k)}(t, \psi) > b_l, 
\end{cases} 
\label{ch4_f14}
\end{eqnarray}
where the functions 
\begin{eqnarray}
u_{l, {\rm st}}^{(k)}(t, \psi) &=& 
\dfrac{S(t)}{\hbar \lambda_u} {\rm Im} \left\langle \chi^{(k)}(t), {\bf H}_l \psi \right\rangle,
\qquad l = \overline{1,m} 
\label{ch4_f15}
\end{eqnarray} 
are obtained from the condition $\dfrac{\partial H}{\partial u} = 0$. 
\end{remark}

In the beginning of 1990s, the 1st order Krotov method, 
related to the conditions $\lambda_u > 0$, $S(t) \equiv 1$
in the criterion (\ref{ch2_f4}), was considered   
in the articles \cite{Tannor_Kazakov_Orlov_1992} (D.J.~Tannor, V.A.~Kazakov, V.N.~Orlov, 1992) and 
\cite{Somloi_Kazakov_Tannor_article_1993} (J.~Soml\'{o}i, V.A.~Kazakov, D.J.~Tannor, 1993),
and further in the books \cite[pp. 253--259]{KrotovVF_book_NewYork_1996} (V.F.~Krotov, 1996)
and \cite[\S 16.2.2]{TannorD_book_IntroQM_2007} (D.J.~Tannor, 2007).   
In the articles~\cite{KrotovVF_Dokl_AN_2008, KrotovVF_AiT_article_2009} 
(V.F.~Krotov, 2008, 2009) both cases $\lambda_u = 0$ and $\lambda_u > 0$ 
were analyzed. If $\lambda_u > 0$, then a trade-off 
between minimizing $\mathcal{F}$ and the regularizer is important. 
 
Another version of the Krotov method is obtained with 
the following regularizer. Consider the cost functional $J(v)$ with $\lambda_u = 0$ in (\ref{ch2_f4})
and  the following regularized cost criterion:
\begin{eqnarray}
\tilde J(v, v^{(k)}) &=& J(v) + \Gamma(\gamma_u, v, v^{(k)}) \to \min, \label{ch3_f31} \\
\Gamma(\gamma_u, v, v^{(k)}) &=& \gamma_u \int\limits_0^T 
\dfrac{\| u(t) - u^{(k)}(t) \|^2}{S(t)} dt, \qquad \gamma_u > 0. 
\label{ch3_f32}
\end{eqnarray} 
The  modified cost functional $\tilde J(v, v^{(k)})$
is separate for each iteration of the Krotov method, because it 
depends on the current approximation $v^{(k)}$. 

\begin{lemma}
	Suppose a process $\widehat{v} \in \mathcal{D}$ improves a given process $v^{(k)} \in \mathcal{D}$ 
	for the regularized functional (\ref{ch3_f31}), (\ref{ch3_f32}),
	i.e. $\tilde J(\widehat{v}, v^{(k)}) < \tilde J(v^{(k)}, v^{(k)}) = J(v^{(k)})$.
	Then the process $\widehat{v}$ improves $v^{(k)}$ also for 
	the initial functional $J$, i.e. $J(\widehat{v}) < J(v^{(k)})$. 	
\end{lemma}

To prove this Lemma, we will argue from the opposite,
that is, suppose that $J (\widehat{v}) \geq J(v^{(k)})$ simultaneously
with $\tilde J(\widehat{v}, v^{(k)}) < \tilde J(v^{(k)}, v^{(k)}) = J(v^{(k)})$.
Substituting $\widehat{v}$ in (\ref{ch3_f31}), (\ref{ch3_f32}),
we have $\tilde J(\widehat{v}, v^{(k)}) = J(\widehat{v}) + \widehat{c}$,
 $\widehat{c} = \Gamma(\gamma_u, \widehat{v}, v^{(k)})$,
where $J (\widehat{v}) = \tilde J^{(k)} (\widehat{v}) - \widehat{c}$.
It is clear that $\widehat{c} > 0$ if $\widehat{u}(t) \not \equiv u^{(k)}(t)$.
Based on the assumptions that $J(\widehat{v}) \geq J(v^{(k)})$ and
$\tilde J(\widehat{v}, v^{(k)}) < \tilde J(v^{(k)}, v^{(k)}) = J(v^{(k)})$
we obtain
\begin{eqnarray*}
	J(\widehat{v}) &=& \tilde J(\widehat{v}, v^{(k)}) - \widehat{c} \geq
	J(v^{(k)}) > \tilde J(\widehat{v}, v^{(k)}).
\end{eqnarray*}
Thus, we obtain that $\tilde J(\widehat{v}, v^{(k)}) - \widehat{c} > 
\tilde J(\widehat{v}, v^{(k)})$, i.e. $\widehat{c} < 0$. The latter contradicts
the fact that $\widehat{c} > 0$ if $\widehat{u}(t) \not \equiv u^{(k)}(t)$.
If $\widehat{u}(t) \equiv u^{(k)} (t)$, then there is no improvement for $v^{(k)}$
for $\tilde J$. Thus, it turns out that $J(\widehat{v}) < J(v^{(k)})$.
This finishes the proof.

Let in the problem (\ref{ch2_f1}) --- (\ref{ch2_f4})  
$Q = [a_1, b_1] \times \dots \times [a_m, b_m]$ and $\lambda_u = 0$.
Consider the problem of improving the process $v^{(k)} \in \mathcal{D}$.
Using (\ref{ch3_f31}), (\ref{ch3_f32}), one obtains
the corresponding version of the Krotov method, where the formula (\ref{ch4_f14})
is used, and the components of the vector $u^{(k)}_{{\rm st}}(t,\psi)$ are defined  
not by (\ref{ch4_f15}), but as follows:
\begin{eqnarray}
u^{(k)}_{l, {\rm st}}(t,\psi) &=& u_l^{(k)}(t) + 
\dfrac{S(t)}{\hbar \gamma_u} {\rm Im} \left\langle \chi^{(k)}(t), {\bf H}_l \psi \right\rangle,
\qquad l = \overline{1,m}, \qquad \gamma_u > 0. 
\label{ch4_f16}
\end{eqnarray} 
The formula (\ref{ch4_f16}) is obtained from the condition 
\begin{eqnarray*}
	\dfrac{\partial H}{\partial u_l} &=& 
	2 {\rm Re} \left\langle \chi^{(k)}(t), -\dfrac{i}{\hbar} {\bf H}_l \psi \right\rangle
	- \dfrac{2 \gamma_u \left( u_l - u_l^{(k)}(t) \right)}{S(t)} = 0.  
\end{eqnarray*} 

The Krotov method with (\ref{ch3_f31}) 
is used for solving OCPs for Schr\"{o}dinger equation long, e.g., in the article 
\cite{Koch_Palao_Kosloff_et_al_article_2004} (C.P.~Koch, J.P.~Palao, R.~Kosloff, F. Masnou-Seeuws, 2004).
In the article \cite{Sklarz_Tannor_article_2002} (S.E.~Sklarz, D.J.~Tannor, 2002)
the Krotov method combined with  such regularizer was used for OCP with 
the Gross--Pitaevskii equation that will be discussed below in this review. 
The 1st and 2nd order versions of the Krotov method are used 
in combination with this regularization in a number of
articles since 2002, including, for example,
\cite{Palao_Kosloff_Koch_article_2008, Ndong_Koch_article_2010, Eitan_Mundt_Tannor_article_2011, 
Reich_Ndong_Koch_article_2012, Ndong_Koch_Sugny_article_2014, CPKoch_2016_OpenQS,
Basilewitsch_Marder_Koch_2018}. 

The functional (\ref{ch3_f31}), (\ref{ch3_f32}) is designed to regulate distance to the current approximation $u^{(k)}$. Such regularization  
is used in the theory of optimal control, e.g., \cite[p. 61]{Srochko_book_2000} (V.A.~Srochko, 2000). 

\subsection{Krotov method for the Liouville--von Neumann equation} 

The Krotov method was applied for solving in OCPs for open quantum systems, including for systems of the form (\ref{ch2_f10})~\cite{Bartana_Kosloff_Tannor_2001, Goerz_dissertation_Kassel_2015, 
	Reich_dissertation_Kassel_2015, CPKoch_2016_OpenQS, 
	Basilewitsch_Schmidt_Sugny_et_al_2017, Goerz_Motzoi_Whaley_Koch_2017, 
	Goerz_Jacobs_article_2018}. 
The Liouville--von Neumann equation (\ref{ch2_f9}) with a controlled Hamiltonian
follows from (\ref{ch2_f10}) for the dissipator $\mathcal{L}(\rho) \equiv 0$.

Consider the problem (\ref{ch2_f3}), (\ref{ch2_f9}), (\ref{ch2_f12}) for $\lambda_u = 0$ and $\lambda_{\rho} > 0$. The conjugate system is the following:
\begin{eqnarray}
\dfrac{d\sigma^{(k)}(t)}{dt} &=& -\dfrac{i}{\hbar} \left[ {\bf H}[u^{(k)}(t)], \sigma^{(k)}(t) \right] - \lambda_{\rho} D(t),
\qquad \sigma^{(k)}(T) = \rho_{\rm target}. 
\label{ch4_f20}
\end{eqnarray} 
We briefly describe  the application of the 1st order Krotov method for this problem. Consider the following 
iterative process, where $v^{(k)} = (\rho^{(k)}, u^{(k)})$ and 
$v^{(k+1)} = (\rho^{(k+1)}, u^{(k+1)})$ are input and output admissible 
processes, correspondingly, at $k$th iteration of the method. 
\begin{enumerate}
	\item Compute the solution $\sigma^{(k)}$  of the Cauchy problem (\ref{ch4_f20}).
	\item Find the solution $\rho^{(k+1)}$ of the Cauchy problem 
	\begin{eqnarray}
	\dfrac{d \rho^{(k+1)}(t)}{dt} &=& -\dfrac{i}{\hbar} \Big[ {\bf H}[\widetilde{u}^{(k)}(t, \rho^{(k+1)}(t))], 
	\rho^{(k+1)}(t) \Big],\qquad
	\rho^{(k+1)}(0) = \rho_0.  
	\end{eqnarray}
	Find the control $u^{(k+1)}$  defined by 
	\begin{eqnarray}
	u_l^{(k+1)}(t) &=& \widetilde{u}^{(k)}_l(t,
	\rho^{(k+1)}(t)) = \nonumber \\ 
	&:=& \begin{cases}
	a_l, & \quad u^{(k)}_{l, {\rm st}}(t,\rho^{(k+1)}(t)) < a_l, \\
	b_l, & \quad u^{(k)}_{l, {\rm st}}(t,\rho^{(k+1)}(t)) > b_l, \\
	u^{(k)}_{l, {\rm st}}(t,\rho^{(k+1)}(t)), \quad &
	u^{(k)}_{l, {\rm st}}(t,\rho^{(k+1)}(t)) \in [a_l, b_l],
	\end{cases}  
	\end{eqnarray}
	where
\begin{eqnarray}
u^{(k)}_{l, {\rm st}}(t,\rho^{(k+1)}(t)) = u^{(k)}_l(t) &+& 
\dfrac{S(t)}{\gamma_u \hbar} {\rm Im} \Big( \Tr\Big\{ \sigma^{(k)}(t)  
\Big( {\bf H}_l \rho^{(k+1)}(t) - \nonumber \\
&& - \rho^{(k+1)}(t) {\bf H}_l \Big) \Big\} \Big), 
\label{ch4_f19}
\end{eqnarray}
	$l = \overline{1,m}$, and $\gamma_u > 0$.
\end{enumerate}  

\begin{theorem} 
For the problem (\ref{ch2_f3}), (\ref{ch2_f9}), (\ref{ch2_f12}), 
where $\lambda_u = 0$ and $\lambda_{\rho} > 0$, the method 
(\ref{ch4_f20}) --- (\ref{ch4_f19}) provides  sequence of processes 
$\{v^{(k)} \} \subset \mathcal{D}$ such that $J(v^{(k+1)}) \leq J(v^{(k)})$. Moreover, if  
\begin{eqnarray*}
\int\limits_0^T \max\limits_{u \in Q} H \left(t, \sigma^{(k)}(t), \rho^{(k)}(t), u \right)dt
&\neq& \int\limits_0^T  H\left(t, \sigma^{(k)}(t), \rho^{(k)}(t), u^{(k)}(t)\right)dt,  
\end{eqnarray*}
then the sequence $\{v^{(k)}\}$ is improving.
\end{theorem}

\subsection{Zhu--Rabitz and Maday--Turinici methods. The connection with the 1st order Krotov method}

Сonsider OCP (\ref{ch2_f1}) --- (\ref{ch2_f4}) for $O \geq 0$, 
$m=1$, $Q = \mathbb{R}$, $\lambda_u > 0$, $S(t) \equiv 1$, $\lambda_{\psi} = 0$, 
and the Hamiltonian ${\bf H} = {\bf H}_0 - {\bm \mu} u(t)$,
where ${\bm \mu}$ is the dipole moment operator.
Taking into account the specifics of this OCP, 
in the article~\cite{Zhu_Rabitz_article_1998} (W.~Zhu, H.A.~Rabitz, 1998), 
a method for nonlocal improvements was proposed
which we call {\it the Zhu--Rabitz method}.
 
The Pontryagin function for this problem is
$H(t, \chi, \psi, u) = 2 {\rm Re} \left\langle \chi, -\dfrac{i}{\hbar} {\bf H}[u] \psi \right\rangle -
\lambda_u u^2$. Because $H$ is quadratic in $u$, 
the condition $\dfrac{\partial H}{\partial u} = 0$ gives  
\begin{eqnarray}
\widetilde{u}(t, \chi, \psi) &=& -\dfrac{1}{\lambda_u \hbar} {\rm Im}
\left \langle \chi, {\bf} {\bm \mu} \psi \right\rangle. 
\label{ch4_f21}
\end{eqnarray}

Consider the following 
iterative process, where $v^{(k)} = (\psi^{(k)}, u^{(k)})$ and 
$v^{(k+1)} = (\psi^{(k+1)}, u^{(k+1)})$ are input and output admissible 
processes, correspondingly, at $k$th iteration.
\begin{enumerate}
	\item Compute the function $\chi^{(k + 1)}$ solving 
	the Cauchy problem 
	\begin{eqnarray} 
	\dfrac{d\chi^{(k+1)}(t)}{dt} &=& -\dfrac{i}{\hbar}
	\left( {\bf H}_0 - {\bm \mu} \widetilde{u}(t, \chi^{(k+1)}(t), 
	\psi^{(k)}(t)) \right) \chi^{(k+1)}(t), \label{ch4_f22} \\
	\chi^{(k+1)}(T) &=& O \psi^{(k)}(T),  
	\end{eqnarray}  
	obtained by the substitution $\widetilde{u}(t, \chi^{(k+1)}(t), \psi^{(k)}(t))$
	to the conjugate system (\ref{ch4_f7}) instead of $u^{(k)}(t)$ for $\lambda_{\psi} = 0$.
	\item Find the function $\psi^{(k+1)}$ which satisfies the Cauchy problem
	\begin{eqnarray}
	\dfrac{d\psi^{(k+1)}(t)}{dt} &=& -\dfrac{i}{\hbar}
	\left( {\bf H}_0 - {\bm \mu} \widetilde{u}(t, \chi^{(k+1)}(t), 
	\psi^{(k+1)}(t)) \right) \psi^{(k+1)}(t), \label{ch4_f23} \\ 
	\psi^{(k+1)}(0) &=& \psi_0.  
	\end{eqnarray} 
	Find the control $u^{(k+1)}$ using the formula (\ref{ch4_f21}) as 
	\begin{eqnarray} 
u^{(k+1)}(t) &=& \widetilde{u} \left(t, \chi^{(k+1)}(t), \psi^{(k+1)}(t)\right) := \nonumber \\
&:=&
-\dfrac{1}{\lambda_u \hbar} {\rm Im}
\left \langle \chi^{(k+1)}(t), {\bm \mu} \psi^{(k+1)}(t) \right\rangle. 
\label{ch4_f24} 
\end{eqnarray}
\end{enumerate} 

\begin{theorem} 
In the problem (\ref{ch2_f1}) --- (\ref{ch2_f4}), consider $O \geq 0$, $m=1$, $Q = \mathbb{R}$, 
$\lambda_u > 0$, $S(t) \equiv 1$, $\lambda_{\psi} = 0$, 
and the Hamiltonian ${\bf H} = {\bf H}_0 - {\bm \mu} u(t)$.
The Zhu--Rabitz method (\ref{ch4_f21}) --- (\ref{ch4_f24}) gives a process 
$v^{(k+1)} = (\psi^{(k+1)}, u^{(k+1)}) \in \mathcal{D}$
such that $J(v^{(k+1)}) \leq J(v^{(k)})$. 
\end{theorem}

The Zhu--Rabitz method (\ref{ch4_f22}) --- (\ref{ch4_f24}) is provided in
the article~\cite{Zhu_Rabitz_article_1998}, where also 
numerical results are presented which show 
that this method 
gives successive improvements, and on the first 
iterations $J(v)$ is improved  faster than in the subsequent iterations.
In \cite{Sundermann_VivieRiedle_article_1999} 
(K.~Sundermann, R.~de~Vivie-Riedle, 1999), the Zhu--Rabitz method
was applied for the shape function $S(t) = \sin^2(\pi t/T)$.

In the article~\cite{Maday_Turinici_New_formulations__2003} (I.~Maday, G.~Turinici, 2003) 
general method was proposed which includes as particular cases 
the 1st order Krotov method and the Zhu--Rabitz method. 
We call this method as {\it the Maday--Turinici method}. It works as follows.

Fix two parameters: $\delta, \eta \in [0,2]$.  
\begin{enumerate}
	\item Find the solution $\psi^{(k+1)}$ of the Cauchy problem
\begin{eqnarray}
\dfrac{d \psi^{(k+1)}(t)}{dt} &=& 
-\dfrac{i}{\hbar} {\bf H}[\widetilde{u}(t,\psi^{(k+1)}(t); \delta)] \psi^{(k+1)}(t),
\qquad \psi^{(k+1)}(0) = \psi_0.  
\label{ch4_f26}
\end{eqnarray}
Find the control $u^{(k+1)}$ defined by  
\begin{eqnarray} 
u^{(k+1)}(t) &=& \widetilde{u}(t, \psi^{(k+1)}(t); \delta) = \nonumber \\
&:=& (1-\delta) u^{(k)}(t) -
\dfrac{\delta}{\lambda_u \hbar} {\rm Im} \left\langle \chi^{(k)}(t), 
{\bm \mu} \psi^{(k+1)}(t) \right\rangle, \label{ch4_f25}  
\end{eqnarray}
where the functions $u^{(k)}$ and $\chi^{(k)}$ 
are from the previous iteration. 
	\item Find the function $\chi^{(k+1)}$ as the solution
of the Cauchy problem 
\begin{eqnarray} 
\dfrac{d \chi^{(k+1)}(t)}{dt} &=& 
-\dfrac{i}{\hbar} {\bf H}[\widetilde{u}(t, \chi^{(k+1)}(t); \eta)] \chi^{(k+1)}(t),
\nonumber \\
\chi^{(k+1)}(T) &=& O \psi^{(k+1)}(T) \label{ch4_f28}
\end{eqnarray} 
where  
\begin{eqnarray} 
\widetilde{u}(t, \chi^{(k+1)}; \eta) &:=& (1-\eta) \widetilde{u}(t, \psi^{(k+1)}(t); \delta) - \nonumber \\
&&- \dfrac{\eta}{\lambda_u \hbar} {\rm Im}\left\langle \chi^{(k+1)}, {\bm \mu}  \psi^{(k+1)}(t) \right\rangle. \label{ch4_f27}  
\end{eqnarray}  
\end{enumerate} 
 
\begin{theorem}
In the problem (\ref{ch2_f1}) --- (\ref{ch2_f4}), 
consider $O \geq 0$, $m=1$, $Q = \mathbb{R}$, $\lambda_u > 0$, $S(t) \equiv 1$, $\lambda_{\psi} = 0$, 
and the Hamiltonian ${\bf H} = {\bf H}_0 - {\bm \mu} u(t)$.  
Then, for any $\eta, \delta \in [0, 2]$ the Maday--Turinici method 
(\ref{ch4_f26}) --- (\ref{ch4_f27})
provides $J(v^{(k+1)}) \leq J(v^{(k)})$. 
\end{theorem}

The parameters $(\delta, \eta)$ in the Maday--Turinici 
method (\ref{ch4_f25}) --- (\ref{ch4_f28})   
should be correctly chosen. The choice $(\delta, \eta) = (1,0)$ determines 
a version of the 1st order Krotov method, and the choice
$(\delta, \eta) = (1,1)$ leads to the Zhu--Rabitz method. As noted 
in~\cite{Maday_Turinici_New_formulations__2003}, the Maday--Turinici method can 
be better than the 1st order Krotov method or the Zhu--Rabitz
method. The Maday--Turinici method was also developed for the density matrix 
evolution~\cite{Ohtsuki_Turinici_Rabitz_article_2004} (Y.~Ohtsuki, G.~Turinici, H.A.~Rabitz, 2004).

Both the Zhu--Rabitz and the Maday--Turinici methods exploit 
the regularizer $\lambda_u \int\limits_0^T (u(t))^2 dt$, $\lambda_u > 0$ 
in the cost functional. This regularizer implies that the Pontryagin function
is quadratic in $u$ that is convenient for maximizing the Pontryagin function. 
On other hand, if the original cost functional $J$ has no  
term of the type $\lambda_u \int\limits_0^T (u(t))^2 dt$, 
then a trade-off between minimizing the terminal and integral parts of
$J$ is important.

\subsection{Krotov method and other methods in numerical experiments for controlled Schr\"{o}dinger and Liouville--von Neumann equations}

Based on a number of publications since the early 1990s, we outline 
the applications of the 1st order Krotov method and 
other methods in the problem (\ref{ch2_f1}) --- (\ref{ch2_f4}),
where $\psi(t) \in L^2$ or $\psi(t) \in \mathbb{C}^n$.
For the considered OCP, gradient methods 
in control space (the steepest descent method, conjugate gradient method)
are used for a long time~\cite{Gross_Neuhauser_Rabitz_article_1992, Szakacs_Amstrup_Gross_Kosloff_Rabitz_Lorincz_1994}. 

\begin{definition}
The following iterative method is called the steepest descent method for the problem
(\ref{ch2_f1}) --- (\ref{ch2_f4}):
\begin{eqnarray} 
u^{(k)}(t; \beta) &=& 
u^{(k)}(t) + \beta \dfrac{\partial H}{\partial u}\left(t, \chi^{(k)}(t), 
\psi^{(k)}(t), u^{(k)}(t) \right), \qquad \beta >0, 
\label{ch4_f29_2} \\
u^{(k+1)}(t) &=& u^{(k)}(t; \beta = \widehat{\beta}), \qquad
\widehat{\beta} = 
{\rm arg}\min\limits_{\beta > 0} J\left( \psi^{(k)}(\cdot; \beta), u^{(k)}(\cdot; \beta) \right), 
\label{ch4_f29} 
\end{eqnarray}
where $\left(\psi^{(k)}, u^{(k)}\right)$ is a given process that should be improved;
$H(t, \chi, \psi, u)$ is the Pontryagin function (\ref{ch4_f3}); 
$\psi^{(k)}(\cdot; \beta)$ is the solution of the Cauchy problem (\ref{ch2_f1}) 
with control $u^{(k)}(\cdot; \beta)$.
\end{definition}

In the steepest descent method, the complexity of one iteration 
is essentially determined by multiple integration of the Schr\"{o}dinger equation 
for a family of controls $\left\{ u^{(k)}(\cdot; \beta),~~ \beta > 0 \right\}$, which is necessary
to find the value $\widehat{\beta}$ which provides 
a variation of control with the highest possible decrease of $J$ on this iteration.

In addition to (\ref{ch2_f1}) --- (\ref{ch2_f4}) one can set    
spectral constraints on the control. For such problem a modified steepest descent method was developed
~\cite{Gross_Neuhauser_Rabitz_article_1992} (P.~Gross, D.~Neuhauser, H.A.~Rabitz, 1992),
where, in contrast to  (\ref{ch4_f29_2}), the following formula with Fourier transformations
is used:
\begin{eqnarray*}
u^{(k+1)}(t; \beta) &=& u^{(k)}(t) + \nonumber \\
&& + \dfrac{\beta}{2\pi} \int\limits_{-\infty}^{\infty} \left[\int\limits_0^T 
\dfrac{\partial H}{\partial u}\left(t, \chi^{(k)}(t), 
\psi^{(k)}(t), u^{(k)}(t) \right) e^{-i \omega t}dt\right] g(\omega) e^{i \omega t} d\omega,  
\end{eqnarray*}
where $\beta > 0$ and
\begin{eqnarray*}
g(\omega) &=& 
\begin{cases}
1, & \omega_{\min} \leq |\omega| \leq \omega_{\max},\\
0, & |\omega| < \omega_{\min},~~ |\omega| > \omega_{\max}.
\end{cases}    
\end{eqnarray*} 
In the article~\cite{Jirari_Hekking_Buisson_article_2009} (2009),
this modification of the steepest descent method 
was adapted for solving OCP for the Liouville--von Neumann equation (\ref{ch2_f9}). 

The article~\cite{Somloi_Kazakov_Tannor_article_1993} (J.~Soml\'{o}i, V.A.~Kazakov, 
D.J.~Tannor, 1993) uses the 1st order Krotov method 
for OCPs (\ref{ch2_f1}) --- (\ref{ch2_f4}) with $\lambda_u > 0$, $S(t) \equiv 1$, and
$\lambda_{\psi} = 0$ for modelling controlled dissociation of an iodine molecule 
${\rm I}_2$ by the Schr\"{o}dinger equation (\ref{ch2_f1}) with $M = 2$
(two electronic states): 
\begin{eqnarray*}
	\dfrac{d \psi(t)}{dt} &=& -\dfrac{i}{\hbar} 
	\begin{pmatrix}
		{\bf H}_{\rm gr}& -\mu~u(t)\\
		-\mu~u(t)& {\bf H}_{\rm ex}
	\end{pmatrix} \psi(t), \qquad \psi(0) = \psi_{0}. 
\end{eqnarray*}
The operators ${\bf H}_{\rm gr}$, ${\bf H}_{\rm ex}$ and functions $\psi_{\rm gr}$, 
$\psi_{\rm ex}$ describe the ground and excited electronic states;
$\mu$ means equal dipole operators between two electronic states
of iodine molecule (in the general case different dipole operators 
$\mu_{\rm gr, ex}$ and $\mu_{\rm ex, gr}$ are considered); 
each electronic state has Morse type potential 
$V_{j} = D_{{\rm e},j} \{1 - \exp[-\beta_j(r - r_{{\rm e},j})] \}^2$
with some values of the parameters. Here 
$D_{{\rm e},j}$ is the dissociation energy; $r$ is the 
nuclear distance between two atoms; $r_{{\rm e},j}$ is the equilibrium nuclear distance; 
$j \in \{\rm gr, ex \}$. 
It was shown \cite{Somloi_Kazakov_Tannor_article_1993}, that the Krotov method: 
a)~can provide macro-steps compared 
to local improvements produced by the steepest descent method;
b)~is free of expensive operation of finding variational parameter;
c)~can improve $J$ with less computational cost compared to the steepest descent method. Fig.~4 and 6 
in \cite[p. 92]{Somloi_Kazakov_Tannor_article_1993} show 
the dependence of the maximized probability of dissociation on the number 
of solved Cauchy problems (graphs are plotted for up to 20 Cauchy problems) and
demonstrate an advantage of the Krotov method over the gradient method.
Few first iterations of the 1st order Krotov method 
give the main contribution and then the rate of changing the values of $J$ decreases 
(the target population of 99~\% at time $T$ is obtained).
The Husimi transform was applied for the analysis of the optimized control 
in the time-frequency domain. 

The paper~\cite{Szakacs_Amstrup_Gross_Kosloff_Rabitz_Lorincz_1994}
(T.~Szak\'{a}cs, B.~Amstrup, P.~Gross, R.~Kosloff, H.A.~Rabitz, A.~L\"{o}rincz, 1994)
considers the Schr\"{o}dinger equation for two
electronic states for modelling controlled blocking of the molecular bond for iodide CsI.
The 1st order Krotov method is used in combination 
with the method of conjugate gradient in the form of Fletcher--Reeves
in the functional space: first the Krotov method, and then switching to
the Fletcher--Reeves method, starting with the result of the Krotov method. 

In the article~\cite{Zhu_Rabitz_article_1998} (W.~Zhu, H.A.~Rabitz, 1998),
the problem (\ref{ch2_f1}) --- (\ref{ch2_f4}) with the Hilbert space $\mathcal{H} = L^2(\mathbb{R})$, 
the observable 
$O(x) = (\gamma_0 \sqrt{\pi}) e^{-\gamma_0^2 (x - x')^2}$,
$S(t) \equiv 1$, $m  = 1$, $Q = \mathbb{R}$, $\lambda_{\psi} = 0$ was considered.
The goal is to localize the wave packet at a given location $x'$ (its value was taken equal to 2.5) according to
the operator $O$ that is related to the Morse potential
$V(x) = D_0 \left(e^{-\beta(x-x_0)}-1 \right)^2 - D_0$ of the O--H bond where
$D_0$, $\beta$, $x_0$ have some values. The numerical results 
show that the Zhu--Rabitz method in the first few iterations converges to near 80~\%
of its converged value. The article~\cite{Maday_Turinici_New_formulations__2003} (I.~Maday, G.~Turinici, 2003)
considers the same OCP for two values of $x'$ (2.5 and 1.821) and 
represents the comparative results computed with three cases of the Madey--Turinici method.
The pairs $(\delta, \eta)= (1,1)$ and $(\delta, \eta) = (1,0)$ in the method represent 
the Zhu--Rabitz method and the 1st order Krotov method, correspondingly. The third case
is $(\delta, \eta) = (2,0)$. Fig.~2 and~3 in~\cite{Maday_Turinici_New_formulations__2003} 
show first ten iterations of the computations 
for both values of $x'$. According to these figures,
the case $(\delta,\eta)= (1,0)$ is slower or almost the same in comparison to the case $(\delta, \eta) = (2,0)$, 
and these both cases are better than the case  $(\delta,\eta)= (1,1)$. 

In the paper~\cite{Sola_Santamaria_Tannor_article_1998}
(I.~Sola, I.~Santamaria, D.J.~Tannor, 1998), an OCP for the Schr\"{o}dinger equation,
related to the Morse potential with several energy levels, was considered, 
and multi-photon excitations were investigated. 
Fig.~1 in~\cite{Sola_Santamaria_Tannor_article_1998}
shows the comparative results and better performance 
(first 40 iterations are considered)
of the 1st order Krotov method vs three gradient methods, 
including the conjugate gradient method, 
in the problem for maximizing the probability to select the target state.

In the article~\cite{Ho_Rabitz_PhysRev_2010} (T.-S.~Ho, H.A.~Rabitz, 2010)
the problem of maximizing the mean $\left\langle O_T \right\rangle = 
\left\langle \psi(T), O_T \psi(T) \right\rangle$ with respect to the
Schr\"{o}dinger equation is considered and TBQCP method is presented.
For the target observable $O_T$, a time-dependent Hermitian operator $O(t)$
such that $O(t) \geq 0$, $\dfrac{dO(t)}{dt} = 0$ is considered. 
For some OCP related to the Morse potential, Fig.~2, 4 --- 7 in \cite{Ho_Rabitz_PhysRev_2010} 
show that Ho---Rabitz TBQCP method can be faster than the 1st order Krotov method.

The paper~\cite{Palao_Kosloff_Koch_article_2008} (J.P.~Palao, R.~Kosloff, C.P.~Koch, 2008) 
considers the problem (\ref{ch2_f1}) --- (\ref{ch2_f4}) with $\lambda_{\psi} \geq 0$
for a model of the vibrations in a rubidium molecule ${\rm Rb}_2$, where three
electronic states are taken. The goal is to transfer population,
which is initially in level $v=0$ of the electronic ground state,
to level $v=1$ of the same electronic state at time $T$, 
using Raman-like transitions using levels in the
$^1\Sigma_u^+$ excited state, but without populating levels in the upper
electronic state $^1\Pi_g$ during all the time. Fig.~2 
in~\cite{Palao_Kosloff_Koch_article_2008} shows the values 
of the maximized normalized cost functional $J_{\rm norm}$ and the average
value of population in the allowed subspace, $I_P$, as functions of the number
of iterations both with and without state constraint. For both cases,
the 1st order Krotov method, which uses the regularization (\ref{ch3_f31}),
gives a population transfer larger than 99.9~\%. However, as shown at Fig.~2
and~3 in~\cite{Palao_Kosloff_Koch_article_2008}, it is required to use  
the state constraint for avoiding populating the forbidden subspace, and the Krotov method
successfully solves the problem with this constraint. The payment for
the constraint's usage is significant increase in the amount of required
iterations of the Krotov method in comparison to the case without the constraint:
500 vs 17 iterations.

The article~\cite{Kumar_Malinovskaya_2011} (P.~Kumar, S.A.~Malinovskaya, V.S.~Malinovsky, 2011) 
considers the Schr\"{o}digner equation which describes 
a three-level $\Lambda$-system controlled by the pump and Stokes fields. 
The article shows the comparative results computed 
using the 1st order Krotov method,
the Zhu--Rabits method, and the conjugate gradient method, 
which works in the functional space, 
for two OCPs: first, for a complete population transfer,
and, second, for maximizing coherence between two given energy levels
(1st and 3rd). The article shows that all three methods 
are able to find solutions of these OCPs, 
including the cases with $\lambda_{\psi} > 0$ (states constraints).

Further, consider the results for quantum control computed 
by solving OCPs with real-valued states which represent real and imaginary parts of 
complex-valued $\psi(t)$. 

\begin{figure}[h!]
	\centering
	\includegraphics[scale=0.5]{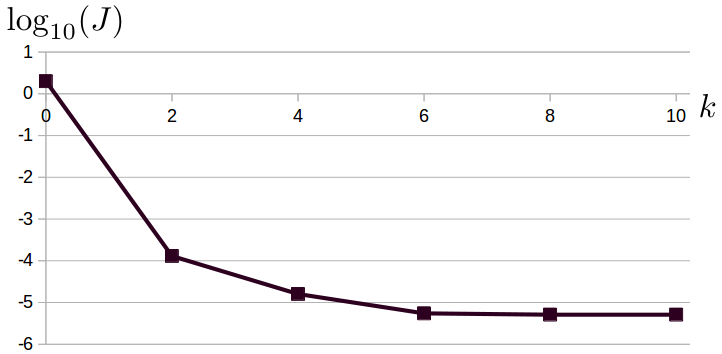} 
	\caption{Illustration of the Krotov method in the OCP for rotations of a molecule.}
\end{figure} 
The works~\cite{Boussaid_Caponigro_Chambrion_IEEE_2012}
(N.~Boussa\"\i d, M.~Caponigro, T.~Chambrion, 2012) 
and~\cite{Trushkova_SaratovUniv_article_2013} (E.A.~Trushkova, 2013)
consider the Schr\"{o}dinger equation with states from $L^2(\Omega; \mathbb{C})$ which 
describes rotations of a planar molecule, and an approximate system 
with states in $\mathbb{C}^n$ where $n=22$. This leads to OCP
(\ref{ch2_f7_add}), (\ref{ch2_f7}) with real-valued states $y(t) \in \mathbb{R}^{44}$, 
where $y_j(t) = {\rm Re} z_j(t)$ and $y_{2j}(t) = {\rm Im} z_j(t)$,
$j  = \overline{1,n}$. For this OCP the article~\cite{Trushkova_SaratovUniv_article_2013} uses the Krotov method 
with $\varphi^{(k)}(t,y)$ obtained from the 
discrete analogue of the Cauchy problem
\begin{eqnarray*}
	H\left(t, \dfrac{\partial \varphi^{(k)}(t,y)}{\partial y}, y, u^{(k)}(t)\right) + 
	\dfrac{\partial \varphi^{(k)}(t,y)}{\partial t} = 0, 
	\qquad \varphi^{(k)}(T,y) = -F(y) 
\end{eqnarray*}
considered in the article~\cite{TrushkovaEA_AiT_2011} (E.A.~Trushkova, 2011). 
Control is restricted by $| u(t) | \leq \frac{1}{3}$ 
according to the operation ${\rm arg}\max\limits_{u}$ in (\ref{ch3_f9}). 
Based on the data from~\cite{Trushkova_SaratovUniv_article_2013} 
Fig.~1 shows the logarithmic dependence of $J$ on the number of iterations.  
For the initial iteration $J = 2$, for 4th iteration $J \approx 1.6 \cdot 10^{-5}$, for 10th
iteration $J \approx 5.1 \cdot 10^{-6}$. 

In the article~\cite{Krotov_Morzhin_Trushkova_AiT_2013}
(V.F.~Krotov, O.V.~Morzhin, E.A.~Trushkova, 2013),
a method, which incorporates the 1st order Krotov method, 
was suggested for the problem (\ref{ch3_f7}), (\ref{ch3_f6}) with 
$\mathcal{F}(y(T)) = -\left\langle y(T), M y(T) \right\rangle$, $M \geq 0$
(and for  a more general class of OCPs).
The idea is to consider a such generalized OCP 
which allows {\it impulsive control} and {\it discontinues state function $y(t)$}. 
An impulsive control combines usual piecewise continuous control function $u(t)$
and impulsive actions, which are used at some time instants. At each such time instant
the system's trajectory is described in the state space $\mathbb{R}^n$ using
some special dynamic system. Due to coupling the solution $y(t)$
of the initial system, which is integrated between the time instances of impulsive actions,
and the solutions of the special system, which is considered at the mentioned time instances, 
we obtain the trajectory which is continuous in the state space.
Sequential improvements of control are build for the generalized OCP.   
After solving this OCP, its solution can be approximated by processes which are 
admissible in the initial OCP. The numerical results are provided for
an OCP for a quantum system defined by the Schr\"{o}dinger equation (\ref{ch2_f1})
with the Landau-Zener Hamiltonian and states $\psi(t) \in \mathbb{C}^2$. For this OCP, its equivalent 
with states $y(t) \in \mathbb{R}^4$ is considered. The Table~4
in~\cite{Krotov_Morzhin_Trushkova_AiT_2013} shows fast decreasing  
of the cost functional in the generalized OCP. 

\section{Krotov method for controlling unitary dynamics and ensembles of solutions of the Schr\"{o}dinger equation}

\subsection{Krotov method of the 2nd order with constraints on quantum states}

In the articles~\cite{Sklarz_Tannor_article_2002} 
(S.E.~Sklarz, D.J.~Tannor, 2002) and~\cite{Reich_Ndong_Koch_article_2012} 
(D.M.~Reich, M.~Ndong, C.P.~Koch, 2012) the 2nd order Krotov method,
which uses (\ref{ch3_f26}) for defining $\Sigma^{(k)}$, 
was developed for OCPs for quantum systems, including 
for modelling control of Bose-Einstein condensate dynamics. 

Consider the problem (\ref{ch2_f34}),
(\ref{ch2_f33}), (\ref{ch2_f3}) (ensemble of solutions of the Schr\"{o}dinger equation) for
$\lambda_{\psi} \geq 0$ without spectral constraints. The Pontryagin function 
for the regularized cost functional $\tilde J(v, v^{(k)})$  (\ref{ch3_f31}), (\ref{ch3_f32}) is 
\begin{eqnarray*}
H\left(t, \{q_j \}, \{ \psi_j \}, u \right) &=& 
-2 {\rm Re} \sum\limits_{j=1}^n \Big\langle q_j, \dfrac{i}{\hbar} {\bf H}[u] \psi_j \Big\rangle -
\gamma_u \| u - u^{(k)}(t) \|^2,  
\end{eqnarray*}
where $\psi_j, q_j \in \mathcal{H}$. Based on 
the publications~\cite{Reich_Ndong_Koch_article_2012}
(D.M.~Reich, M.~Ndong, C.P.~Koch, 2012),
\cite{Goerz_Gualdi_Reich_Koch_et_al_article_2015} 
(M.H.~Goerz, G.~Gualdi, D.M.~Reich, C.P.~Koch, F.~Motzoi, K.B.~Whaley, J.~Vala, M.M.~M\"{u}ller, S.~Montangero, T.~Calarco, 2015), 
we describe the method.

Consider the set of $n$ conjugate systems:
\begin{eqnarray} 
\dfrac{d\chi_j^{(k)}(t)}{dt} &=& -\dfrac{i}{\hbar} {\bf H}[u^{(k)}(t)] \chi_j^{(k)}(t) +
\lambda_{\psi} D(t) \psi_j^{(k)}(t), 
\label{ch6_f4} \\
\chi_j^{(k)}(T) &=& 
-\dfrac{\partial}{\partial \psi_j} \mathcal{F}\left(\{ \psi_j^{(k)}(T) \}_{j = \overline{1,n}} \right), 
\qquad j = \overline{1, n}.
\label{ch6_f4_add}
\end{eqnarray}

For the problem (\ref{ch2_f33}), (\ref{ch2_f34}), (\ref{ch2_f3}), consider the following iterative process, where $v^{(k)} = \left(\{ \psi_j^{(k)}\}, u^{(k)}\right)$ and $v^{(k+1)} = \left(\{ \psi_j^{(k+1)}\}, u^{(k+1)}\right)$ are input and output admissible processes, correspondingly, at $k$th iteration of the method. 
\begin{enumerate}
\item Compute the solutions $\chi_j^{(k)}$, $j = \overline{1,n}$ of $n$ Cauchy problems (\ref{ch6_f4}), (\ref{ch6_f4_add}). 
\item Find the solutions $\psi^{(k+1)}_j$, $j = \overline{1,n}$ of $n$ Cauchy problems  
\begin{eqnarray}
\dfrac{d\psi^{(k+1)}_j(t)}{dt} &=& 
-\dfrac{i}{\hbar} {\bf H} \Big[\widetilde{u}^{(k)}\big(t, \{ \psi_j^{(k+1)}(t) \}_{j= \overline{1,n}} \big)\Big] 
\psi_j^{(k+1)}(t), \nonumber \\ 
\psi_j^{(k+1)}(0) &=& \psi_{j,0}.  
\end{eqnarray} 
Find the control $u^{(k+1)}$ defined by
\begin{eqnarray}
u^{(k+1)}(t) &=& \widetilde{u}^{(k)}\left(t, \{\psi_j^{(k+1)}(t) \}_{j = \overline{1,n}} \right) = \nonumber \\
&:=& {\rm arg}\max\limits_{u \in Q} 
H\bigg(t,  \big\{ \chi_j^{(k)}(t) + \nonumber \\
&&+ \dfrac{1}{2} \Sigma^{(k)}(t) \big(\psi_j^{(k+1)}(t) - 
\psi_j^{(k)}(t) \big) \big\}, \left\{ \psi_j^{(k+1)}(t) \right\}_{j = \overline{1,n}}, u \bigg),
\label{ch6_f2}
\end{eqnarray}
where the function $\Sigma^{(k)}$ is defined in (\ref{ch3_f26}) with some 
$\alpha,~\beta < 0$, $\gamma > 0$. 
\end{enumerate} 

\begin{theorem}
Consider for the OCP (\ref{ch2_f33}), (\ref{ch2_f34}), (\ref{ch2_f3})  
the problem of improving the process $v^{(k)} = \left( \{ \psi_j^{(k)} \}_{j = \overline{1,n}}, 
u \right) \in \mathcal{D}$. Then the method (\ref{ch6_f4}) --- (\ref{ch6_f2}) provides such process $v^{(k+1)}$ that $J(v^{(k+1)}) \leq J(v^{(k)})$. 
\end{theorem}
 
The article~\cite{Reich_Ndong_Koch_article_2012} (D.M.~Reich, M.~Ndong, C.P.~Koch, 2012)
provides numerical results obtained with the 2nd order Krotov method 
for an OCP with the terminant $\mathcal{F}\left( \{ \psi_j(T) \}_{j= \overline{1,n}} \right)$  (\ref{ch2_f37}) which is a polynomial of 8th degree with respect to $\psi_j$ \cite[pp. 7--8] {Muller_Reich_Murphy_et_al_article_2011}. As shown in Fig.~1 in~\cite{Reich_Ndong_Koch_article_2012}, the efficiency of the Krotov method depends substantially on~$\gamma_u$.   

\subsection{Krotov method with constraints on spectrum of the control}

It can happen that spectrum of the control 
contains components which are not desirable for practical
implementation. Following the article~\cite{Palao_Reich_Koch_PhysRev_2013} 
(J.P.~Palao, D.M.~Reich, C.P.~Koch, 2013),
we describe the modification of the Krotov method
for problems (\ref{ch2_f34}), (\ref{ch2_f33}), 
(\ref{ch2_f3}) for $\lambda_{\psi} = 0$, $Q = \mathbb{R}$ with spectral constraints on control. 

At $k$th iteration, consider the process $v^{(k)} \in \mathcal{D}$ 
and the following functional which takes into account both
the regularizer (\ref{ch3_f31}), (\ref{ch3_f32}) and frequencies
$\omega_m$, $m = \overline{1,M}$, for which the filtration is done: 
\begin{eqnarray} 	
J_{\rm spec}(v; v^{(k)}) &=& 
\int\limits_0^T \Big[ \gamma_u \dfrac{\left( u(t) - u^{(k)}(t) \right)^2}{S(t)} + \nonumber \\
&&+ \dfrac{1}{2\pi} \int\limits_0^T \left(u(t) - u^{(k)}(t)\right) K(t-t') 
\left(u(t') - u^{(k)}(t') \right) dt' \Big] dt. \label{ch6_f6}
\end{eqnarray}
Here Gaussian kernel is
\begin{eqnarray}
K(t - t') &=& \sum\limits_{m = 1}^M \lambda_{{\rm spec},m} \sqrt{2\pi \sigma_m^2} 
\cos\left[ \omega_m (t - t') \right] 
\exp\left[ - \dfrac{\sigma_l^2 (t - t')^2}{2} \right]. \label{ch6_f6_add1}
\end{eqnarray}
The values of $\gamma_u$ and $\lambda_{{\rm spec},m}$, $\sigma_m$ ($m = \overline{1,M}$) 
are given, and $S(t)$ is some shape function. 
 
Setting $t = t'$, we have $K(0) = \sum\limits_{m=1}^M
\lambda_{{\rm spec},m} \sqrt{2 \pi \sigma^2_m}$.  
The Fourier transform of $K(t - t')$ is
\begin{eqnarray*}
	\overline{K}(\omega) &=& \sum\limits_{m = 1}^M \dfrac{\lambda_{{\rm spec},m}}{2} 
	\left[ \exp\left( -\dfrac{(\omega - \omega_m)^2}{2 \sigma_m^2} \right) + 
	\exp\left( -\dfrac{(\omega + \omega_m)^2}{2 \sigma_m^2} \right) \right]. \label{ch6_f6_add2}
\end{eqnarray*}

Let the Hamiltonian ${\bf H}$ 
be linear in $u$. Consider the following iterative 
formula$^8$\footnote{In the formula (\ref{ch6_f8}), 
there is the minus sign before the last summand, 
in contrast to the corresponding formulas~(8), (11) 
in~\cite{Palao_Reich_Koch_PhysRev_2013}. At the same time, the kernel $K(t-t')$ 
in the formula~(6b) 
in~\cite{Palao_Reich_Koch_PhysRev_2013}
contains the minus sign before the summation sign, which is absent in the formula 
(\ref{ch6_f6_add1}). Thus, they compensate each other.}:
\begin{eqnarray} 
u^{(k+1)}(t) &=& \widetilde{u}^{(k)}\left(t, \{ \psi_j^{(k+1)}(t), 
j = \overline{1, n} \}\right) = \nonumber \\
&:=& u^{(k)}(t) + \dfrac{S(t)}{\gamma_u \hbar} {\rm Im} 
\Bigg[ \sum\limits_{j=1}^n \left\langle \chi_j^{(k)}(t), 
\dfrac{\partial {\bf H}}{\partial u} \psi_j^{(k+1)} \right\rangle + \nonumber \\
&&+ \dfrac{1}{2} \sum\limits_{j=1}^n 
\left\langle \left(\psi_j^{(k+1)}(t) - \psi_j^{(k)}(t)\right), 
\Sigma^{(k)}(t) \dfrac{\partial {\bf H}}{\partial u} 
\left( \psi_j^{(k+1)}(t) - \psi_j^{(k)}(t) \right) \right\rangle \Bigg] - \nonumber \\
&&- \sum\limits_{m=1}^M \dfrac{\lambda_{{\rm spec},m} S(t)}{2\pi \gamma_u} \sqrt{2\pi \sigma_m^2}
\int\limits_0^T \cos\left[ \omega_m (t - t') \right] \times \nonumber \\
&&\times \exp\left[ -\dfrac{\sigma_m^2 (t - t')^2}{2} \right] 
\left( u(t') - u^{(k)}(t') \right) dt',
\qquad k \geq 0, 
\label{ch6_f8}
\end{eqnarray}
where, at $k$th iteration, 
$u^{(k)}$ and $u^{(k+1)}$ are the current and the next approximations
correspondingly; $\psi_j^{(k+1)}$, $\overline{1, n}$  
are the solutions of the Cauchy problems (\ref{ch2_f34}) with control 
$u(t) = \widetilde{u}^{(k)}\left(t, \{ \psi_j^{(k+1)}(t), \overline{1, n} \}\right)$; 
$\chi^{(k)}_j$, $\overline{1, n}$ are the solutions of the Cauchy problems
(\ref{ch6_f4}), (\ref{ch6_f4_add}) with $\lambda_{\psi} = 0$. 

Due to the linearity of ${\bf H}[u]$ in $u$, the formula (\ref{ch6_f8}) can be
represented as an integral 
Fredholm equation for the increment $\Delta u = u^{(k+1)} - u^{(k)}$:
\begin{eqnarray}
\Delta u(t) &=& I(t) + \beta \int\limits_0^T \mathcal{K}(t,t') \Delta u(t') dt', \label{ch6_f9}
\end{eqnarray}
where
\begin{eqnarray} 
I(t) &=& \dfrac{S(t)}{\gamma_u \hbar} {\rm Im} 
\Bigg[ \sum\limits_{j=1}^n \left\langle \chi_j^{(k)}(t), 
\dfrac{\partial {\bf H}}{\partial u} \psi^{(k+1)}_j(t) \right\rangle + \nonumber \\
&&+ \dfrac{1}{2} \sum\limits_{j=1}^n 
\left\langle \left(\psi_j^{(k+1)}(t) - \psi_j^{(k)}(t)\right), 
\Sigma^{(k)}(t) \dfrac{\partial {\bf H}}{\partial u} 
\left(\psi_j^{(k+1)}(t) - \psi_j^{(k)}(t) \right) \right\rangle \Bigg], \label{ch6_f10} \\
\mathcal{K}(t,t') &=& - \sum\limits_{l=1}^L \dfrac{\lambda_{spec,l} S(t)}{2\pi \gamma_u} 
\sqrt{2\pi \sigma_l^2}
\cos\left[ \omega_l (t - t') \right] \exp\left[ -\dfrac{\sigma_l^2 (t - t')^2}{2} \right].
\label{ch6_f10_add}
\end{eqnarray} 
Formula (\ref{ch6_f10}) includes unknown functions $\psi_j^{(k+1)}$, 
$j = \overline{1, n}$ corresponding to the desired control 
$u^{(k+1)}$ for which the spectral constraints are given. In~\cite{Palao_Reich_Koch_PhysRev_2013} 
the following approach for solving the equation (\ref{ch6_f9}) --- (\ref{ch6_f10_add}) was suggested:
1)~set the function $\Sigma^{(k)} \equiv 0$;
2)~find an improving process $\widehat{v}^{(k+1)} = 
\left(\{ \widehat{\psi}_j^{(k+1)} \}_{j=1}^n, \widehat{u}^{(k+1)} \right)$
for the current process $v^{(k)}$ without spectral 
constraints on control (in (\ref{ch6_f6_add1}) set all $\lambda_{{\rm spec},l} = 0$);
3)~in (\ref{ch6_f10}) substitute the computed functions 
$\widehat{\psi}_j^{(k+1)}$, $j = \overline{1,n}$ instead of unknown functions 
$\psi_j^{(k+1)}$, $j = \overline{1,n}$ for obtaining approximation $\widehat{I}$ for $I$;
4)~solve the equation
\begin{eqnarray}
\Delta u(t) &=& \widehat{I}(t) + \beta \int\limits_0^T \mathcal{K}(t,t') \Delta u(t') dt' 
\label{ch6_f11}
\end{eqnarray}
using the method of degenerate kernels known in the theory of 
integral equations~\cite{AtkinsonKE_IntEq_book_1997} as described in \cite[p. 3]{Palao_Reich_Koch_PhysRev_2013}.

The need to solve the Fredholm equation (\ref{ch6_f9}) or
its simplified form (\ref{ch6_f11}) complicates each iteration 
of the Krotov method. 

\subsection{Maday--Turinici and Krotov methods with modified quality criteria and smoothing of control}

In the articles~\cite{Maximov_et_al_article_2008, Maximov_Salomon_Turinici_Nielsen_article_2010} 
(I.I.~Maximov, N.C.~Nielsen, J.~Salomon, Z.~To\v{s}ner, G.~Turinici, 2008, 2010), 
and also~\cite{Schirmer_Fouquieres_2011} (S.G.~Schirmer, P.~de~Fouquieres, 2011), 
the Maday--Turinici method was applied for the OCP (\ref{ch2_f3}), (\ref{ch2_f21}), (\ref{ch2_f24}) 
considering the unitary operator.
   
Not for every terminant (\ref{ch2_f25}) --- (\ref{ch2_f28}) it is correct to use the 
1st order Krotov method or the Maday--Turinici method. Generalizing the regularized terminant, which was proposed in the paper~\cite{Maximov_et_al_article_2008}, we consider the following regularized cost functional to be minimized:
\begin{eqnarray}
\tilde J_X(U, u; M) &=& \mathcal{F}_X(U(T)) - \Tr \{ U^{\dagger}(T) M U(T )\} +
\lambda_u \int\limits_0^T \dfrac{\| u(t) \|^2}{S(t)} dt, 
\label{ch6_f11_add1}
\end{eqnarray} 
where $M \geq 0$ is some symmetric matrix with real numbers. 

In (\ref{ch6_f11_add1}), consider $\mathcal{F}_O(U(T)) = -\Tr\{ O U(T) \rho_0 U^{\dagger}(T) \}$ 
defined by (\ref{ch2_f26}), and $Q = \mathbb{R}^m$.
Based on~\cite{Maximov_et_al_article_2008}, we describe the 
$k$th iteration of the Maday--Turinici method using (\ref{ch6_f11_add1}). Set some $\delta, \eta \in [0, 2]$.
\begin{enumerate}
\item Compute the solution $U^{(k+1)}$ of the Cauchy problem
\begin{eqnarray}
\dfrac{d U^{(k+1)}(t)}{dt} &=& -\dfrac{i}{\hbar} {\bf H}[\widetilde{u}(t, B^{(k)}(t); \delta)] U^{(k+1)}(t),
\qquad U^{(k+1)}(0) = \mathbb{I}.  
\end{eqnarray}
Find the control $u^{(k+1)}$ defined by
\begin{eqnarray} 
u^{(k+1)}_l(t) &=& \widetilde{u}_l(t, U^{(k+1)}(t); \delta) = \nonumber \\ 
&:=& (1-\delta) u_l^{(k)}(t) +
\dfrac{\delta}{\lambda_u \hbar} {\rm Im} \left\langle B^{(k)}(t), {\bf H}_l U^{(k+1)}(t) \right\rangle, 
\end{eqnarray}
where $l = \overline{1,m}$, the functions $u_l^{(k)}$ and $B^{(k)}$ are from the previous iteration. 
\item Find the function $B^{(k+1)}$ as the solution of the Cauchy problem 
\begin{eqnarray} 
\dfrac{d B^{(k+1)}(t)}{dt} &=& -\dfrac{i}{\hbar} {\bf H}[\widetilde{u}(t,B^{(k+1)}(t); \eta)] B^{(k+1)}(t), 
\label{ch6_f16}\\
B^{(k+1)}(T) &=& O U^{(k+1)}(T) \rho_0 + U^{(k+1)}(T) M, 
\end{eqnarray}
where  
\begin{eqnarray} 
\widetilde{u}_l(t, B^{(k+1)}; \eta) &:=& (1-\eta) \widetilde{u}_l(t, U^{(k+1)}(t); \delta) + \nonumber \\
&=& \dfrac{\eta}{\lambda_u \hbar} {\rm Im}\left\langle B^{(k+1)}, {\bf H}_l U^{(k+1)}(t) \right\rangle, 
\qquad l = \overline{1,m}. \label{ch6_f15}  
\end{eqnarray}  
\end{enumerate}  

In the article~\cite{Maximov_Salomon_Turinici_Nielsen_article_2010} only the case
of $(\delta, \eta) = (1,0)$ is considered and $M = \kappa \mathbb{I}$, $\kappa>0$.
This case means the 1st order Krotov method 
with regularizations with respect to $u$ and $U(T)$. In addition, 
it was suggested to make smoothing for the computed improving control
$u^{(k+1)}$. The vector function   
$u_{\rm smooth}^{(k+1)}(\cdot; \alpha)$ is constructed with the components 
\begin{eqnarray*}
u_{{\rm smooth}, l}^{(k+1)}(\cdot; \alpha) &=& (1- \alpha) u_l^{(k+1)} + 
\alpha F\left( u_l^{(k+1)} \right), \qquad l = \overline{1, m}
\end{eqnarray*}
for some $\alpha \in [0,1]$ $\forall l$, where 
$F$ is a frequency filter, for example, with direct and inverse 
Fourier transformations. Suppose that $u^{(k+1)}$ for $\alpha= 0$ provides
an improvement. Then 
starting from $\alpha = 1$ and decreasing $\alpha$, one can search for such 
a smoothed control which also gives and improvement for $(U^{(k)}, u^{(k)})$. This modification 
is called {\it the smooth Krotov method}. Search for a 
suitable $\alpha \in [0, 1]$ increases the complexity of each iteration.

\subsection{Applications in numerical experiments for controlling unitary dynamics}

The articles~\cite{Palao_Kosloff_2002, Palao_Kosloff_article_2003, 
Palao_Kosloff_Koch_article_2008, Palao_Reich_Koch_PhysRev_2013, 
Muller_Reich_Murphy_et_al_article_2011, Reich_Ndong_Koch_article_2012, 
Goerz_Whaley_Koch_article_2015, Goerz_Gualdi_Reich_Koch_et_al_article_2015} 
consider OCPs for ensembles of the solutions
$\{ \psi_j(t),~j= \overline{1,n} \}$ of the Schr\"{o}dinger equation using
the 1st and 2nd order Krotov methods, including constraints
on the states $\{ \psi_j(t) \}$ and on the spectrum of control.

As noted in~\cite{Palao_Kosloff_article_2003} 
(J.P.~Palao, R.~Kosloff, 2003), there are important questions:
a)~on existence of the solution for the problem of the type  
(\ref{ch2_f34}), (\ref{ch2_f33}), (\ref{ch2_f3}) 
taking into account the feasibility of the computed controls 
in practice and scalability with increasing $n$;
b)~on complexity of computations for finding such control. 

In the article \cite[pp. 7, 8]{Reich_Ndong_Koch_article_2012} (D.M.~Reich, M.~Ndong, C.P.~Koch, 2012), 
Fig.~1 illustrates the first 40 iterations 
of the 1st and 2nd order versions of the Krotov method, 
which were applied for an OCP with the terminant being a polynomial 
of 8th degree with respect to $\{ \psi_j \}$. The iterative process 
is sensitive to $\gamma_u$ in the regularization of the type 
(\ref{ch3_f31}), (\ref{ch3_f32}) and also to setting $\Sigma^{(k)}$ by (\ref{ch3_f26}). 
In particular, for $\gamma_u = 0.133$ the 1st order Krotov method computes fast improvements 
at the first two iterations, but after that the method gives the degraded values of $\mathcal{F}$.
For the same $\gamma_u$, 40 iterations of the 2nd order Krotov method give
the terminant's value equal only near to 0.01. For $\gamma_u = 0.4$, 
both 1st and 2nd order versions of the Krotov method show sufficiently 
good results: the terminant reaches near $10^{-5}$. It means the importance
of the 2nd order Krotov method and usage of an appropriate $\gamma_u$ 
for such nonlinear terminants. 

In~\cite{Goerz_Whaley_Koch_article_2015} (M.H.~Goerz, K.B.~Whaley, C.P.~Koch, 2015) 
a multistage optimization scheme is discussed where 
on the first stage a reduction of an OCP
to the problem of minimizing the corresponding function $\widetilde{\mathcal{F}}$
is performed, and after that the solution of the parameterized problem is taken 
as an initial approximation for the Krotov method.

Consider control parametrization using trigonometric functions.  
The articles~\cite{Goerz_Gualdi_Reich_Koch_et_al_article_2015, 
Goerz_Whaley_Koch_article_2015, Goerz_Motzoi_Whaley_Koch_2017}, 
along with using the Krotov method,   
use the approach based on, first, considering controls in some class
of trigonometric functions parametrized by some amount of parameters 
and, second, reducing an OCP to minimizing the corresponding
function $\widetilde{\mathcal{F}}$ of these parameters. CRAB 
method~\cite{Caneva_Calarco_Montangero_2011} uses the following parameterization: 
\begin{eqnarray}
u(t) &=& u_{\rm guess}(t) \left(1 + S(t) \sum\limits_{j=1}^N 
\bigg( a_j \sin(\omega_{j} t) + b_j \cos(\omega_{j} t)\bigg) \right), 
\label{ch6_f18}
\end{eqnarray}
where $u_{\rm guess}(t)$ is some initial approximation. This formula uses 
the Fourier basis to reflect the physical nature of the control. The corresponding objective function is
$\widetilde{\mathcal{F}}(a_j,~b_j,~\omega_j~|~j=\overline{1,N})$. To reduce the dimension 
of the problem it was suggested to consider 
$\omega_j = 2\pi j (1 + r_j)/T$, where $r_j \in [-0.5, 0.5]$ is a random number chosen with the uniform distribution. In~\cite{Goerz_Gualdi_Reich_Koch_et_al_article_2015, 
Goerz_Whaley_Koch_article_2015, Goerz_Motzoi_Whaley_Koch_2017},
the Nelder--Mead method was applied for minimizing the objective function.
The Krotov method requires differentiability of $\mathcal{F}$, but
the approach with reducing an OCP to the problem
of minimizing the corresponding $\widetilde{\mathcal{F}}$ with  the Nelder-Mead method
does not require finding the gradient $\nabla \widetilde{\mathcal{F}}$. 

Usage of (\ref{ch6_f18}) narrows the search space in an OCP. 
However, an appropriate parameterization can be useful for obtaining 
analytical formulas for the control, which can be then improved using the Krotov method. 
As noted in~\cite{Goerz_Whaley_Koch_article_2015}, in the case 
of a nonlinear terminant $\mathcal{F}$ it is possible to have 
the situation when the Krotov method, considered for some
given initial approximation, stumbles upon ``plateau''
and is not efficient enough in the sense that several thousand 
iterations can give insufficient results. In this case, it is useful 
to have some pre-optimization with OCP reduction to the corresponding 
finite-dimensional optimization followed by application of the 
Nelder--Mead method. 

In the articles~\cite{Palao_Kosloff_Koch_article_2008} 
(J.P.~Palao, R.~Kosloff, C.P.~Koch, 2008) and 
\cite{Palao_Reich_Koch_PhysRev_2013} (C.P.~Palao, D.M.~Reich, C.P.~Koch, 2013) 
the Krotov method is used together with taking into account constraints on states 
$\{ \psi_j(t) \}$ and on the spectrum of $u$. 
The quantum Fourier transform ($W_{\rm QFT}$ gate), which is 
based on the unitary transformation in the model 
with three electronic states for a molecule ${\rm Rb}_2$, is considered.
The constraint on states is used to specify that the upper electronic 
state $^1 \Pi_g$) is forbidden. Optimization of control for a unitary transformation 
under states constraints is obtained successfully by the 1st order Krotov method.
Fig.~6 in~\cite{Palao_Kosloff_Koch_article_2008} 
shows that 50 iterations of the method are not sufficient for avoiding populating the
forbidden subspace, and 500 iterations of the method provide almost zero population
in the forbidden subspace during all the time. State constraints 
lead to inhomogeneous equations (\ref{ch6_f4}), with respect to which we note the article~\cite{Ndong_TalEzer_Kosloff_Koch_2009}
(M.~Ndong, H.~Tal-Ezer, R.~Kosloff, C.P.~Koch, 2009)
devoted to a Chebychev propagator for such equations.  
The effectiveness of the Krotov method (\ref{ch6_f6}) --- (\ref{ch6_f11}),
which provides improvements of control with filtering its spectrum,
is illustrated in the article~\cite{Palao_Reich_Koch_PhysRev_2013}. 

A modification of the Krotov method was used to estimate time and gate complexity 
of generating multi-qubit unitary operators~\cite{Koike_Okudaira_2010}
(T.~Koike, Y.~Okudaira, 2010). As target operators, 
the unitary operator which realizes $N$-qubit quantum Fourier transform $W_{\rm QFT}$
and a certain unitary operator $W_{\rm f}$ which does not have an apparent 
symmetry were considered. The operator $W_{\rm QFT}$ has a polynomial gate complexity and  
$W_{\rm f}$ was constructed so that it is expected to have exponential complexity. 
The fidelity ${\cal F}_W$ is defined by (\ref{ch2_f25}), where $W=W_{\rm QFT}$ or $W=W_{\rm f}$. 
The modified Krotov scheme was implemented to obtain solutions of the 
fidelity-optimal and time-optimal quantum computation theory. As a result, the time 
complexity for the $W_{\rm QFT}$ was found to be linear in the number of qubits, 
while the time complexity for the $W_{\rm f}$ is found to be exponential. 

The article~\cite{Maximov_et_al_article_2008} (I.I.~Maximov, Z.~To\v{s}ner, N.C.~Nielsen, 2008) 
describes the method incorporating the Maday--Turinici method and
regularization for $U(T)$ (see (\ref{ch6_f11_add1}) --- (\ref{ch6_f15}) with
$M = \kappa \mathbb{I}$ and $\kappa>0$) and applications of this method
for OCPs related to nuclear magnetic resonance, dynamic nuclear polarization.
Different values of the parameters $\delta, \eta \in [0,2]$ were used. 
For a model of  two-spin Hermitian coherence transfer, Fig.~4 in~\cite{Maximov_et_al_article_2008}
illustrates the results of this method for the same 
initial approximation $u^{(0)}$ and for different $\delta, \eta \in [0,2]$.
This figure shows how many iterations are needed to achieve 
a change of the cost functional below $10^{-4}$. Almost the same results 
can be achieved in 40 --- 50 iterations for some pairs $(\delta, \eta)$,
in 100 --- 200 iterations for some other pairs, etc. 
For example, for $(\delta,~\eta) = (1, 0)$, which
corresponds to a version of the Krotov method,
the amount of iterations is 60. For a model of coherence transfer, 
Fig.~13 in~\cite{Maximov_et_al_article_2008} shows
that increasing number of spins gives faster growing
of the processor time for GRAPE, in comparison
to the Maday--Turinici method combined with the 
regularized terminant. In the case of five spins, 
the difference in complexity is 3.8 times. However, 
note the effectiveness of the Maday--Turinici method
depends on the parameters $\lambda_u$, $\delta$, $\eta$, 
and, in addition, on $\kappa$. 

\section{Krotov method of the 2nd order for controlling Bose--Einstein condensate governed by the Gross--Pitaevskii equation} 

\subsection{Nonlinear dynamics and the 2nd order Krotov method}  

In the problem (\ref{ch2_f1}) --- (\ref{ch2_f4})   
the dynamic equation (Schr\"{o}dinger equation) is linear with respect to $\psi$, 
and the terminant $\mathcal{F}$ is defined for a positive semi-defined 
operator $O$. As noted in Section~4, under these conditions, one can use the Krotov method 
with a linear function $\varphi(t, \psi) = 2{\rm Re} \left \langle 
\chi(t), \psi \right\rangle$, the Zhu--Rabitz method, 
and the Maday--Turinici method. 
In the problem (\ref{ch2_f13}), (\ref{ch2_f3}), (\ref{ch2_f19}) 
the dynamic equation (Gross--Pitaevskii equation) 
is non-linear in $\psi$. Therefore the 2nd order 
(linear-quadratic function $\varphi$) version of the Krotov method is needed.

In~\cite{Sklarz_Tannor_article_2002} (S.E.~Sklarz, D.J.~Tannor, 2002), 
an important step was done: the 2nd order Krotov method was extended to OCPs 
with controlled the Gross--Pitaevskii equation. The article 
\cite{Jager_Reich_Goerz_et_al_2014} (G.~J\"{a}ger, D.M.~Reich, M.H.~Goerz, C.P.~Koch, U.~Hohenester, 2014)
also is devoted to applying the Krotov method for optimization of controls for Bose-Einstein 
condensate.

Consider OCP for the Gross--Pitaevskii equation (\ref{ch2_f13}) 
together with the cost functional $J(v)$ with $\lambda_u = 0$, 
$\lambda_{du} = 0$ given in (\ref{ch2_f19}). 
For each iteration of the Krotov method, we consider 
the regularized cost functional $\tilde J(v,v^{(k)})$
given in (\ref{ch3_f31}). The Pontryagin function is 
\begin{eqnarray}
H(t, q, \psi, u) &=& 2{\rm Re} \left\langle q, 
-\dfrac{i}{\hbar} \left( K + V(\cdot, u) + \kappa | \psi |^2 \right) \psi \right\rangle -
\gamma_u \dfrac{\| u - u^{(k)}(t) \|^2}{S(t)}, \label{ch5_f4}
\end{eqnarray} 
where $q, \psi \in \mathcal{H}$,
$u \in Q$, $\lambda_u > 0$. Similar to Definition~\ref{Def3.6} (the formula (\ref{ch3_f24})), consider
the following definition.
 
\begin{definition}
The function $\varphi(t,\psi)$ is called linear-quadratic if it has the following form: 
\begin{eqnarray}
\varphi(t,\psi) &=& \langle \chi(t), \psi \rangle_{L^2} + 
\langle \psi, \chi(t) \rangle_{L^2} + \nonumber \\
&& + \dfrac{1}{2} \left\langle \psi - 
\psi^{(k)}(t), \Sigma(t) (\psi - \psi^{(k)}(t)) \right\rangle_{L^2},
\label{ch5_f1} 
\end{eqnarray} 
where $\psi^{(k)}$ is the solution of the Cauchy problem (\ref{ch2_f13}) for $u = u^{(k)}$;  
$\chi$ and $\Sigma$ are some continuous functions.
\end{definition}  

For the OCP (\ref{ch2_f13}), (\ref{ch2_f3}), (\ref{ch2_f19}) 
with $Q = \mathbb{R}$,  $\lambda_u = 0$, 
$\lambda_{du} = 0$ consider the problem of improving the process 
$v^{(k)} = \left(\psi^{(k)}, u^{(k)} \right) \in \mathcal{D}$  
according to the regularized cost functional $\tilde J(v, v^{(k)})$ 
(\ref{ch3_f31}), (\ref{ch3_f32}). Based on~\cite{Sklarz_Tannor_article_2002, Jager_Reich_Goerz_et_al_2014}, 
we formulate the following iterative process, where $v^{(k)} = (\psi^{(k)}, u^{(k)})$ and 
$v^{(k+1)} = (\psi^{(k+1)}, u^{(k+1)})$ are input and output admissible processes, correspondingly,
at $k$th iteration of the method.   
\begin{enumerate}
	\item Compute the matrix function $\Sigma^{(k)}(t)$ according to  
	the formula (\ref{ch3_f24}) with some values of $\alpha, \beta < 0$, $\gamma > 0$,
	and find the solution $\chi^{(k)}$ of the Cauchy problem
	\begin{eqnarray}
	\dfrac{d\chi^{(k)}(t)}{dt} &=&  
	- \dfrac{i}{\hbar} \Big( K + V[u^{(k)}(t)] + 2 \kappa \left| \psi^{(k)}(t) \right|^2 \Big) \chi^{(k)}(t) + \nonumber \\
	&& + i \kappa \left( \psi^{(k)}(t) \right)^2 \chi^{\ast (k)}(t), \label{ch5_f5} \\
	\chi^{(k)}(T) &=& -\dfrac{\partial \mathcal{F}}{\partial \psi^{\ast}(T)} (\psi^{(k)}(T));
	\label{ch5_f6}
	\end{eqnarray} 
	\item Find the control $u^{(k+1)}$ by 
	\begin{eqnarray}
	u^{(k+1)}(t) &=& u^{(k)}(t) + 
	\dfrac{S(t)}{\gamma_u \hbar} {\rm Im} \Bigg[ \left\langle \chi^{(k)}(t),
	\dfrac{\partial V}{\partial u}\bigg|_{u^{(k+1)}(t)} ~\psi^{(k+1)}(t) \right\rangle + \nonumber \\
	&& + \dfrac{1}{2} \left\langle \psi^{(k+1)}(t) - \psi^{(k)}(t), \Sigma^{(k)}(t)
	\dfrac{\partial V}{\partial u}\bigg|_{u^{(k+1)}(t)}~ \psi^{(k+1)}(t) \right\rangle \Bigg],   
	\label{ch5_f11}
	\end{eqnarray} 
	where the function $\psi^{(k+1)}$ is the solution of the Cauchy problem (\ref{ch2_f13}) with control $u = u^{(k+1)}$.  
\end{enumerate}  

\begin{theorem} For the OCP (\ref{ch2_f13}), (\ref{ch2_f3}), (\ref{ch2_f19})
with $Q = \mathbb{R}$, $\lambda_u = 0$, $\lambda_{du} = 0$ the method (\ref{ch5_f1}) --- (\ref{ch5_f11}),
which uses the regularization (\ref{ch3_f31}), (\ref{ch3_f32}) with $\gamma_u > 0$, provides the property
$J(v^{(k+1)}) \leq J(v^{(k)})$.  
\end{theorem}  

\begin{remark}
Formulas (\ref{ch2_f14}), (\ref{ch2_f16}) --- (\ref{ch2_f18}) 
describe substantially different potentials in the Gross--Pitaevskii equation. 
The potential (\ref{ch2_f14}) is linear in $u$.
Unlike (\ref{ch2_f14}), control in the potential (\ref{ch2_f18}) 
enters in a polynomial form. Formula (\ref{ch5_f11}) is obtained from the condition
\[
\dfrac{\partial}{\partial u} H\left(t, \chi^{(k)}(t) + \dfrac{1}{2}\Sigma^{(k)}(t) \psi^{(k+1)}(t),
\psi^{(k+1)}(t), u \right) = 0,
\]
where $H$ is defined by $(\ref{ch5_f4})$. 
If $V$ depends linearly on $u$, then the r.h.s.
of (\ref{ch5_f11}) does not contain $u^{(k+1)}(t)$, i.e. it is easy to compute $u^{(k+1)}$
by (\ref{ch5_f11}). If $V$ depends nonlinearly on $u^{(k+1)}$,
then the complexity of (\ref{ch5_f11}) is completely different in comparison to the linear case.  
\end{remark} 
 
In the article~\cite{Sklarz_Tannor_article_2002}, it was noted that the function $\Sigma^{(k)}$ 
can either be computed from the solution of a special Cauchy problem 
(by analogy with~\cite{Krotov_Feldman_IzvAN_article_1983}),
or can be specified according to the formula (\ref{ch3_f26}). The computations were performed 
involving the Krotov method with the regularization
(\ref{ch3_f31}), (\ref{ch3_f32}). In~\cite{Jager_Reich_Goerz_et_al_2014} the function $\Sigma^{(k)}$ 
is determined by (\ref{ch3_f26}). In the dissertation 
\cite[p. 34] {JagerG_thesis_2015} (G.~J\"{a}ger, 2015) 
it is noted that the Krotov method with
linear-quadratic $\varphi^{(k)}$ drastically depends on the 
values of $\alpha$, $\beta$, $\gamma$ in (\ref{ch3_f26}).
As discussed in \cite[p. 5]{Sklarz_Tannor_article_2002}, one can start 
from $\Sigma^{(0)} \equiv 0$ with subsequent decreasing
$\alpha,~ \beta$ and increasing $\gamma$. If (\ref{ch5_f11}) does not provide improvement of the process $v^{(k)}$, 
then we have to adjust $\alpha$, $\beta$, $\gamma$ and repeat the computations.
	  	  
In~\cite{Jager_Reich_Goerz_et_al_2014} and other publications, an important tool  
is the regularization   (\ref{ch3_f31}), (\ref{ch3_f32}) 
with parameter $\gamma_u$. In \cite[p. 86]{JagerG_thesis_2015}, this parameter 
is called ``step'' in the context of the Krotov method. 

For the criterion (\ref{ch2_f15}),
the transversality condition (\ref{ch5_f6}) is \cite{Sklarz_Tannor_article_2002}:
\begin{eqnarray*} 
\chi^{(k)}(T) &=& -{\rm Re}\big[\psi^{(k)}(T)\big]  
+\dfrac{1}{2} \big|\psi^{(k)}(T)\big| \left\langle \cos(\theta^{(k)}(T)) \right\rangle
\left( \dfrac{\psi^{(k)}(T)}{(\psi^{\ast (k)}(T))} + 3 \right).  
\end{eqnarray*}
The transversality condition for the terminant $\mathcal{F}_{\psi_{\rm target}}(\psi(T))$ 
(\ref{ch2_f5}) is  
\begin{eqnarray}
\chi^{(k)}(T) &=& \left\langle \psi_{\rm target}, 
\psi^{(k)}(T) \right\rangle \psi_{\rm target}. \label{ch5_f9}
\end{eqnarray}
In the article~\cite{Jager_Reich_Goerz_et_al_2014}  
the terminant (\ref{ch2_f5}) is considered, and the conjugate system
is written for the function $p(t) = i \chi(t)$. The condition (\ref{ch5_f9}) has the form  
$p^{(k)}(T) = i \left\langle \psi_{\rm target}, \psi^{(k)}(T) \right\rangle \psi_{\rm target}$.
 
For OCPs where $V$ depends on $u$  nonlinearly, in~\cite{Jager_Reich_Goerz_et_al_2014}, the following simplifications
for (\ref{ch5_f11}) were suggested: a)~consider $\Sigma^{(k)} \equiv 0$  
(in \cite[p. 8]{Jager_Reich_Goerz_et_al_2014}, for such variant of the Krotov method
there is a description how to solve the equation derived from (\ref{ch5_f11}));
b)~use $\dfrac{\partial V}{\partial u}\bigg|_{u^{(k)}(t)}$ 
instead of $\dfrac{\partial V}{\partial u}\bigg|_{u^{(k+1)}(t)}$. Taking into account 
both simplifications, consider 
\begin{eqnarray} 
u^{(k+1)}(t) &\approx& u^{(k)}(t) + 
\dfrac{S(t)}{\gamma_u \hbar} {\rm Im} \bigg[ \left\langle \chi^{(k)}(t),
\dfrac{\partial V}{\partial u}\bigg|_{u^{(k)}(t)} ~\psi^{(k+1)}(t) \right\rangle \bigg] 
\label{ch5_f12}
\end{eqnarray}
instead of (\ref{ch5_f11}). Then this formula was applied, with the remark that for sufficiently small values of the
parameter $\kappa>0$ in the Gross--Pitaevskii equation the control $u$ 
moderately varies from one iteration to another. Judging by the computational 
results described in~\cite{Jager_Reich_Goerz_et_al_2014}, the Krotov method 
in the simplified version with (\ref{ch5_f12}) was successful. 
The difficulty due to nonlinearity of $V$ in $u$ 
complicates the application of the   method for controlling Bose--Einstein condensate.

\subsection{Krotov and GRAPE methods in numerical experiments for controlling Bose--Einstein condensate}

First, we explain the basics of the GRAPE 
method~\cite{Khaneja_Reiss_Kehlet_SchulteHerbruggen_Glaser_2005}
(N.~Khaneja, T.~Reiss, C.~Kehlet, T.~Schulte-Herbr\"{u}ggen, S.J.~Glaser, 2005)
which is often used to solve OCPs for quantum systems. 
In this method: 1)~control $u: [0,T] \mapsto \mathbb{R}$ is represented as a 
piecewise constant function 
\begin{eqnarray} 
u(t) &=& c_j, \quad t \in [t_j, t_{j+1}), \quad t_j = j T/N, \quad j = \overline{0, N}, \label{ch5_f13} 
\end{eqnarray}
where $c_j \in \mathbb{R}$, the control is $c = [c_0, \dots, c_N]$ 
and $\Delta t = T/N$ is the discretization step; 
2)~the OCP is reformulated as the problem of minimizing some function   
$\widetilde{\mathcal{F}}(c)$. Then gradient optimization methods are used. 
GRAPE is also used~\cite{deFouquieres_Schirmer_Glaser_Kuprov_2011},~\cite{Jager_Reich_Goerz_et_al_2014} 
with the BFGS (Broyden--Fletcher--Goldfarb--Shanno) method.
Other finite-dimensional optimization methods are also used, 
including methods which do not require differentiation of the objective function:
the Nelder--Mead method (downhill simplex method), 
particle swarm optimization, differential evolution methods. 

For the OCP with linear potential (\ref{ch2_f14}) and the criterion 
(\ref{ch2_f15}), the article~\cite{Sklarz_Tannor_article_2002}
(S.E.~Sklarz, D.J.~Tannor, 2002) shows the results of the 2nd order
Krotov method, which uses (\ref{ch3_f26}) for computing the function 
$\Sigma^{(k)}$. As shown at Fig.~3 in~\cite{Sklarz_Tannor_article_2002},
the solution was found in near 30 iterations with sequential improvements.
Fig.~4 in~\cite{Sklarz_Tannor_article_2002} represents the phase 
$\theta(T)$ related to the optimized control.

In~\cite{Jager_Reich_Goerz_et_al_2014} 
(G.~J\"{a}ger, D.M.~Reich, M.H.~Goerz, C.P.~Koch, U.~Hohenester, 2014) 
two OCPs were considered for the Gross--Pitaevskii 
equation of a Bose--Einstein condensate. 
The goal in the first OCP is to realize splitting of the Bose-Einstein condensate 
at time $T$, so that the wave function at time $T$ should correspond to the ground state 
of the two-well potential. The second OCP is for shaking the condensate, 
where the goal for the anharmonic single-well potential $V(x-u(t))$  is 
to move the condensate from the ground state $V$, where the system is 
at the time $t=0$, to the first excited state. For the OCP with splitting Bose-Einstein condensate,  
this OCP was solved independently by: a)~using the Krotov method in its simplified version 
using (\ref{ch5_f12}); b)~applying the GRAPE-BFGS method with $H_1$-regularization
(\ref{ch2_f20}). Fig.~1~(b,c) in~\cite{Jager_Reich_Goerz_et_al_2014} represents the density
$n(x,t) = |\psi(x,t)|^2$ on the space--time plane, and shows
that splitting of the Bose-Einstein condensate into two parts is achieved. 
Fig.~2 in~\cite{Jager_Reich_Goerz_et_al_2014} shows that the Krotov method 
using (\ref{ch5_f12}) gives sequential improvements, the efficiency of which depends 
on the parameter $\kappa$ in the Gross--Pitaevskii equation. 
For $\kappa = \pi/2$ the Krotov method using (\ref{ch5_f12})
is faster than GRAPE-BFGS-${\rm H}_1$ in terms of the number of solved 
Cauchy problems: by the Krotov method the terminant $\mathcal{F} < 10^{-4}$, and the complexity  
equals to solving 100 Cauchy problems. For the OCP with shaking the Bose-Einstein condensate, 
as seen in Fig.~4 in~\cite{Jager_Reich_Goerz_et_al_2014}: a)~solving this 
OCP turned out to be significantly more difficult (hundreds of Cauchy problems) 
in comparison with solving OCP for splitting the condensate; 
b)~the Krotov method using (\ref{ch5_f12}) for the problem of shaking the Bose-Einstein condensate 
yields consistent improvements, and for $\kappa = 2\pi$ 
it shows much better efficiency than GRAPE-BFGS-${\rm H}_1$. 
Thus, the 1st order Krotov method using (\ref{ch5_f12}) in the described OCP 
can be successful for some values of the parameters $\kappa$ and $\gamma_u$.
 
\section{Conclusions}

Starting its development in the late 1970s,
quantum control has now become a large interdisciplinary direction
of great importance for the science
and technology of present and future. This field is related to diverse 
areas of mathematics (e.g., differential equations, optimal control,
functional analysis, group theory, differential geometry, 
finite-dimensional optimization, algebra), physics and chemistry 
(control of molecular dynamics, nuclear magnetic resonance, laser 
chemistry, etc.), quantum computing and 
information~\cite{Butkovsky_Samoilenko_book_1984_1990, Krasnov_Shaparev_Shkedov_book_1989, 
Rice_Zhao_book_2000, Bandrauk_Delfour_Bris_Edt__book_AMS_2003, DAlessandro_Directions_2003, 
Brumer_Shapiro_book_2003, TannorD_book_IntroQM_2007, Letokhov_book_2007, DAlessandro_book_2007, Fradkov_book_2007, 
Brif_Chakrabarti_Rabitz_article_2010, Dong_Petersen_2010, Wiseman_Milburn_book_2010, 
Altafini_Ticozzi_article_IEEE_2012, Bonnard_Sugny_2012, Gough_article_2012, CongS_book_2014,
Dong_Wu_Yuan_Li_Tarn_2015, Glaser_Boscain_Calarco_et_al_2015, CPKoch_2016_OpenQS, Borzi_book_2017}.

Formulation of a quantum control problem includes defining a suitable 
mathematical model of the controlled system and cost functional subject 
to maximization or minimization. A mathematical model should 
effectively describe the controlled dynamics of the real quantum system. 
Modelling should involve description of space of quantum states of the system and their controlled evolution equation which can be Schr\"odinger 
or Liouville--von Neumann equation for a closed quantum system, 
Gross--Pitaevskiy equation for wave function of Bose--Einstein 
condensate, master equation with control for an open quantum system.

A cost functional includes terminant $\mathcal{F}$ which represents the 
control goal, and additional integral terms used to impose 
constraints on control, admissible states,   or to improve quality 
of optimization methods. Terminant for maximizing the expectation
of target observable $O$ of the quantum system  at a finite moment $T$ is 
$\mathcal{F}=\left\langle \psi(T), O \psi(T) \right\rangle$;
for minimizing distance to the target state  $\mathcal{F}=\left\| 
\psi(T) - \psi_{\rm target} \right\|^2$, etc. 
The cost functional can 
include integral terms $\lambda_u \int\limits_0^T \dfrac{\| u(t) \|^2}{S(t)} dt$ 
and $\lambda_{\psi} \int\limits_0^T \left\langle \psi(t), D(t) \psi(t) \right\rangle dt$
to restrict control and quantum states, correspondingly. Along with the integral constraint
on $u$, the pointwise constraint $u(t) \in Q \subset \mathbb{R}^m$ can be used. Moreover,
one can consider (in)equality $\int\limits_0^T \| u(t) \|^2 dt \{=, \leq\} E$.   
The final time moment $T$ is either fixed or free. 

Success in finding quantum optimal control depends on both the art of formulating the optimization model
and the art of applying optimization methods. For OCPs for quantum systems
the following optimization methods are used: a)~methods which operate in 
the control functional space, in particular, methods based on Pontryagin maximum principle and Krotov 
method; b)~methods based on reducing OCP to a finite-dimensional
optimization problem via some parametrization of the control
(for example, GRAPE uses piecewise-constant functions,
CRAB uses trigonometric functions);
c)~hybrid methods (e.g., combination of reduction to a final-dimensional optimization 
problem and the Krotov method). The efficiency 
of solving an OCP depends, in particular, on the quality of numerical solving ODEs and PDEs.

The Krotov method has been widely used for quantum control. 
This method was applied to control of molecular dynamics, realization of quantum gates, 
control of Bose-Einstein condensate, control of nuclear magnetic resonance, etc.

The efficiency of the Krotov method depends on 
the specific of an OCP, the way for defining 
the improving function $\varphi^{(k)}$, regularizers, and adjusting the method's parameters. For example, if the potential $V$ 
depends nonlinearly on $u(t)$ in the Gross--Pitaevskii equation (see Section 6), 
then usage of the Krotov method is more difficult 
in comparison to the linear case. The function $\varphi^{(k)}$ 
can be linear  or linear-quadratic with respect to the quantum state. If the function $\Sigma^{(k)}$ is defined using 
the formula (\ref{ch3_f26}), then it is needed to adjust the parameters 
$\alpha, \beta < 0$ and $\gamma>0$ for providing the control improvement. 

The Krotov method is a {\it nonlocal improvement method}, which in contrast to the methods of 
{\it local improvements} (e.g., steepest descent, 
conditional gradient, etc.) are not limited to small control variations and
do not contain a cost expensive procedure to find the best parametrized variation. 
In the articles~\cite{Tannor_Kazakov_Orlov_1992, Somloi_Kazakov_Tannor_article_1993} 
(D.J.~Tannor, V.A.~Kazakov, V.N.~Orlov, J.~Soml\'{o}i) from the early 1990s, the Krotov method with linear function $\varphi^{(k)}$
was applied for the OCP (\ref{ch2_f1}) --- (\ref{ch2_f4}) with 
$\lambda_{\psi} = 0$, and was shown to give macrosteps for control improvements that
can be faster than the steepest descent method. 
Ho---Rabitz TBQCP method~\cite{Ho_Rabitz_PhysRev_2010} 
can be faster than the 1st order Krotov method.  

In a number of publications, including~\cite{Sklarz_Tannor_article_2002,
Reich_Ndong_Koch_article_2012, Goerz_Whaley_Koch_article_2015}, the Krotov method is used with a linear-quadratic 
function $\varphi^{(k)}$ where $\Sigma^{(k)}$ is defined by
(\ref{ch3_f26}). At the same time, in the 
article~\cite{Krotov_Feldman_IzvAN_article_1983}
(V.F.~Krotov, I.N.~Feldman, 1983) the 
function $\Sigma^{(k)}$ is defined as the solution of some special Cauchy problem (\ref{ch3_f28}), 
(\ref{ch3_f29}). In the paper~\cite{Sklarz_Tannor_article_2002} 
(S.E.~Sklarz, D.J.~Tannor), both methods for determining $\Sigma^{(k)}$ are mentioned, although the
computations use the method with the formula (\ref{ch3_f26}).
As it is mentioned in~\cite{Konnov_Krotov_article_1999} for an OCP
beyond quantum control, the method with (\ref{ch3_f28}), 
(\ref{ch3_f29}) is more cost expensive than the version with (\ref{ch3_f26}),
but the last variant can be less effective. Thus, for quantum OCPs
it is important to compare the efficiency of both methods.

Closed quantum systems are approximations for real quantum systems, and 
in practice there is often unavoidable influence of the external 
environment. The Krotov method has been successfully applied also to 
OCPs for open quantum systems~\cite{Kazakov_Krotov_AiT_article_1987, 
Bartana_Kosloff_Tannor_2001, Goerz_dissertation_Kassel_2015, 
Reich_dissertation_Kassel_2015,
CPKoch_2016_OpenQS, Basilewitsch_Schmidt_Sugny_et_al_2017,
Goerz_Motzoi_Whaley_Koch_2017, Goerz_Jacobs_article_2018, 
Basilewitsch_Marder_Koch_2018}. 
This area of research is of high practical interest and has to be considered elsewhere.

\end{document}